\documentclass[a4paper, 12pt]{article}

\usepackage{amsmath}
\usepackage{array}
\usepackage[affil-it]{authblk}
\usepackage[english]{babel}
\usepackage{bookmark}
\usepackage{booktabs}
\usepackage{caption}
\usepackage{cleveref}
\usepackage{datetime2}
\usepackage{enumitem}
\usepackage{geometry}
\usepackage{graphicx}
\usepackage{hyperref}
\usepackage[version=4]{mhchem}
\usepackage{multirow}
\usepackage[numbers]{natbib}
\usepackage{nicematrix}
\usepackage{rerunfilecheck}
\usepackage{tablefootnote}
\usepackage{tabularx}
\usepackage{textcomp}
\usepackage{siunitx}

\crefname{figure}{Figure}{Figures}
\crefname{table}{Table}{Tables}
\crefname{equation}{Equation}{Equations}
\DeclareSIUnit\angstrom{\text {Å}}

\begin{document}
\pdfoutput=1

% Main Text
% Main

% Title, Authors and Affiliations
\newpage
\title{Convolutional Neural Networks and Volcano Plots:
Screening and Prediction of Two-Dimensional Single-Atom Catalysts
for CO\textsubscript{2} Reduction Reactions}

\author[1]{Haoyu Yang}
\author[2]{Juanli Zhao}
\author[3]{Qiankun Wang}
\author[2]{Bin Liu}
\author[4]{Wei Luo}
\author[1,*]{Ziqi Sun}
\author[5,*]{Ting Liao}

\affil[1]{School of Chemistry and Physics, Queensland University of Technology,
George Street, Brisbane, QLD 4000, Australia}
\affil[2]{School of Materials Science and Engineering, Shanghai University, Shanghai, China}
\affil[3]{School of Computer Science and Engineering, Southeast University, Nanjing, China}
\affil[4]{College of Materials Science and Engineering, Institute of Functional Materials,
Donghua University, Shanghai, China}
\affil[5]{School of Mechanical Medical and Process Engineering, Queensland University of Technology,
George Street, Brisbane, QLD 4000, Australia}
\affil[*]{Corresponding author: ziqi.sun@qut.edu.au}
\affil[*]{Corresponding author: t3.liao@qut.edu.au}

\date{February 06, 2024}
\maketitle

% Abstract
\newpage
\begin{abstract}
    Single-atom catalysts (SACs) have emerged as pivotal frontiers for catalyzing a myriad of chemical reactions,
    yet the diverse combinations of active elements and support materials, the nature of coordination environments,
    elude traditional methodologies in searching optimal SAC systems with superior catalytic performance.
    Herein, by integrating multi-branch Convolutional Neural Network (CNN) analysis models
    to hybrid descriptor based activity volcano plot,
    two-dimensional (2D) SAC system composed of diverse metallic single atoms anchored on six type of 2D supports,
    including graphitic carbon nitride (g-C\textsubscript{3}N\textsubscript{4}), nitrogen-doped graphene,
    graphene with dual-vacancy, black phosphorous, boron nitride (BN), and C\textsubscript{2}N,
    are screened for efficient CO\textsubscript{2}RR.
    Starting from establishing a correlation map between the adsorption energies of
    intermediates and diverse electronic and elementary descriptors,
    sole singular descriptor lost magic to predict catalytic activity, including d-band center.
    Deep learning method utilizing multi-branch CNN model therefore was employed,
    using 2D electronic density of states (eDOS) as input to predict adsorption energies.
    Hybrid-descriptor enveloping both C- and O-types of CO\textsubscript{2}RR intermediates
    was introduced to construct volcano plots and limiting potential periodic table,
    aiming for intuitive screening of catalyst candidates for efficient CO\textsubscript{2} reduction to CH\textsubscript{4}.
    The eDOS occlusion experiments were performed to unravel individual orbital contribution to adsorption energy.
    To explore the electronic scale principle governing practical engineering catalytic CO\textsubscript{2}RR activity,
    orbitalwise eDOS shifting experiments based on CNN model were employed.
    The study involves examining the adsorption energy and,
    consequently, catalytic activities while varying supported single atoms.
    This work offers a tangible framework to inform both theoretical screening and experimental synthesis,
    thereby paving the way for systematically designing efficient SACs.
\end{abstract}

% Introduction
\newpage
\section{Introduction}

    % Figure 1: Catalyst Performance Analysis Pipeline
    \begin{figure}[htbp]
        \centering
        \includegraphics[width=\textwidth]{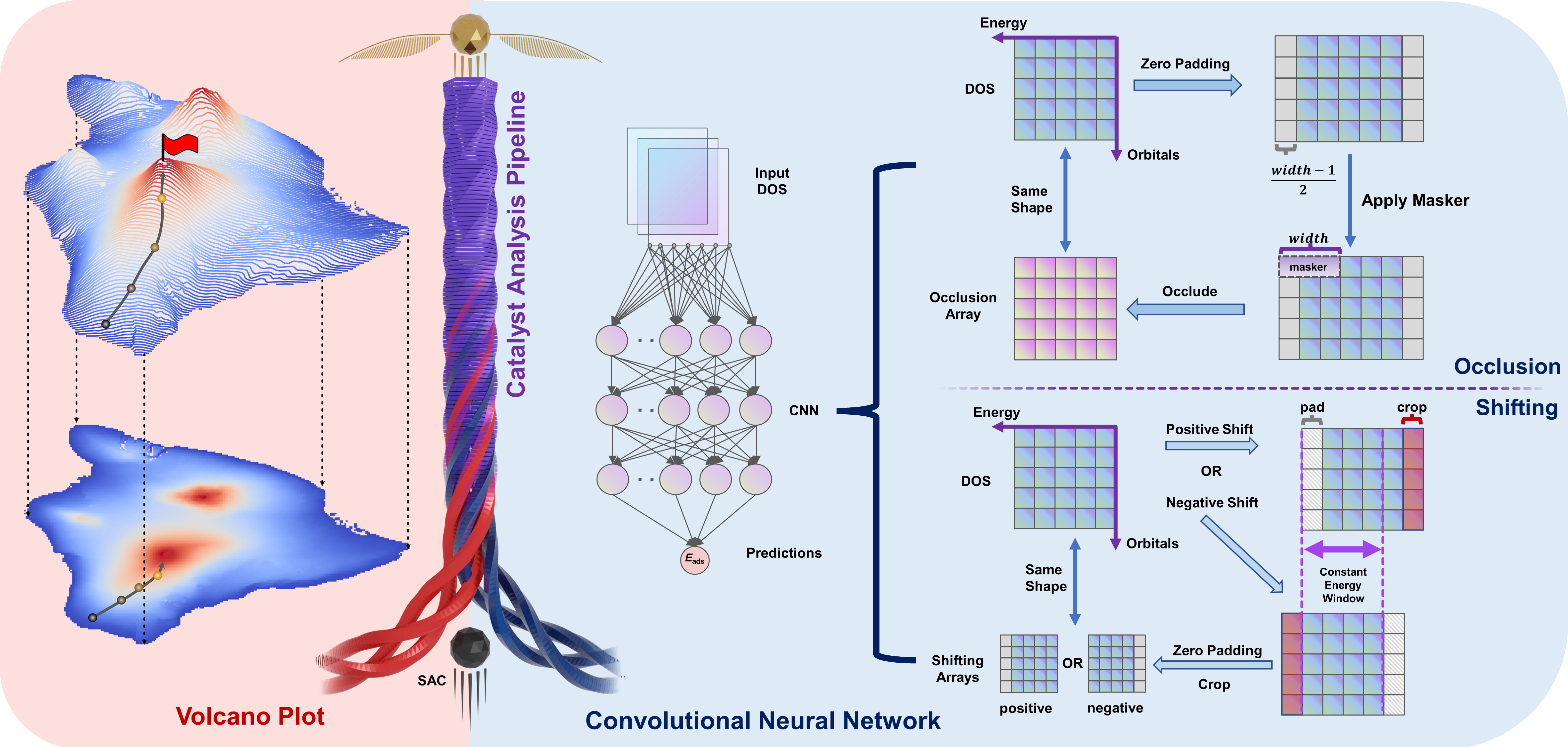}
        \caption{\textbf{Catalyst performance analysis pipeline integrating volcano plots and CNN.}
        The figure demonstrates the integrated usage of volcano plots for
        predictive assessment of existing catalysts, and the utility of CNNs
        to predict and modulate adsorption energies \(E_{\text{ads}}\) using eDOS as input.
        eDOS occlusion and shifting experiments are conducted to extract chemical information from CNN model.}
        \label{main_fig1:pipeline}
    \end{figure}

    Electrochemical CO\textsubscript{2}RR stands as a promising avenue for long-term
    seasonal energy storage \cite{dinh2018co2} and mitigating excess CO\textsubscript{2} in the atmosphere.

    Despite decades of research, metallic copper remains a leading CO\textsubscript{2}RR electrocatalyst
    for hydrocarbon production \cite{osella2023co2}, albeit burdened by substantial overpotential
    and low faradic efficiency and poor operational stability \cite{chen2019identifying, liu2021co2}.
    In this regard, supported SACs have displayed remarkable activity and selectivity
    across various catalytic reactions \cite{wang2018heterogeneous, yang2018atomically},
    leveraging their unique electronic properties and coordination environment.
    To date, experimental studies have primarily focused on two-electron reduction products
    due to the facile release of gaseous CO \cite{cai2021insights, ju2017understanding, ren2019isolated}.
    Consequently, there is a pressing need to modulate the electronic structure of single-atom sites
    to enable hydrocarbon production at reduced overpotential.

    The challenge towards effective catalysts design lies in understanding the reaction mechanisms.
    Experimental determination of these mechanisms remains challenging due to
    the complexity of in situ spectroscopic detection of reaction intermediates \cite{zhao2021revisiting}.
    Computational modeling, particularly methods based on quantum mechanics (QM) such as density functional theory (DFT),
    allows for the profiling the energetics of intermediates and unveils structure-activity relationships
    for identified active sites \cite{feaster2017understanding, carter2008challenges},
    contributing to understanding of reaction mechanisms.
    However, the high computational cost of QM methods restricts the
    accessible catalyst space \cite{jinnouchi2017predicting, cuenya2015nanocatalysis, goldsmith2018machine}.

    In this context, researchers have explored descriptor-based methods
    to circumvent QM calculations and establish direct correlations
    between target properties and electronic factors, commonly referred to as descriptors or features.
    Traditional methods of crafting descriptors, as seen in the d-band model of Hammer and Nørskov \cite{hammer1995electronic},
    necessitate meticulous selection of chemical properties based on chemical knowledge
    and are inherently case-dependent \cite{kajita2017universal}.
    Particularly in the case of SACs, the relationship is intricate,
    rendering manual crafting inadequate \cite{han2021single, thirumalai2018investigating}.
    To streamline this featurization process, we employed CNNs,
    which autonomously derive a hierarchical representation directly from the eDOS
    \cite{tran2015learning, socher2012convolutional, NIPS2012_c399862d}.
    This approach accurately maps to the activity space, negating the need for manual crafting of descriptors
    and providing a more efficient method to understand and
    establish the structure-mechanism-activity relationship for electrochemical reactions.

    In this study, we introduced a catalyst performance analysis pipeline (\cref{main_fig1:pipeline})
    integrating a customized CNN model and volcano plots.
    Our CNN model accurately estimated the adsorption energies, the activity descriptor,
    from the eDOS of 2D materials supported single metal catalysts.
    Evaluations on the adsorption energies of nine intermediates along CO\textsubscript{2}RR
    and competing hydrogen evolution reaction (HER) as well
    yielded prediction mean absolute errors (MAEs) at the level of 0.1 eV,
    which is well aligned with the precision of conventional DFT calculations
    \cite{kirklin2015open, lejaeghere2016reproducibility, wellendorff2015benchmark}.
    This result validated the capability of current CNN model to accurately capture intricate spatial information from eDOS.
    Building on this, we introduced an orbitalwise eDOS occlusion experiment was adopted afterwards
    inspired from the occlusion technique widely used in computer vision and pattern recognition applications.
    Systematic eDOS experiment was performed to elucidate orbitalwise contributions to
    the adsorption energies at different energy levels,
    which produced consistent outcome from Crystal Orbital Hamilton Population (COHP) analysis
    and real-space wavefunction visualization.
    All these results justified the proficiency of CNN model on identifying
    the dominant orbitals for adsorbate-substrate interactions.
    Furthermore, orbital-wise eDOS shifting experiment, exploring the impact of orbital-wise eDOS shifting
    on the adsorption behavior of studied catalysts were performed.

    Simultaneously, we incorporated the volcano plots into the catalysis performance analysis pipeline.
    Besides offering intuitive analysis of catalyst activities (represented by the elevation of volcanoes),
    volcano plots points to the directions on how effort can be exerted to identify promising catalyst candidates.
    Built upon scaling relations \cite{peterson2012activity}, we introduced a hybrid-descriptor scheme
    incorporating both C- and O-centered species.
    The scaling coefficient of determination ($R^2$) was increased by 0.1387 using hybrid descriptor,
    which demonstrate an improved prediction accuracy compared to the well documented single descriptor based method.

    In summary, we introduced a catalyst performance analysis pipeline by
    integrating CNN and volcano plot analysis, using CO\textsubscript{2}RR as an example proof of concept.
    The CNN, which correlates eDOS with activity descriptors, establishes a robust electronic structure-activity relationship.
    Notably, it is capable to predict the influence of eDOS disturbances,
    caused by either alloying, strain, or defects \cite{norskov2011density, bera2017density, kusada2019emergence},
    to modulate the catalytic activity of materials,
    which will offer insights into the intrinsic mechanism for rationalized catalyst optimization strategies.
    Advancing the current volcano plot method,
    the involvement of CNN models in catalyst performance analysis pipeline holds enormous promise
    for advanced exploration of catalysts with superior performance.

\newpage
\section{Results}

% Results Part One: Adsorption energy prediction from eDOS with CNN
\subsection{Adsorption energy prediction from electronic density of states with convolutional neural network}

    % Figure 2: CNN for eDOS
    \begin{figure}[htbp]
        \centering
        \includegraphics[width=\textwidth]{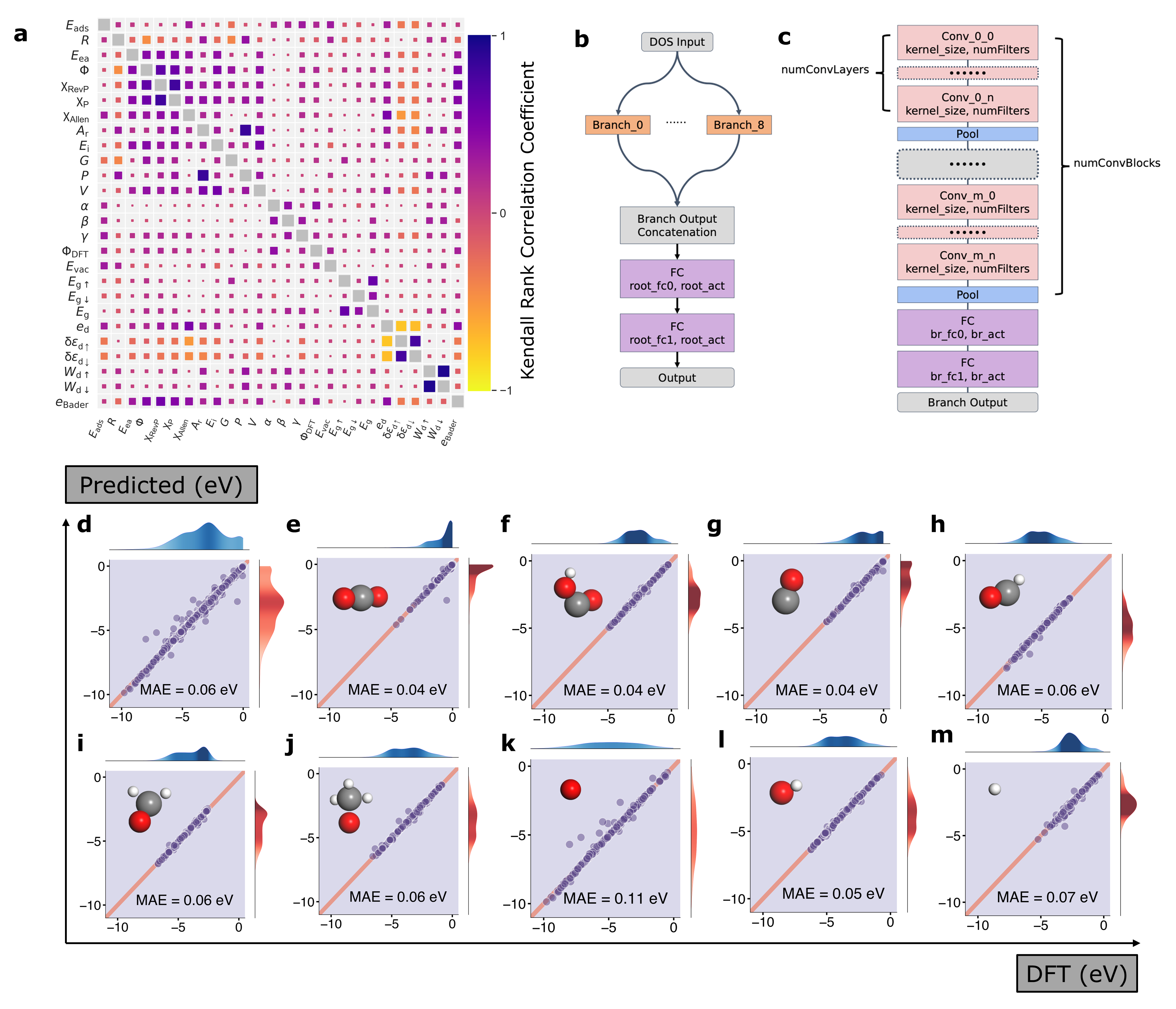}
        \caption{\textbf{CNN for electronic density of states.}
        \textbf{a}. Kendall rank correlation coefficient map illustrating the correlation
        between adsorption energy and other electronic and elementary descriptors.
        Definitions of the notations are provided in
        Supplementary \cref{supp_table15:descriptor_notations}.
        \textbf{b}-\textbf{c}. Architectural of (\textbf{b}) the root and
        (\textbf{c}) each branch of the CNN model.
        The architecture stays consistent across individual branches.
        \textbf{d}-\textbf{m}. Parity plots demonstrating the performance of the
        CNN model across nine adsorbates:
        (\textbf{d}) all adsorbates, (\textbf{e}) CO\textsubscript{2}, (\textbf{f}) COOH,
        (\textbf{g}) CO, (\textbf{h}) CHO, (\textbf{i}) CH\textsubscript{2}O,
        (\textbf{j}) OCH\textsubscript{3}, (\textbf{k}) O, (\textbf{l}) OH, and (\textbf{m}) H.
        The augmentation data are not displayed in these plots, nor included in MAE evaluations.
        Kernel density estimate results are shown along the top and right sides of the main plot,
        representing DFT calculations and CNN model prediction distributions, respectively.}
        \label{main_fig2:cnn_for_eads}
    \end{figure}

\begin{figure}[htbp]
  \centering
  \includegraphics[width=\textwidth]{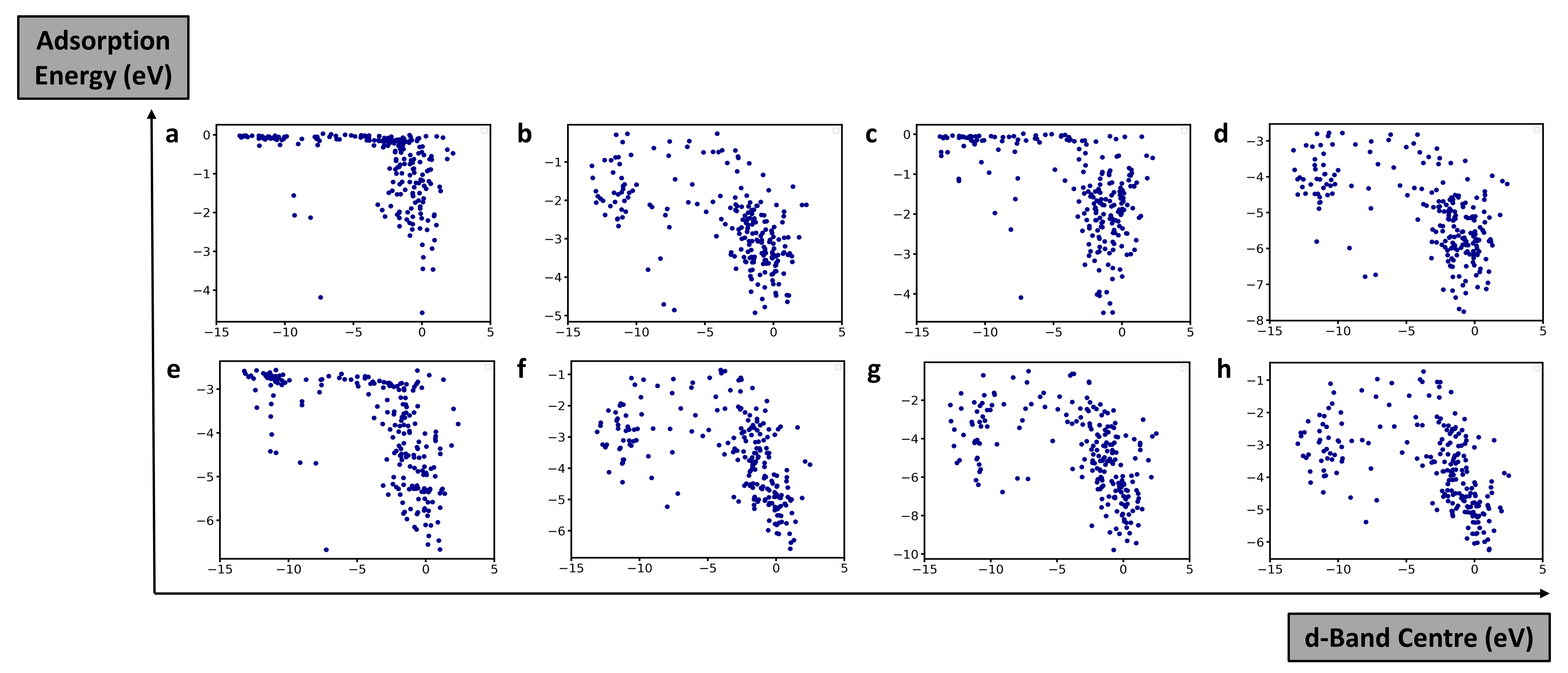}
  \caption{\textbf{Scatter plot of d-Band Center and Adsorption Energy Relationships.}
  This figure delineates the relationship between d-band center and adsorption energies for
  various species: (\textbf{a}) CO\textsubscript{2}, (\textbf{b}) COOH, (\textbf{c}) CO,
  (\textbf{d}) CHO, (\textbf{e}) CH\textsubscript{2}O, (\textbf{f}) OCH\textsubscript{3},
  (\textbf{g}) O, and (\textbf{h}) OH.
  Data from all six substrates investigated in this study are included.}
  \label{supp_fig18:dband_vs_eads}
\end{figure}

    Originating from the pioneering work of Hammer and Nørskov \cite{hammer2000theoretical},
    the d-band center model has emerged as a successful paradigm for employing descriptor to understand and predict the adsorption behavior on transition metal surfaces.
    To investigate alternative electronic and elementary descriptors that may serve as more suitable substitutes for the d-band center model in the context of SACs,
    we performed Kendall rank correlation analysis to account for both linearity and non-linearity \cite{kendall1938new}.
    The correlation coefficient map, as presented in \cref{main_fig2:cnn_for_eads}a, highlighted the d-band center as the most informative descriptor with a Kendall's $\tau$ coefficient of -0.33, and followed by the vacuum level and electronegativity.
    However, d-band center alone does not suffice for accurate predictions of intermediate adsorption energy as illustrated in the scatter plot Supplementary \cref{supp_fig18:dband_vs_eads}, and therefore, refinements are imperative.
    Unlike bulk transition metal surfaces, where the variations in the d-band centre account for the majority of the adsorption energy variations \cite{norskov2011density, takigawa2016machine}, the unique structure of SACs introduces more variables rather than d-band center that are intrinsically correlated with the variations of adsorption energy.
    This finding also interpreted that in some reported prior studies, indicating that depending solely on the d-band center may be insufficient for accurate adsorption energy predictions in the context of SACs \cite{sun2022going, fung2020descriptors, di2022universal, yuan2020descriptor, huang2020rational}.
    As a result, the development of a method tailored specifically for catalysts like SACs becomes essential.

    In our study, we employ CNN, a renowned deep learning method proficient in capturing spatial hierarchy from matrix-like data, such as the eDOS.
    Notably, CNN's autonomy from prior knowledge and human intervention during feature extraction renders it more efficient than traditional machine-learning methods.
    Our approach harnesses the entire 2D eDOS with a multi-branch CNN model.
    \cref{main_fig2:cnn_for_eads}b and \cref{main_fig2:cnn_for_eads}c delineate the model and its branches.
    The workflow involves processing the entire 2D eDOS as input, segregating orbitals into branches, and directing the concatenated outputs through another convolutional block, which generates the adsorption energy prediction.
    Sharing learnable parameters across branches optimizes computational efficiency.
    By leveraging the CNN model, we predicted adsorption energies from the eDOS of supported single atoms, and validated through parity plots (\cref{main_fig2:cnn_for_eads}d-m).
    The model predicts adsorption energies across nine species within the CO\textsubscript{2}RR process and competing HER side reaction, attaining an averaged MAE of 0.06 eV.
    This MAE aligns with the established precision of DFT methods, which typically range between 0.1-0.2 eV concerning experimental measurement \cite{wellendorff2015benchmark, kirklin2015open, lejaeghere2016reproducibility}.
    The full list of prediction errors is available in Supplementary \cref{supp_table18:cnn_mae}.
    These results confirm the CNN model's ability to predict adsorption energies from eDOS input.
    Moreover, they suggest the potential for predicting adsorption energies for SACs using eDOS, although this would demand more sophisticated feature engineering techniques.

    In terms of adsorption energy prediction, the CNN model maintained consistency across diverse elemental compositions and varied adsorption strengths.
    It handled species containing oxygen, hydrogen, and carbon atoms equally well, spanning an adsorption energy range from 0 to 10 eV.
    Significantly, after training the CNN model, making predictions (inferences) scales only linearly with system size and demands considerably less computational resources compared to DFT calculations \cite{chandrasekaran2019solving}.
    Conversely, unlike QM based methods, which could generate reasonably precise energy estimation for individual candidates, CNN relies on extensive datasets to ensure reliable predictions.
    In our case, the original dataset comprised 2052 samples, leading to a CNN model with a validation MAE of 0.3725 eV.
    In pursuit of enhanced prediction accuracy, we proposed a data augmentation method detailed in the Methods section.
    This augmentation resulted in an expanded dataset of 12312 samples, enhancing prediction performance with a reduced validation MAE of 0.1736 eV.
    Comparatively, our CNN model diverges from prior attempts, notably Fung's pioneer work \cite{fung2021machine}, by requiring only the eDOS of supported single metal atom alone and adsorbate.
    This revision eliminated the necessity for a complete set of substrate and adsorbate eDOS and significantly reduced computational cost.
    We foresee the potential in the applicability of our method to diverse species and electrocatalytic reactions, given the availability of comprehensive datasets.
    The strength of our CNN model lies in its freedom from element- or adsorbate-specific parameters, relying solely on eDOS as input.
    This adaptability opens up the possibility for its utilization in a wide array of electrocatalytic studies.

% Results Part Two: Hybrid descriptors enabled volcano plot for catalyst analysis
\subsection{Hybrid descriptors enabled volcano plot for catalyst analysis}

    % Figure 3: Limiting Potential Volcano Plot and Periodic Table
    \begin{figure}[htbp]
        \centering
        \includegraphics[width=0.9\textwidth]{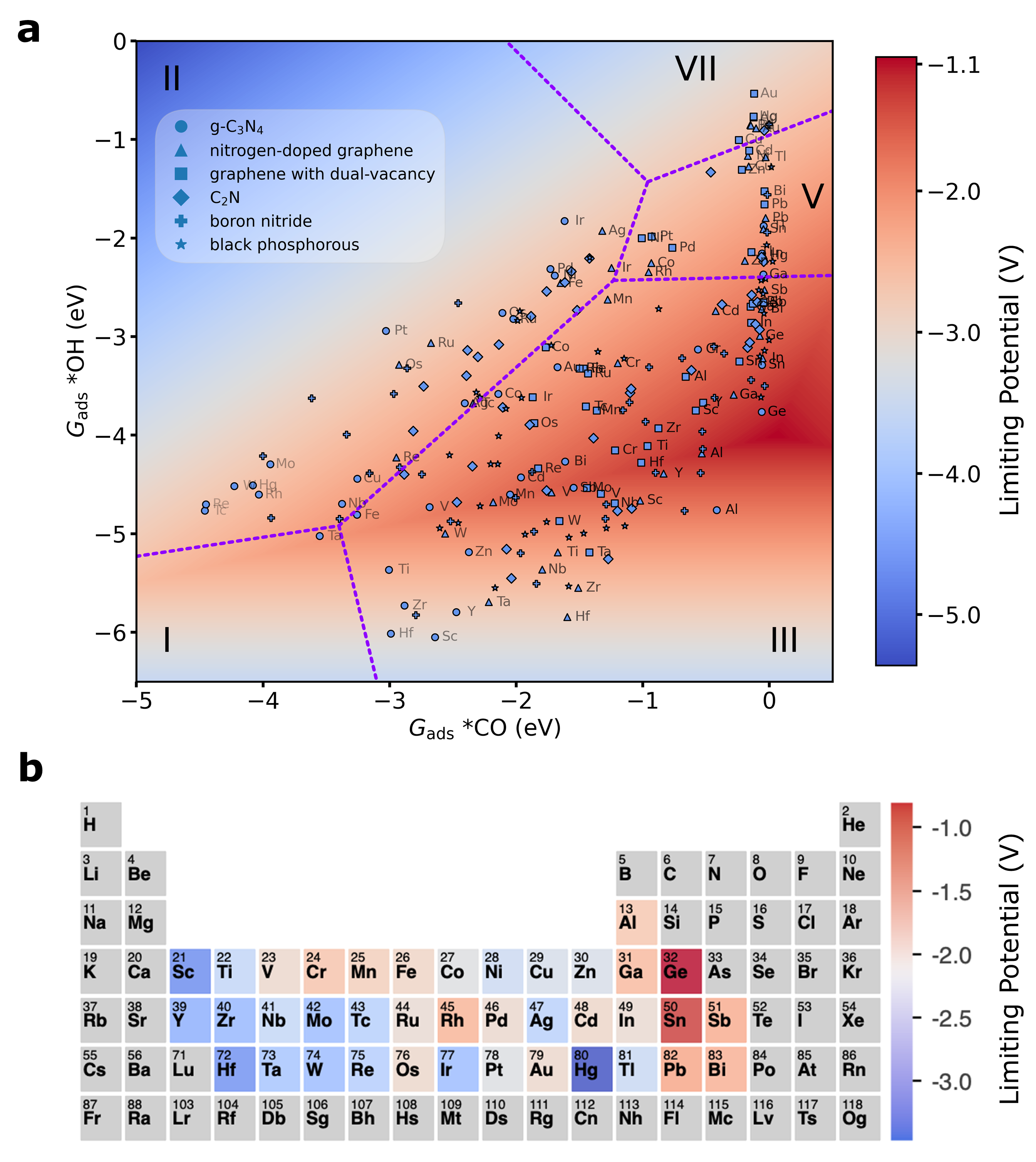}
        \caption{\textbf{Limiting potential volcano plot and periodic table.}
        \textbf{a}. Limiting potential volcano plot computed from hybrid-descriptor-based
        linear scaling relations at potential of -0.17 V vs SHE.
        In the plot, the circle, triangle, square, diamond, plus, and star symbols represent
        metals supported on g-C\textsubscript{3}N\textsubscript{4}, nitrogen-doped graphene,
        graphene with dual-vacancy, black phosphorous, BN, and C\textsubscript{2}N, respectively.
        Dotted purple lines delineate regions representing rate-determining steps,
        with roman numerals denoting the respective RDS for each region.
        The reaction steps are defined in \cref{main_eq1:co2rr1} to \cref{main_eq8:co2rr8}.
        \textbf{b}. Limiting potential periodic table illustrating the theoretical activity of
        single metal atom catalysts supported on g-C\textsubscript{3}N\textsubscript{4}.}
        \label{main_fig3:volcano}
    \end{figure}

    Catalyst activity volcano plots, built upon the scaling relations proposed by Peterson and Nørskov's \cite{peterson2012activity},
    goes beyond assessment of candidatures' energetics only but to provide informative guidance on discovering catalysts candidates with superior performance \cite{balandin1969modern, deutschmann2000heterogeneous}.
    Our endeavor to improve the accuracy of volcano plots leads to the development of the hybrid-descriptor,
    which encompasses both C- and O- bonding types of CO\textsubscript{2}RR intermediates within the scaling relations.
    This adaptation arose from the intuition that most CO\textsubscript{2}RR intermediates are featured with both bonding types rather than any single one in dominance.
    By integrating both bonding featured descriptors into the hybrid approach, we found the $R^2$ was increased by 0.1387.
    Detailed information about the scaling parameters and performance enhancements are provided in Supplementary \cref{supp_table13:scaling_params} and Supplementary \cref{supp_fig10:r2_hyb_des}.
    This improvement to some extent underscored our premise that CO\textsubscript{2}RR intermediates typically exhibit chemical similarities encompassing both O- and C-type descriptors.
    After taking this dual nature into consideration, our hybrid-descriptor method can evaluate the intermediates adsorption energies more accurately.
    Importantly, the introduction of hybrid descriptors is computationally affordable compared to DFT methods,
    as it involves only a set of linear regressions to identify the optimal mixing ratio, as elaborated in the Methods section.

    Expanding upon the revised scaling relations, we introduce an activity volcano plot and a rate-determining step (RDS) volcano plot for the CO\textsubscript{2}RR process, shown in \cref{main_fig3:volcano}a and Supplementary \cref{supp_fig11:rds_volcano}.
    Given that HER is a key competing reaction with CO\textsubscript{2}RR \cite{goyal2020competition}, we also incorporated a selectivity volcano plot, presented in Supplementary \cref{supp_fig12:sel_volc}, to offer a comprehensive analysis of catalyst candidates.
    In adherence to Peterson and Nørskov's original framework \cite{peterson2012activity}, our choice of x and y axes descriptors aligned with $\textit{G}_{\text{ads} \ast \text{CO}}$ and $\textit{G}_{\text{ads} \ast \text{OH}}$.
    Further elaboration on how the limiting potentials were evaluated across the volcano plot can be found in the Methods section.
    Among the candidatures explored, Ge@g-C\textsubscript{3}N\textsubscript{4} emerged as the most promising catalyst, demonstrating theoretical activity at a limiting potential of -1.2126 eV.
    Notably, the observed selectivity trend paralleled the activity trend, suggesting that catalysts exhibiting high theoretical activity are also resistant to HER side reactions.
    The RDS volcano plot (Supplementary \cref{supp_fig11:rds_volcano}) indicated protonation of *CO as the rate-determining step for nearly optimal candidates.
    Moreover, the activity volcano plot predicted an optimal catalyst featuring a theoretical limiting potential of -1.0516 V, at $\textit{G}_{\text{ads} \ast \text{CO}}$ of 0.0589 eV and $\textit{G}_{\text{ads} \ast \text{OH}}$ of -4.0075 eV, positioned at the right-center of the volcano plot.
    Results from RDS volcano plot indicate that protonation of *CO likely serves as the RDS for CO\textsubscript{2}RR.
    Consequently, it should be the focal point for future catalyst designs.
    The activity and selectivity volcano plots imply that enhancing theoretical activity could be achieved by decreasing CO affinity while increasing OH affinity with the catalysts.
    This result suggests a perspective: weak binding of CO to the catalyst surface might facilitate the rotation of C-O backbone from the ``upright'' configuration in *CO to the ``horizontal'' orientation in *CHO \cite{peterson2010copper},
    and therefore facilitating dynamics and protonation of the *CO species.

    The volcano plot offers an intuitive analysis of the reaction energetics and provides insights into the mechanisms of the reaction.
    While the volcano plot provides a broad overview, it lacks the ability to delve into lower-level explanations rooted in electronic structures, where detailed electronic interactions and specific bonding mechanisms remain beyond its scope.
    Despite its limitations, the volcano plot serves as a useful tool for identifying focal elementary steps in the reaction pathway that demand closer attention during catalyst design.
    It also highlights directions for further catalyst refinement, and can also be integrated with the shifting experiment discussed later to optimize existing catalysts.

\begin{figure}[htbp]
  \centering
  \includegraphics[width=\textwidth]{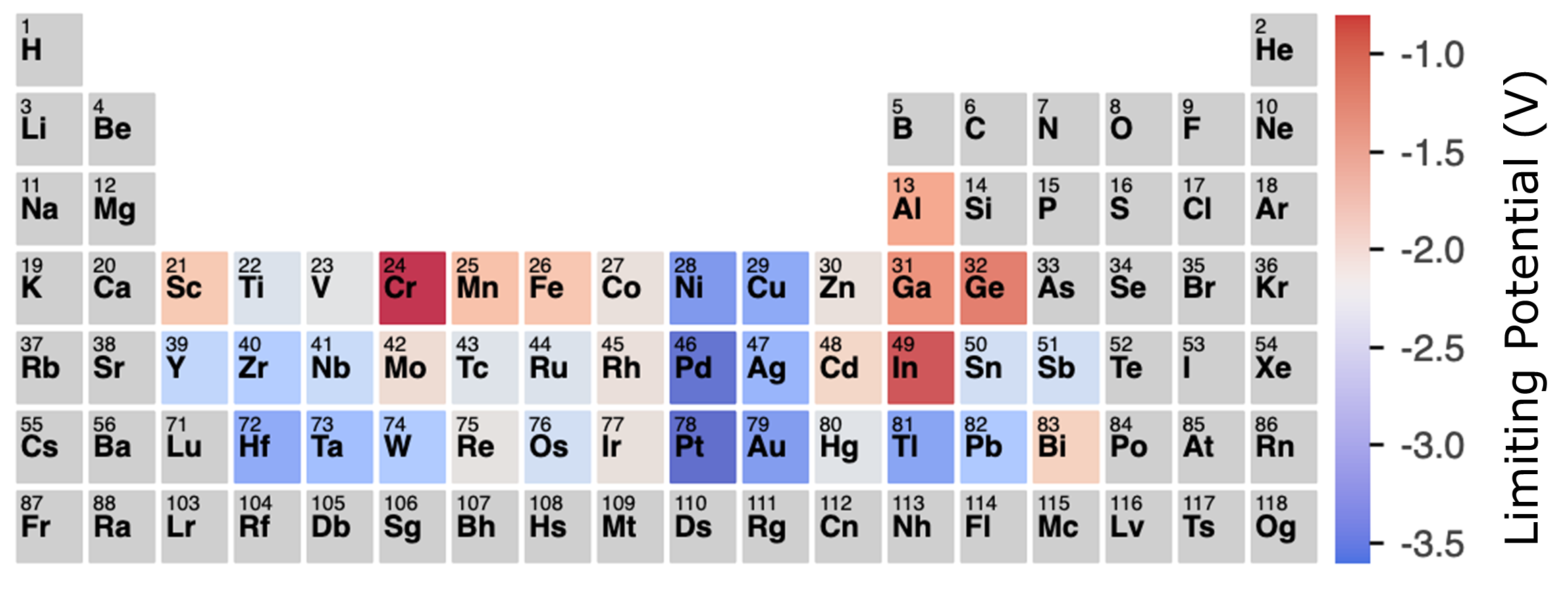}
  \caption{\textbf{Limiting potential periodic table for nitrogen-doped graphene.}}
  \label{supp_fig13:n-gra_ptable}
\end{figure}

\begin{figure}[htbp]
  \centering
  \includegraphics[width=\textwidth]{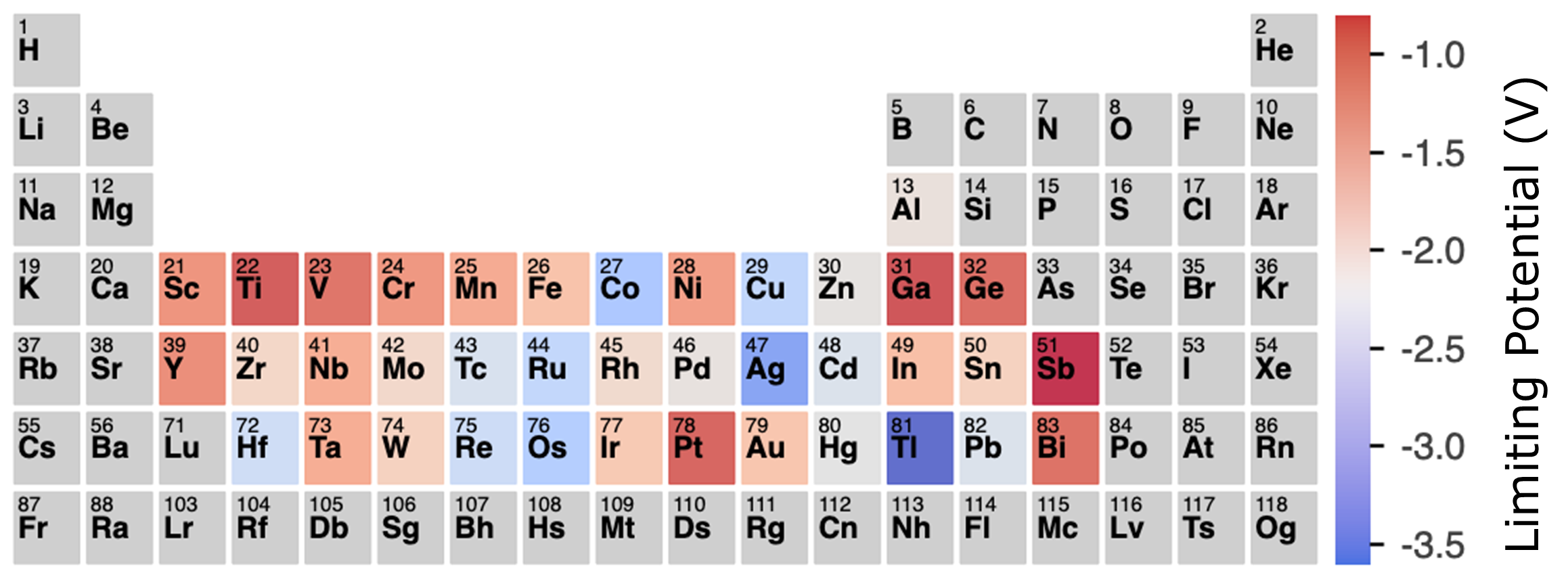}
  \caption{\textbf{Limiting potential periodic table for monolayer C\textsubscript{2}N.}}
  \label{supp_fig17:C2N_ptable}
\end{figure}

    Additionally, we mapped the theoretical limiting potentials of the analyzed catalyst candidates onto periodic tables, shown in \cref{main_fig3:volcano}b and Supplementary \cref{supp_fig13:n-gra_ptable} to Supplementary \cref{supp_fig17:C2N_ptable}.
    Our analyses do not reveal consistent trends within groups or periods, nor did they exhibit evident patterns across substrates.
    These observations suggest intricate interactions between supported metal atoms and substrates.
    The absence of straightforward trends underscores the need for nuanced explorations in understanding these interactions like the CNN model.

% Results Part Three: Validation of CNN predictions
\subsection{Validation of CNN predictions}

    % Figure 4: Validation of CNN Predictions
    \begin{figure}[htbp]
        \centering
        \includegraphics[width=\textwidth]{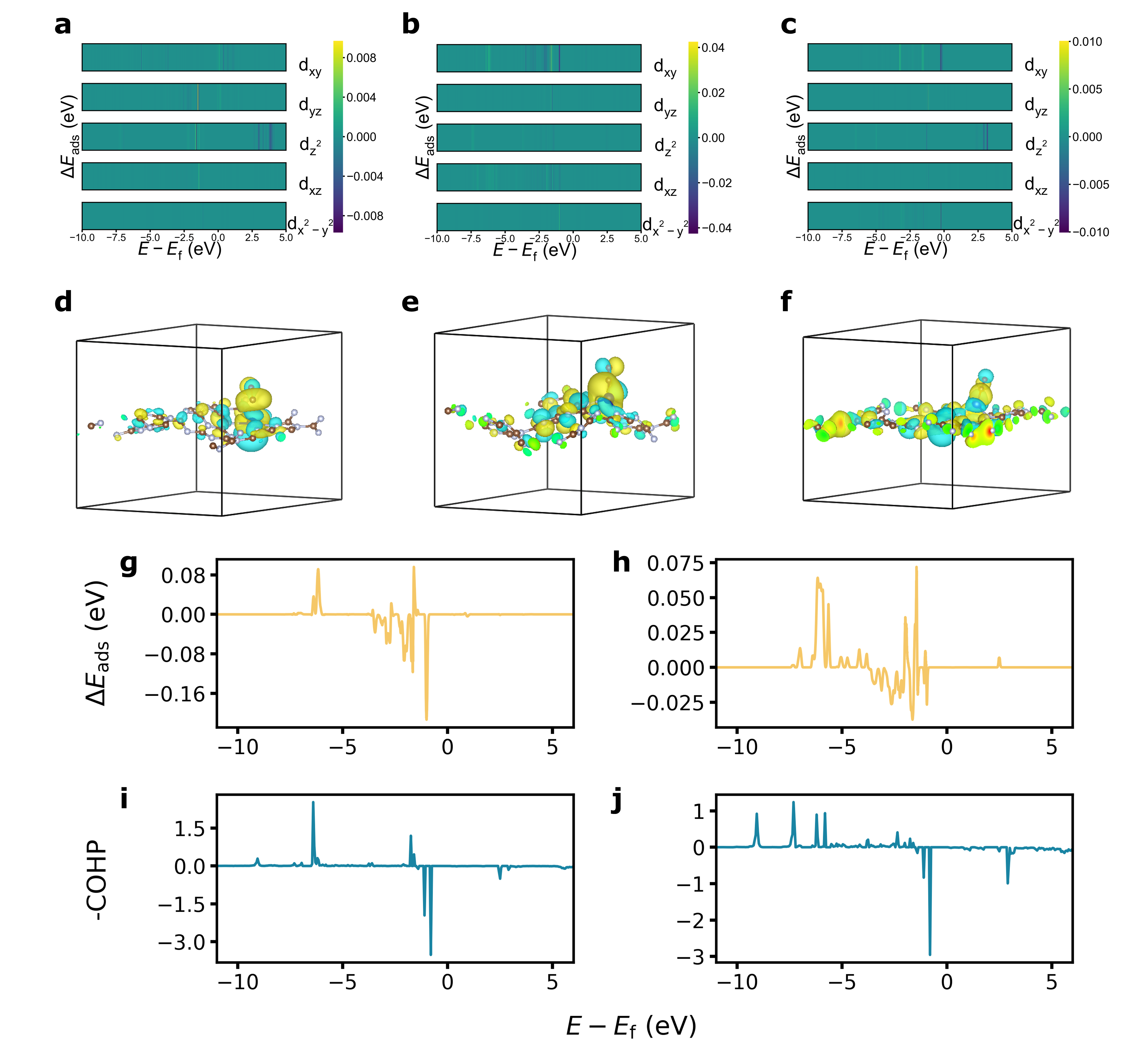}
        \caption{\textbf{Validation of CNN predictions.}
        \textbf{a}-\textbf{c}, The eDOS occlusion experiment results for
        (\textbf{a}) Cr of Cr@g-C\textsubscript{3}N\textsubscript{4},
        (\textbf{b}) Co of Co@g-C\textsubscript{3}N\textsubscript{4} and
        (\textbf{c}) Cu of Cu@g-C\textsubscript{3}N\textsubscript{4}.
        eDOS occlusion experiments were conducted on the initial states of CO adsorption process.
        \textbf{d}-\textbf{f}, Real-space wavefunction visualization for
        (\textbf{d}) Cr $d_{x^2-y^2}$ orbital of Cr@g-C\textsubscript{3}N\textsubscript{4},
        (\textbf{e}) Co $d_{xy}$ orbital of Co@g-C\textsubscript{3}N\textsubscript{4} and
        (\textbf{f}) Cu $d_{xy}$ orbital of Cu@g-C\textsubscript{3}N\textsubscript{4}.
        Wavefunction visualizations were performed on the final states of CO adsorption process.
        The yellow and aqua colors represent areas of positive and negative
        Kohn-Sham orbital coefficients, respectively.
        \textbf{g}-\textbf{j}, Occlusion experiment results for (\textbf{g}) $d_{xy}$ orbital
        and (\textbf{h}) $d_{xz}$ orbital for Co of Co@g-C\textsubscript{3}N\textsubscript{4}.
        COHP analysis for (\textbf{i}) $d_{xy}$ orbital and (\textbf{j}) $d_{xz}$ orbital for
        Co of Co@g-C\textsubscript{3}N\textsubscript{4}, performed between Co and
        all adsorbate atoms in the final states of CO adsorption process.}
        \label{main_fig4:validation}
    \end{figure}

    Historically, interpretability has posed challenges for deep learning methods, often rendering their inner workings akin to black boxes \cite{zhang2018interpretable, zhang2018visual, savage2022breaking}.
    The occlusion technique, widely utilized in Computer Vision \cite{zeiler2014visualizing, kortylewski2020combining, wang2020robust}, involves systematically masking sections of the input matrix to assess their impact on the final prediction score.
    In our work, we introduce an orbitalwise occlusion method to discern individual orbital contribution to adsorption energy within the CNN model, drawing inspiration from Fung's pioneer integration of this technique into eDOS analysis \cite{fung2021machine}.
    In this work, occlusion experiments involve orbitalwise masking specific eDOS, which are then fed into the CNN model to observe resulting disturbance in adsorption energy.
    A visual representation of this process can be found in Supplementary \cref{supp_fig21:occlusion}. Conceptually, this experiment simulates the effect of shielding electronic states, allowing investigation of the potential contribution to adsorption energy from these states.
    Additionally, we attempted to extract the planewave coefficients of Kohn-Sham orbital for wavefunction visualization.
    This validation step was undertaken to confirm the spatial distribution of the orbitals pinpointed by the occlusion experiments.

    To validate our method, we conducted occlusion experiments on Cr@g-C\textsubscript{3}N\textsubscript{4}, Co@g-C\textsubscript{3}N\textsubscript{4} and Cu@g-C\textsubscript{3}N\textsubscript{4}.
    Each of these supported metals possesses 3d subshells with different electronic configurations: half-filled, partially filled and fully filled, allowing us to test the universality of our approach.
    The results of the occlusion experiment on the supported Cr single atom in Cr@g-C\textsubscript{3}N\textsubscript{4} are given in \cref{main_fig4:validation}a.
    The $d_{x^2-y^2}$ orbital emerges as dominant in CO interaction.
    Shielding electrons centered around -2 eV significantly alter the adsorption energy, indicating the potency of this orbital.
    The real-space wavefunction visualization in \cref{main_fig4:validation}d corroborates these findings, highlighting the interaction between the $d_{x^2-y^2}$ orbital of single metal atom and the adsorbate.
    This suggests the $d_{x^2-y^2}$ orbital of supported Co predominantly influence the interaction with CO, and thus shielding electrons from this orbital markedly disturb the adsorption energy.
    Additionally, occlusion experiments were performed on the Co $d_xy$ orbital of Co@g-C\textsubscript{3}N\textsubscript{4} (\cref{main_fig4:validation}b) and the $Cu d_xy$ orbital of Cu@g-C\textsubscript{3}N\textsubscript{4} (\cref{main_fig4:validation}c).
    In both cases, the $d_xy$ orbital was identified as the dominant contributor.
    While identifying the dominant orbital is crucial, these findings also suggest that interactions with adsorbates occurring out-of-plane are influenced by in-plane orbitals as well.
    Importantly, it becomes evident that d orbitals do not universally impact interactions with adsorbates, emphasizing the need for case-specific discussions.

    To validate the chemical relevance of predictions made by CNN model, we conducted COHP analysis, a theoretical bond-detecting tool used to identify bonding and anti-bonding contributions to band-structure energy \cite{deringer2011crystal}.
    In this research, interactions between the supported metal atom and atoms within a 5 \text{\AA} radius were calculated to understand its bonding environment, as detailed in the Methods section.
    The results of the occlusion experiment on the Co $d_xy$ orbital of Co@g-C\textsubscript{3}N\textsubscript{4}, identified as the dominant orbital via occlusion experiments, were compared with the COHP analysis, as shown in \cref{main_fig4:validation}g and \cref{main_fig4:validation}i.
    The occlusion experiment pinpointed strong metal-adsorbate interactions at energy levels of -1 eV and -6.3 eV, which were confirmed by the COHP analysis.
    We then investigated the non-dominant Co $d_xz$ orbital, and the COHP analysis highlighted interactions at -1 eV and -6.5 eV, aligning with the findings from the occlusion experiments.

    In summary, orbitalwise occlusion experiments provide chemical interpretations into deep learning models, enhancing their interpretability.
    Furthermore, these occlusion results are readily understandable to chemists, enabling the identification of orbitalwise contributions to the adsorption process.

% Results Part Four: Shifting experiments for prediction of better catalysts
\subsection{eDOS shifting experiments for prediction of better catalysts}

    % Figure 5: eDOS Shifting Experiments
    \begin{figure}[htbp]
        \centering
        \includegraphics[width=\textwidth]{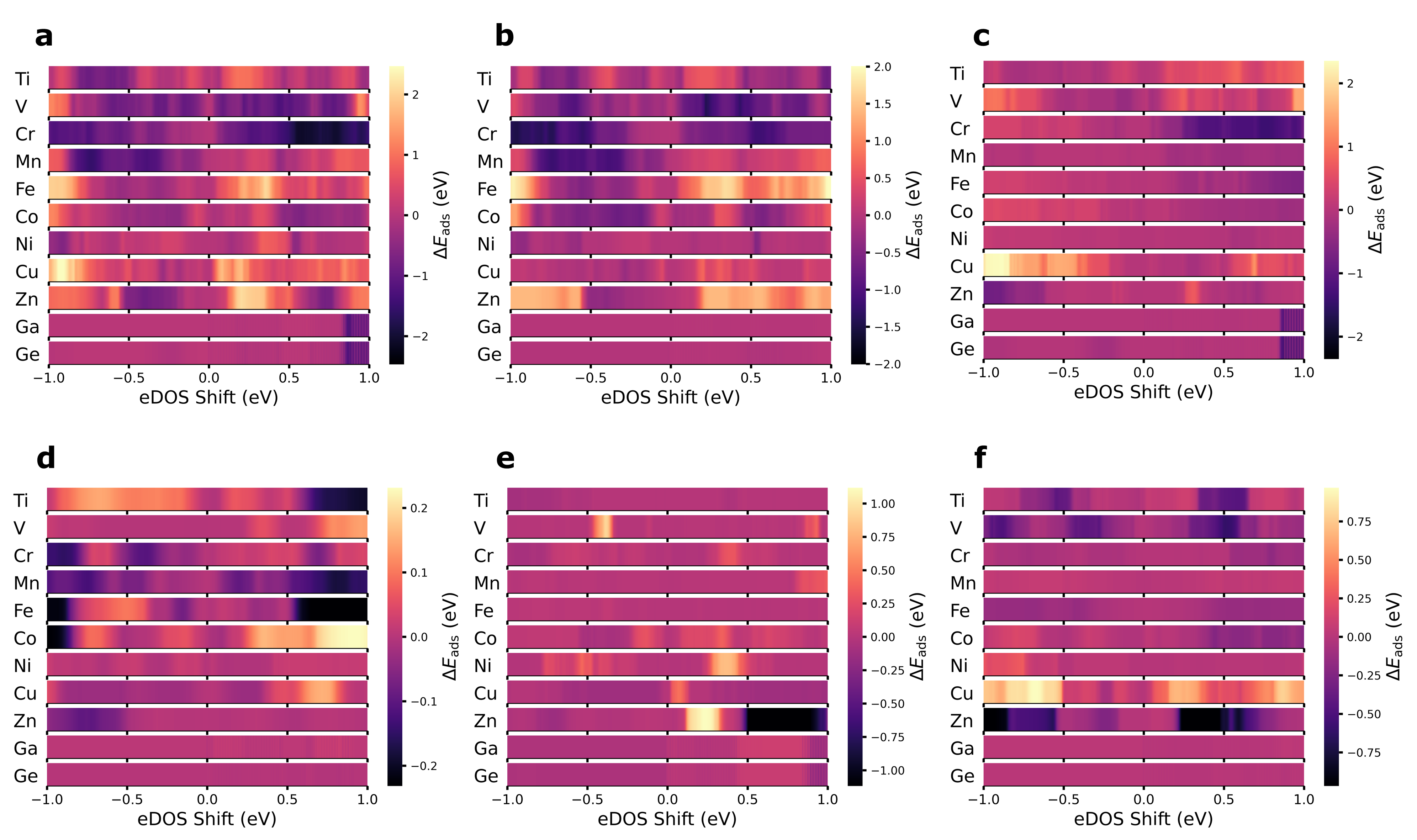}
        \caption{\textbf{eDOS shifting experiments.}
        Impact of orbitalwise eDOS shifting on CO adsorption energy, as predicted by the CNN model,
        for single metal atom catalysts supported on g-C\textsubscript{3}N\textsubscript{4}.
        The disturbances caused by shifting of (\textbf{a}) entire $d$, (\textbf{b}) $d_{xy}$,
        (\textbf{c}) $d_{yz}$, (\textbf{d}) $d_{z^2}$, (\textbf{e}) $d_{xz}$ and
        (\textbf{f}) $d_{x^2-y^2}$ orbitals are illustrated.
        The shifting step size corresponds to the energy resolution of the eDOS,
        and is 0.005 eV in this study.
        A positive eDOS shift indicates a shift towards higher energy levels, and vice versa.}
        \label{main_fig5:shifting}
    \end{figure}

    While CNN models excel in prediction of the activity of catalyst candidates directly from eDOS,
    volcano plots point out directions leading to better catalysts.
    Exploring potential disturbances to adsorption energies and aiming to shift existing catalyst
    candidates closer to the peak of the activity volcano plot presents an intriguing avenue of investigation.
    In this work, we introduce an orbitalwise shifting experiment to explore
    the impact of orbitalwise eDOS shifting on adsorption behavior.
    The experiment involved shifting eDOS along the energy axis,
    and the CNN model was deployed to predict the resultant disturbance
    in adsorption energy for each shifting operation.
    Focusing on all Period 4 metal elements examined in this study, we analyzed the effects of eDOS shifting.
    To induce a moderate disturbance, we selected the energy range of -1 to 1 eV,
    and a shifting resolution of 0.005 eV/step was chosen to align with the eDOS calculation resolution.

    \cref{main_fig5:shifting} illustrates that supported Ga and Ge, the sole p-block metals in our study,
    exhibited no response to d orbital shifting.
    This observation is consistent with their lack of d-electrons in their valence shells,
    reinforcing the reliability of our shifting experiments.
    We also investigated potential disturbance from p orbital shifting,
    as shown in Supplementary \cref{supp_fig31:p_shifting} and found no significant response.
    This implies that Ge@g-C\textsubscript{3}N\textsubscript{4} holds promise as a CO\textsubscript{2}RR electrocatalyst,
    given its high activity and resilience to electronic structure disturbances.
    Nevertheless, fine-tuning its performance through eDOS modulation might pose a challenge.
    In contrast, supported Fe, Cu and Zn display strong sensitivity to eDOS shifting.
    Interestingly, the adsorption energy consistently shifted towards more positive values,
    indicating a weakening of the catalyst-adsorbate interaction,
    regardless of the shifting direction within the investigated energy range.
    This underscores the intricate interplay between the supported metal atom and the substrate,
    emphasizing the necessity of a tool like the CNN model to predict such disturbances.
    Supported Cr exhibited a contrary response to eDOS shifting, strengthening the
    interaction with the CO adsorbate regardless of the shifting direction.
    Supported Ti, V, Co and Ni showed weak response to eDOS shifting,
    suggesting the need for alternative methods beyond manipulating their eDOS
    to regulate their interactions with CO adsorbate.

    In summary, shifting experiments offer a valuable means to predict the potential
    disturbances in adsorption behavior resulting from eDOS shifting,
    a task that is challenging to accomplish directly within the DFT framework.
    However, it's important to note that achieving the predicted shifting
    in actual theoretical or experimental scenarios may not be easily accessible.
    Skillful manipulation of the electronic structure of catalyst candidates would be required,
    presenting a practical challenge.
    In contrast to prior studies on bulk metals \cite{fung2021machine},
    wherein the adsorption energy exhibits a continuous shift with eDOS variations,
    the response of SACs is notably discrete.
    These responses manifest as isolated peaks throughout the shifting range,
    attributed to the distinctive structure of SACs.
    Despite this challenge, having a tool that can predict the potential impact of
    eDOS shifting is advantageous for the theoretical design of SACs.

% Methods
\newpage
\section{Methods}

    \subsection{DFT calculations}
    Spin-polarized density functional theory calculations were performed using
    the Vienna Ab initio Simulation Package (VASP) version 5.4.4 \cite{kresse1996efficient, kresse1996efficiency, kresse1993ab}
    to optimize geometries and determine the electronic structure.
    The Perdew-Burke-Ernzerhof (PBE) functional \cite{kresse1996efficient} within
    the generalized gradient approximation (GGA) \cite{perdew1996generalized}
    was utilized to describe the electron-ion interactions and electronic exchange correlations.
    Kohn-Shan equations were approached employing a plane-wave basis set with a cut-off energy of 440 eV.
    Geometry optimizations were performed using $\Gamma$-Centered K-point meshes of ($2\times2\times1$)
    until the force on each atom fell below \SI{0.02}{\electronvolt\per\angstrom}.
    Electronic structure calculations utilized denser K-point meshes of ($3\times3\times1$)
    and employed self-consistent field (SCF) methods.

    \subsection{Catalyst models}
    Supported SACs were modelled by anchoring individual metal atoms onto various 2D substrates,
    including g-C\textsubscript{3}N\textsubscript{4}, nitrogen-doped graphene,
    graphene with dual-vacancy, black phosphorous, boron nitride, and monolayer C\textsubscript{2}N.
    The specific single metal atoms studied are outlined in Supplementary \cref{supp_fig1:ptable}.
    To prevent interactions between adjacent images along the z-axis, vacuum layers of 15 \text{\AA} were introduced.

    \subsection{CO\texorpdfstring{\textsubscript{2}}{2}RR pathway}
    The C1 production CO\textsubscript{2}RR pathway, as initially described by Peterson et al. \cite{peterson2010copper}
    and later confirmed by Jiao et al. \cite{jiao2017molecular} for
    g-C\textsubscript{3}N\textsubscript{4} based SACs, involves the following elementary steps.
    Asterisks (*) indicate species adsorbed on the SAC surfaces:

    \begin{align}
    \ce{CO2 + H+ + e-          &-> \text{*}COOH}      \label{main_eq1:co2rr1}    \\
    \ce{\text{*}CO + H+ + e-   &-> \text{*}CHO}       \label{main_eq3:co2rr3}    \\
    \ce{\text{*}COOH + H+ + e- &-> \text{*}CO + H2O}  \label{main_eq2:co2rr2}    \\
    \ce{\text{*}CHO + H+ + e-  &-> \text{*}CH2O}      \label{main_eq4:co2rr4}    \\
    \ce{\text{*}CH2O + H+ + e- &-> \text{*}OCH3}      \label{main_eq5:co2rr5}    \\
    \ce{\text{*}OCH3 + H+ + e- &-> \text{*}O + CH4}   \label{main_eq6:co2rr6}    \\
    \ce{\text{*}O + H+ + e-    &-> \text{*}OH}        \label{main_eq7:co2rr7}    \\
    \ce{\text{*}OH + H+ + e-   &-> \text{*} + H2O}    \label{main_eq8:co2rr8}
    \end{align}

    \subsection{Free energy calculations}
    Free energies were computed from electronic energies by considering thermal corrections,
    incorporating normal mode analysis of all degrees of freedom of adsorbates.
    The free energy $\textit{G}$ was calculated using the formula:
    \begin{align}
    G = E + E_{\text{ZPE}} - TS  \label{main_eq9:free_energy}
    \end{align}

    where $\textit{G}$ represents the free energy, $\textit{E}$ denotes electronic energy,
    $\textit{E}_{\text{ZPE}}$ is the zero-point energy, $\textit{T}$ represents temperature in Kevin,
    and $\textit{S}$ stands for entropy.
    Detailed information regarding these corrections can be found in Supplementary \cref{supp_sec2.3_energies}.

    The computational hydrogen electrode (CHE) model \cite{peterson2010copper, norskov2004origin}
    was employed to introduce potential dependence.
    At the reference electrode surface, the reaction
    \begin{align}
    \ce{1/2H2_{(g)}  <-> H+ + e-}  \label{main_eq10:che}
    \end{align}
    is in equilibrium, allowing the calculation of the chemical potential of the electron-proton pair as:
    \begin{align}
    \mu(\mathrm{H}^+) + \mu(e^-) = \frac{1}{2}\mu(\mathrm{H}_{2(\mathrm{g})}) - eU  \label{main_eq11:che_potential}
    \end{align}
    For the external potential $\textit{U}$ applied, the change in free energy ($\Delta \textit{G}$)
    was determined using the equation:
    \begin{align}
    \Delta G_n(U) = \Delta G_n(U=0) + neU  \label{main_eq12:ext_potential_che}
    \end{align}
    Here, $\mu$ represents chemical potential, $\textit{U}$ signifies the external potential applied,
    $\textit{n}$ represents the number of proton-electron pairs transferred,
    and $\textit{e}$ denotes the elementary charge.

    \subsection{Hybrid descriptor for volcano plots}
    Hybrid descriptors incorporate both descriptors within a single equation,
    enhancing the accuracy of scaling relationship with minimal computational overhead,
    as demonstrated in Supplementary \cref{supp_fig10:r2_hyb_des} and Supplementary \cref{supp_table13:scaling_params}.
    Comprehensive analyses are presented Supplementary \cref{supp_sec2.5_scaling}.
    This leads to the formulation of the scaling relation:
    \begin{align}
    G_{\text{ads}} Z =
    k[\theta^* G_{\text{ads}} \ce{CO} + (1-\theta^*) G_{\text{ads}} \ce{OH}] + c_Z  \label{main_eq13:scaling_relation}
    \end{align}
    Here, $\theta$ represents the mixing ratio of the CO descriptor, $\textit{Z}$ denotes any adsorbate,
    and $\textit{k}$ and $\textit{c}_{Z}$ are adsorbate-specific scaling parameters.
    The optimal mixing ratio $\theta$ is determined through iterative exploration across the range,
    utilizing a step length of 0.01.

    \ref{main_eq13:scaling_relation} could be simplified to a more general form:
    \begin{align}
    G_{\text{ads}} Z =
    a_Z G_{\text{ads}} \ce{CO} + b_Z G_{\text{ads}} \ce{OH} + c_Z  \label{main_eq14:general_scaling_relation}
    \end{align}
    Here, $a_Z$, $b_Z$, $c_Z$ are adsorbate-specific parameters.

    Consequently, the limiting potential $\textit{U}_{L}$ is entirely determined
    by two descriptors through the following equation:
    \begin{align}
    U_L = -\frac{\max\{ \Delta G_i \}}{e}  \label{main_eq15:limiting_potential}
    \end{align}
    Where $\Delta G_{i}$ represents the free energy change of reaction step i.

    This approach simplifies the theoretical performance assessment of any catalyst
    within the scope of our research to only two descriptors:
    $G_{\textit{ads} \, \textit{CO}}$ and $G_{\textit{ads} \, \textit{OH}}$.
    This method offers a visual representation of the high-dimensional optimization challenge.
    To aid visualization, a 2D mesh grid is established, enabling vectorized assessments of
    limiting potentials across the descriptor value ranges.
    At each grid point, the limiting potential is computed using \cref{main_eq15:limiting_potential}.

    \subsection{CNN architecture and hyperparameter tuning}
    We designed our CNN architecture drawing inspiration from its successful implementation in
    tasks involving electrocardiogram signal classification \cite{weimann2021transfer},
    as illustrated in \cref{main_fig2:cnn_for_eads}b and \cref{main_fig2:cnn_for_eads}c.
    This multi-branched CNN comprises nine branches tailored to handle eDOS orbitals
    while ensuring a reduced memory footprint.

    The effectiveness of neural network models is intricately tied to the choice of hyperparameters.
    To identify the optimal set of hyperparameters, we employed Hyperband \cite{li2018hyperband},
    a modern and parallelizable bandit-based optimization algorithm.
    The search space for hyperparameters is detailed in Supplementary \cref{supp_table16:hyperparam_space},
    and the resulting optimal hyperparameters are presented in Supplementary \cref{supp_table17:opt_hyperparam}.

    \subsection{Data augmentation and CNN model training}
    To enhance the generalizability and robustness of the our CNN model,
    and achieve comprehensive coverage of the chemical space \cite{DBLP:journals/corr/abs-2112-12542},
    we applied data augmentation techniques to the initial dataset using the following procedures:
        - In addition to the designated ``initial state'' of the adsorption process,
        where the adsorbate positioned 6.5 \text{\AA} above the supported single metal atom,
        five intermediate images were interpolated.
        These images were distributed between the initial and final states at a spacing of 0.5 \text{\AA},
        starting from the initial state side.
        - A diverse set of smearing techniques was employed to determine of partial occupancies $f_{nk}$,
        including the tetrahedron method \cite{blochl1994improved} and the Gaussian method.
        For the Gaussian method, the smearing width was varied within the range of 0.01 eV to 0.1 eV.
        This careful selection of methods was made to guarantee reliable and consistent predictions
        of adsorption energies from the model, regardless of the smearing technique used.

    The augmented dataset comprises 12,312 samples, with adsorption energies having
    a standard deviation of 1.8117 eV and a variance of 3.2821 eV\textsubscript{2}.
    Twenty percent of these samples were allocated to the validation set.
    Throughout the training process, a batch size of 64 was employed to ensure a balanced learning dynamic.

    \subsection{eDOS occlusion experiments}
    Supplementary \cref{supp_fig21:occlusion} visually illustrates the eDOS occlusion
    experiment methodology. To maintain consistent shapes, zero-padding was applied to the
    input eDOS arrays. Focusing on understanding the influence of metals on
    adsorption behavior, a masker with dimensions [width, 1, 1] was employed along
    the metal channel to eliminate cross-orbital interactions.
    Here, ``width'' denotes the masker width, and detailed explanations for determining the width
    are provided in Supplementary \cref{supp_sec3.5_occlusion}.
    Through recursive application of these maskers, variations in predicted
    adsorption energies from the CNN model were recorded, generating an occlusion array.

    \subsection{eDOS shifting experiments}
    Supplementary \cref{supp_fig30:shifting} illustrates the implemented eDOS shifting experiment protocol.
    In this study, the initial electronic density of states underwent controlled shifting along the energy axis.
    This protocol ranges from -1 eV to 1 eV, with a step length of 0.005 eV per image.
    To maintain a uniform energy window, each image underwent cropping and padding procedures.
    The resulted output arrays took the form of [400, 1] arrays for simultaneous orbital shifting
    and [400, 9] arrays for individual orbital shifting.
    The energy range for shifting was chosen to induce a moderate disturbance to adsorption energy.

    \subsection{Chemical bond analysis and real-space wavefunction visualization}
    COHP analysis was performed using the Local-Orbital Basis Suite Towards
    Electronic-Structure Reconstruction (LOBSTER) package
    \cite{deringer2011crystal, koga1999analytical, nelson2020lobster, maintz2013analytic, dronskowski1993crystal} version 4.1.0,
    with the GGA-PBE wavefunctions fitted by S. Maintz \cite{koga1999analytical, maintz2016lobster}.
    A Gaussian smearing for energy integration was employed with a broadening width of 0.05 eV.
    The quantification of bonding strength between the single metal atom and the substrate
    was achieved through the summation of interactions with neighboring atoms
    within a range of 0.5 \text{\AA} to 5.0 \text{\AA} from the single metal atom.
    Orbitals of interest were confirmed through band-structure analysis,
    and real-space wavefunctions were subsequently visualized using VASPKIT \cite{wang2021vaspkit}
    and VESTA \cite{momma2008vesta}.

    \subsection{Machine learning and data analysis environment}
    The CNN model was implemented using TensorFlow \cite{abadi2016tensorflow},
    and hyperparameter optimization was conducted using the Hyperband \cite{li2018hyperband} algorithm
    via KerasTuner \cite{omalley2019kerastuner}.
    For a more detailed overview of our machine learning setup and data analysis environment,
    please refer to Supplementary \cref{supp_sec4_env}.

% Code/Data Availability, Acknowledgements, Author Contributions and Competing Interests
\setcounter{section}{0}
\renewcommand{\thesection}{\Alph{section}}

\newpage
    \section{Code and Data Availability}
    Essential source code and data required to replicate the study's results-including VASP input files,
    structure files of SACs, eDOS arrays and the pretrained CNN model-are available on GitHub repository
    at \url{https://github.com/DanielYang59/cnn4dos}, and can be provided by the corresponding authors upon request.
    Due to their substantial size in terabytes, the complete source data (complete VASP output files and such)
    is not currently hosted on publicly available repositories,
    but can be obtained from the corresponding authors upon request.

    \section{Acknowledgements}
    The QUT eResearch Office provided computational resources and services used in this work.
    This research was undertaken with the assistance of resources and services from
    the National Computational Infrastructure (NCI), supported by the Australian Government.
    It was also supported by resources provided by the Pawsey Supercomputing Research Centre
    with funding from the Australian Government and the Government of Western Australia.
    Insights into deep learning methodologies were contributed by Mr. Yanwei Guan
    from Chongqing University and Mr. Zhipeng He from Assoc. Prof. Chun Ouyang's team at QUT.
    The machine learning and data analysis implementation benefited from
    resources provided by the GitHub, TensorFlow and Stack Overflow communities.
    Dr. Ryky Nelson from RWTH Aachen University offered valuable guidance on
    Crystal Orbital Hamilton Population analysis.

    \section{Author Contributions}
    H.Y. performed DFT calculations, implemented Python code for
    machine learning methodologies and data analysis,
    and drafted and refined the manuscript.
    J.Z. engaged in valuable discussions on DFT calculations.
    Q.W. provided advice on the computer code revisions.
    B.L. provided guidance on DFT calculations and offered research supervision.
    W.L. facilitated resource acquisition and offered research supervision.
    Z.S. initiated the research concept, contributed to manuscript revisions, and supervised the overall research.
    T.L. guided DFT calculations and data analysis, offering continuous supervision
    throughout the research and manuscript drafting.

    \section{Competing Interests}
    The authors declare no competing interests in relation to this research.

% End of Main

% Supplementary Information
\newpage
\setcounter{section}{0}
\renewcommand{\thesection}{\Roman{section}}
% Supplementary Information

% Title
\begin{center}
  \Large{Supporting Information for:  \\
  Convolutional Neural Networks and Volcano Plots:
  Screening and Prediction of Two-Dimensional Single-Atom Catalysts for CO\textsubscript{2} Reduction Reactions}
\end{center}

% SI Section One: Additional details on DFT calculations
\newpage
\section{Additional details on DFT calculations}

\subsection{Catalyst models}
\label{supp_sec2.1_catalysts}

\begin{figure}[htbp]
  \centering
  \includegraphics[width=\textwidth]{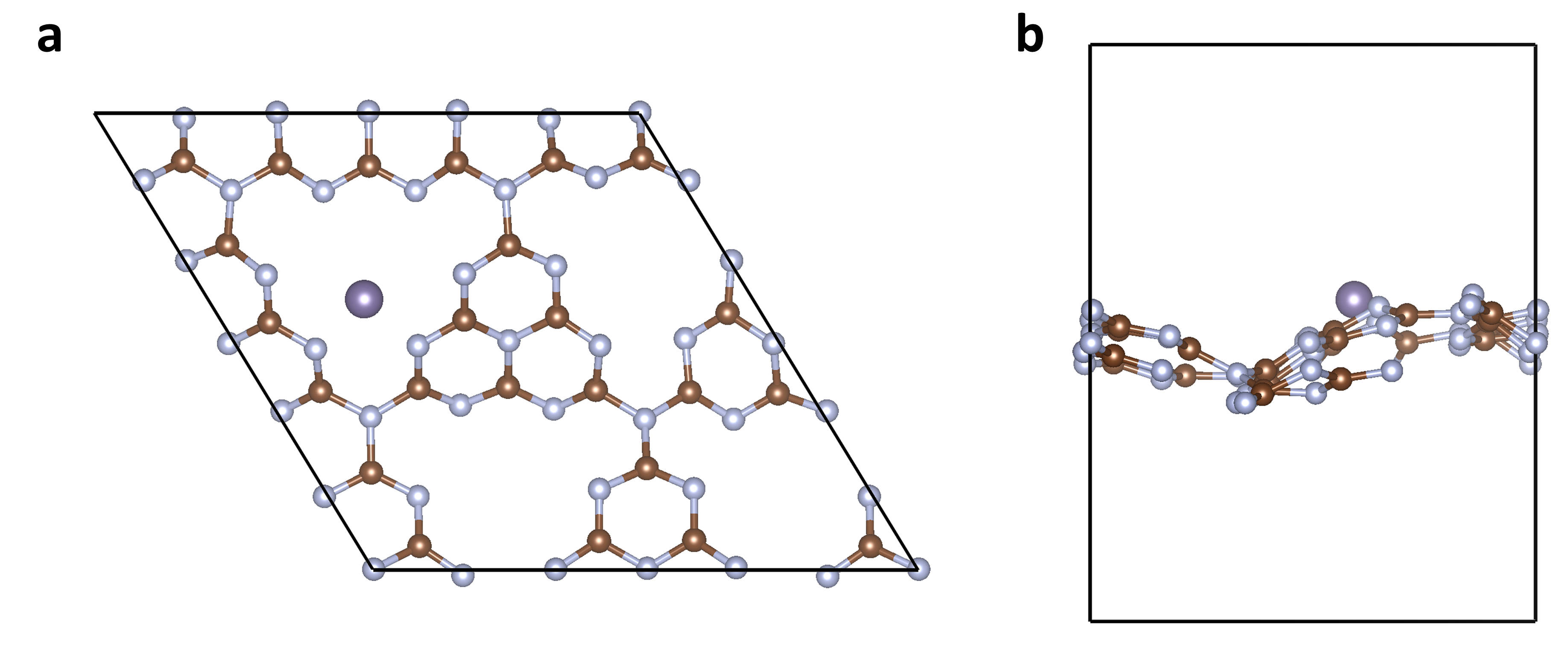}
  \caption{\textbf{Structure of single germanium atom supported on graphitic nitride.}
  (\textbf{a}) View along the z-axis and (\textbf{b}) view along the x-axis
  of the Ge@g-C\textsubscript{3}N\textsubscript{4} structure.
  In the illustrations, silver spheres represent N atoms, brown spheres
  signify C atoms, and purple sphere indicates Ge atom.}
  \label{supp_fig2:Ge_g-C3N4}
\end{figure}

\begin{figure}[htbp]
  \centering
  \includegraphics[width=\textwidth]{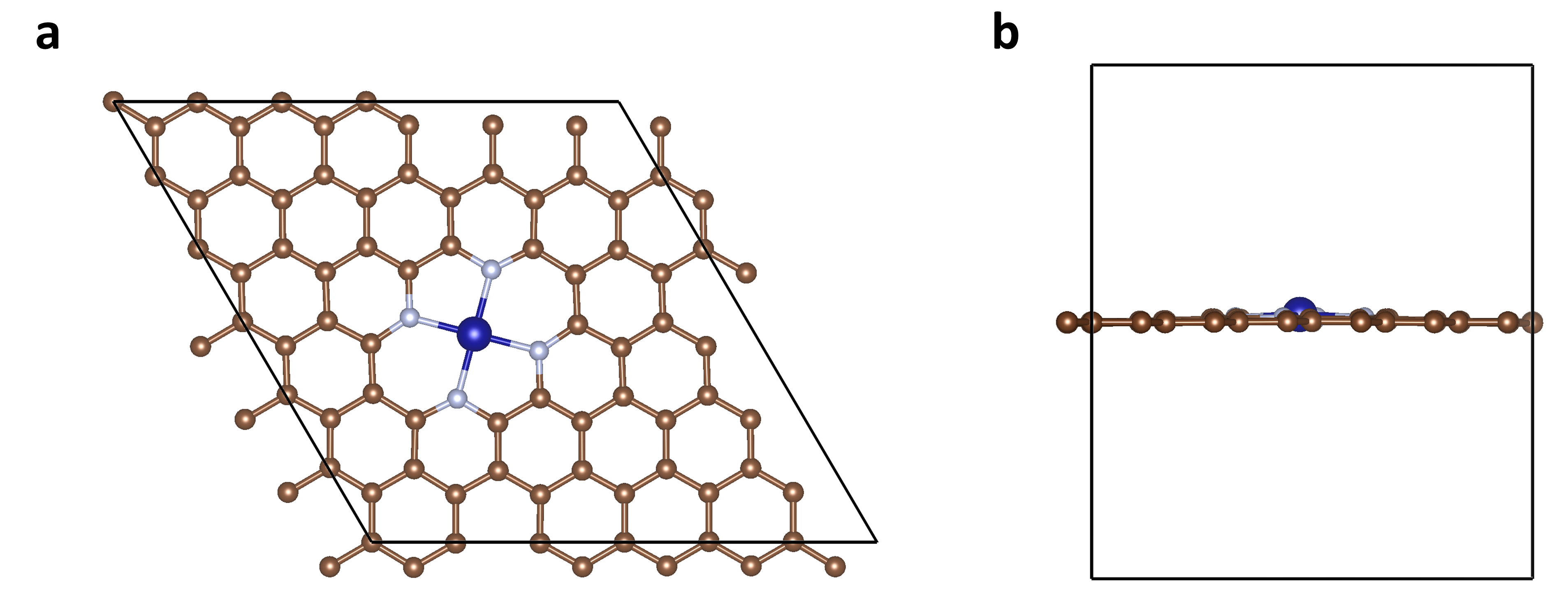}
  \caption{\textbf{Structure of single chromium atom supported on nitrogen-doped graphene.}
  (\textbf{a}) View along the z-axis and (\textbf{b}) view along the x-axis
  of the Cr@nitrogen-doped-graphene structure.
  In the illustrations, silver spheres represent N atoms, brown spheres
  signify C atoms, and indigo sphere indicates Cr atom.}
  \label{supp_fig3:Cr-n-gra}
\end{figure}

\begin{figure}[htbp]
  \centering
  \includegraphics[width=\textwidth]{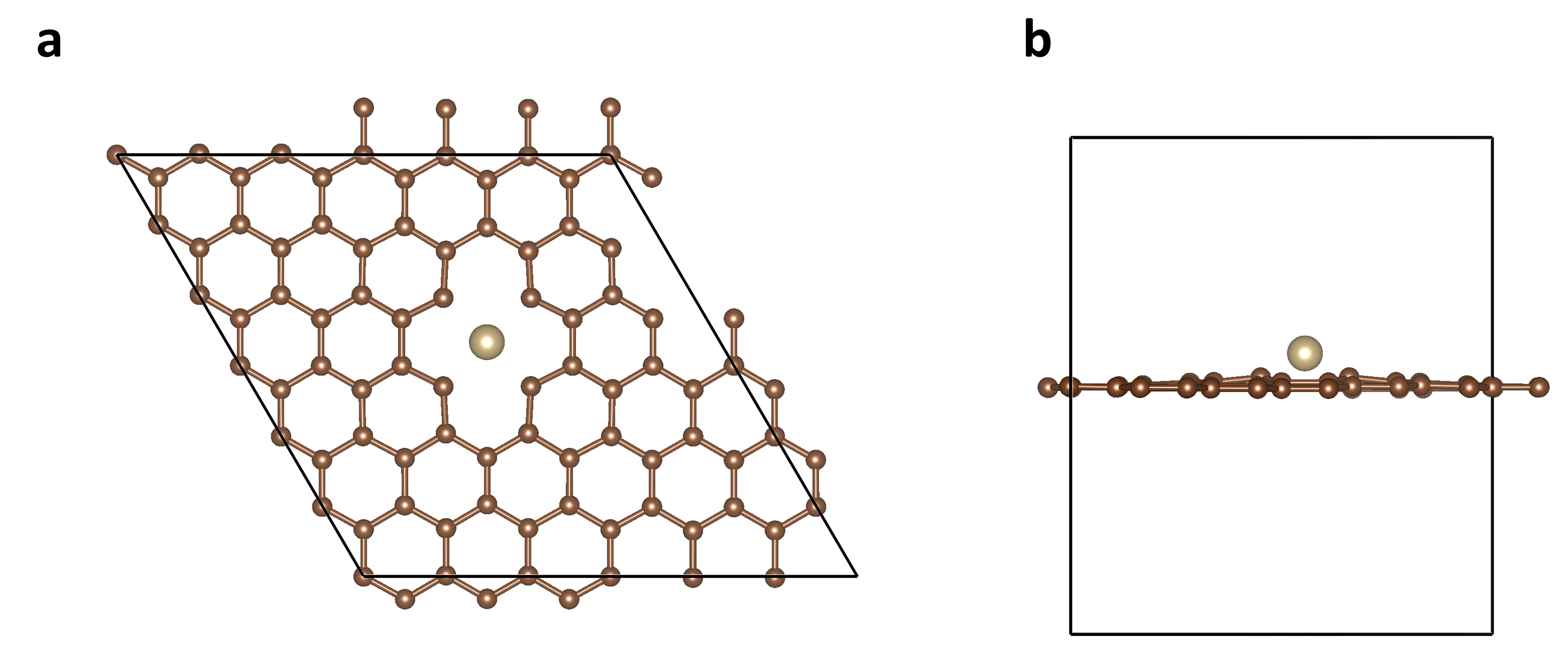}
  \caption{\textbf{Structure of single osmium atom supported on graphene with dual-vacancy.}
  (\textbf{a}) View along the z-axis and (\textbf{b}) view along the x-axis
  of the Os@graphene-with-dual-vacancy structure.
  In the illustrations, brown spheres signify C atoms, and light yellow sphere
  indicates Os atom.}
  \label{supp_fig4:Os-gra-vac}
\end{figure}

\begin{figure}[htbp]
  \centering
  \includegraphics[width=\textwidth]{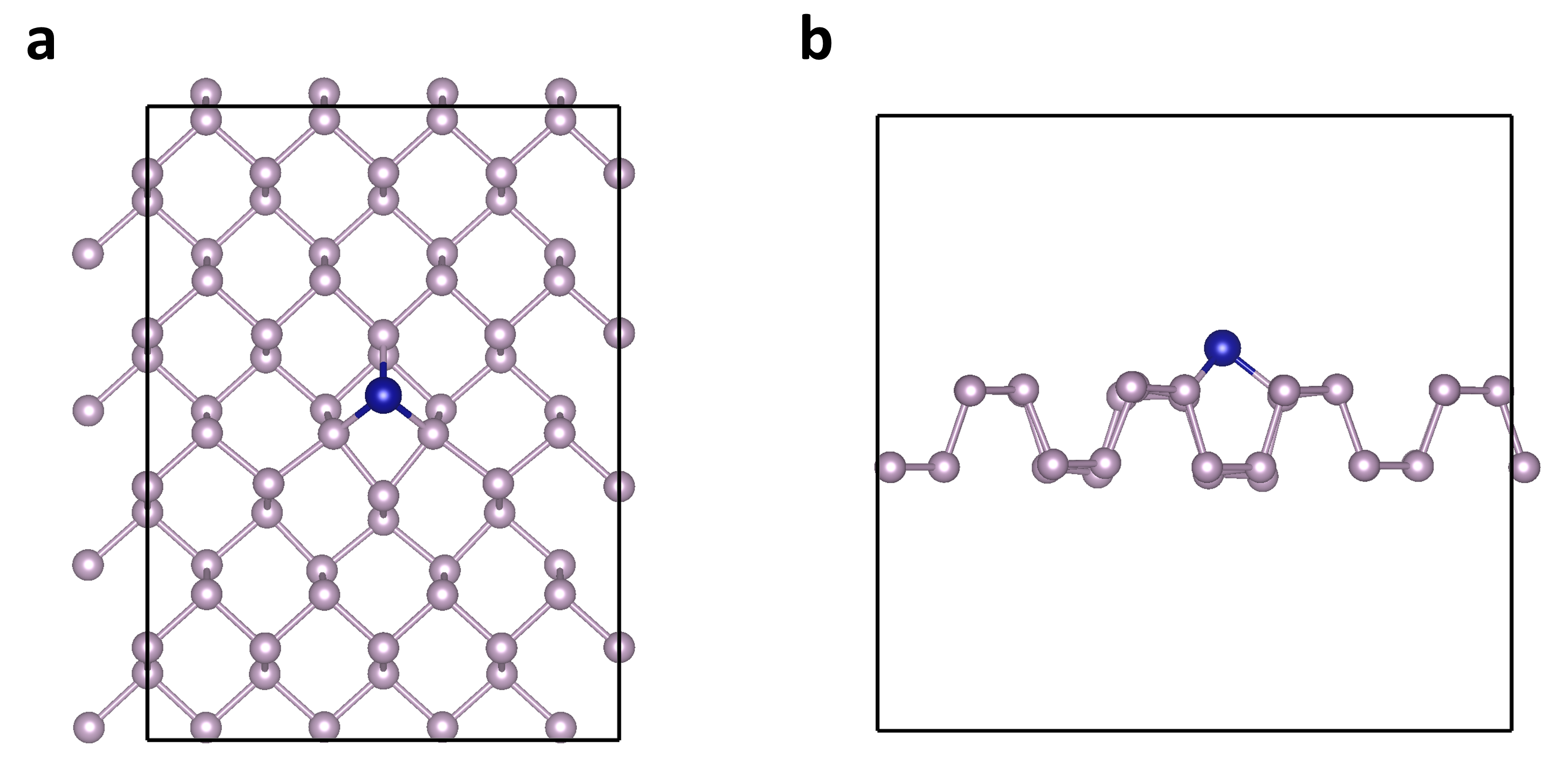}
  \caption{\textbf{Structure of single chromium atom supported on black phosphorous.}
  (\textbf{a}) View along the z-axis and (\textbf{b}) view along the x-axis
  of the Cr@black phosphorous structure.
  In the illustrations, light purple spheres represent P atoms, and indigo sphere
  indicates Cr atom.}
  \label{supp_fig5:Cr-BP}
\end{figure}

\begin{figure}[htbp]
  \centering
  \includegraphics[width=\textwidth]{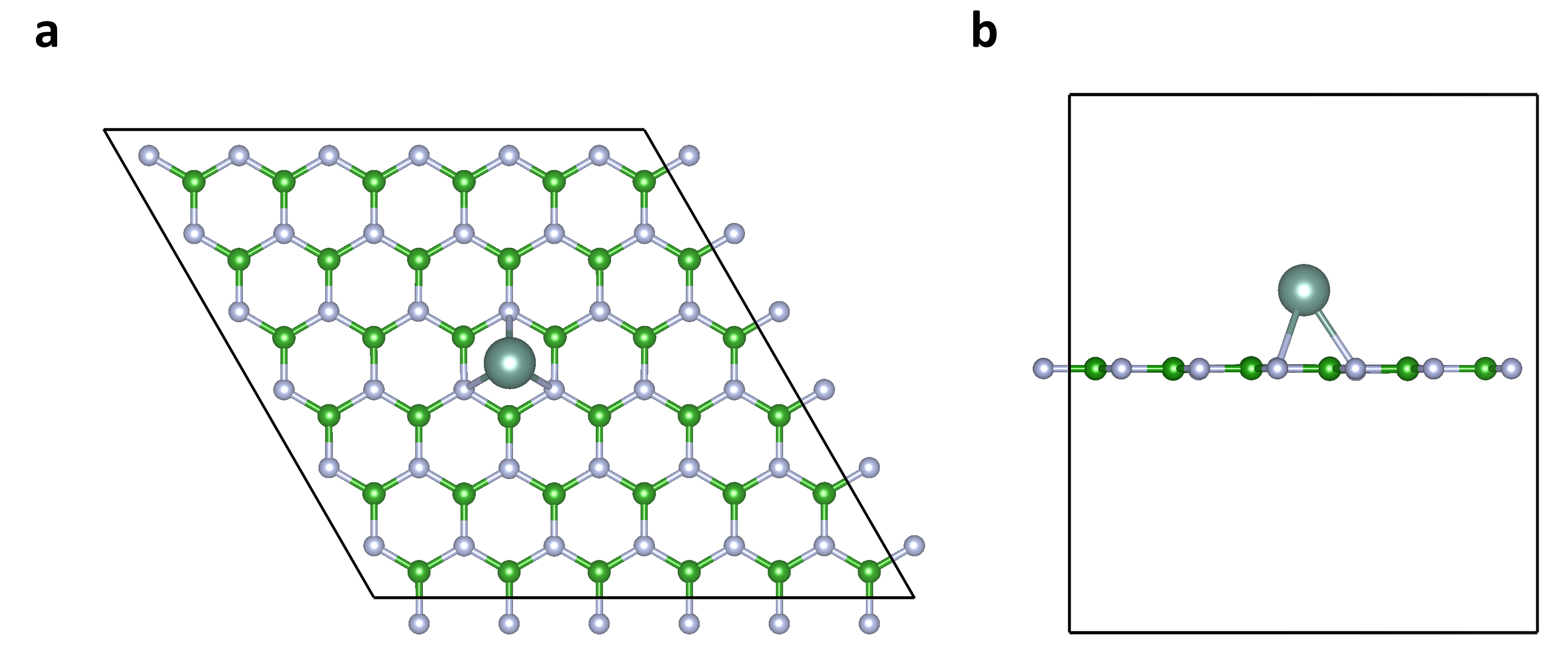}
  \caption{\textbf{Structure of single yttrium atom supported on boron nitride.}
  (\textbf{a}) View along the z-axis and (\textbf{b}) view along the x-axis
  of the Y@boron-nitride structure.
  In the illustrations, silver spheres represent N atoms, green spheres
  signify B atoms, and light green sphere indicates Y atom.}
  \label{supp_fig6:Y-BN}
\end{figure}

\begin{figure}[htbp]
  \centering
  \includegraphics[width=\textwidth]{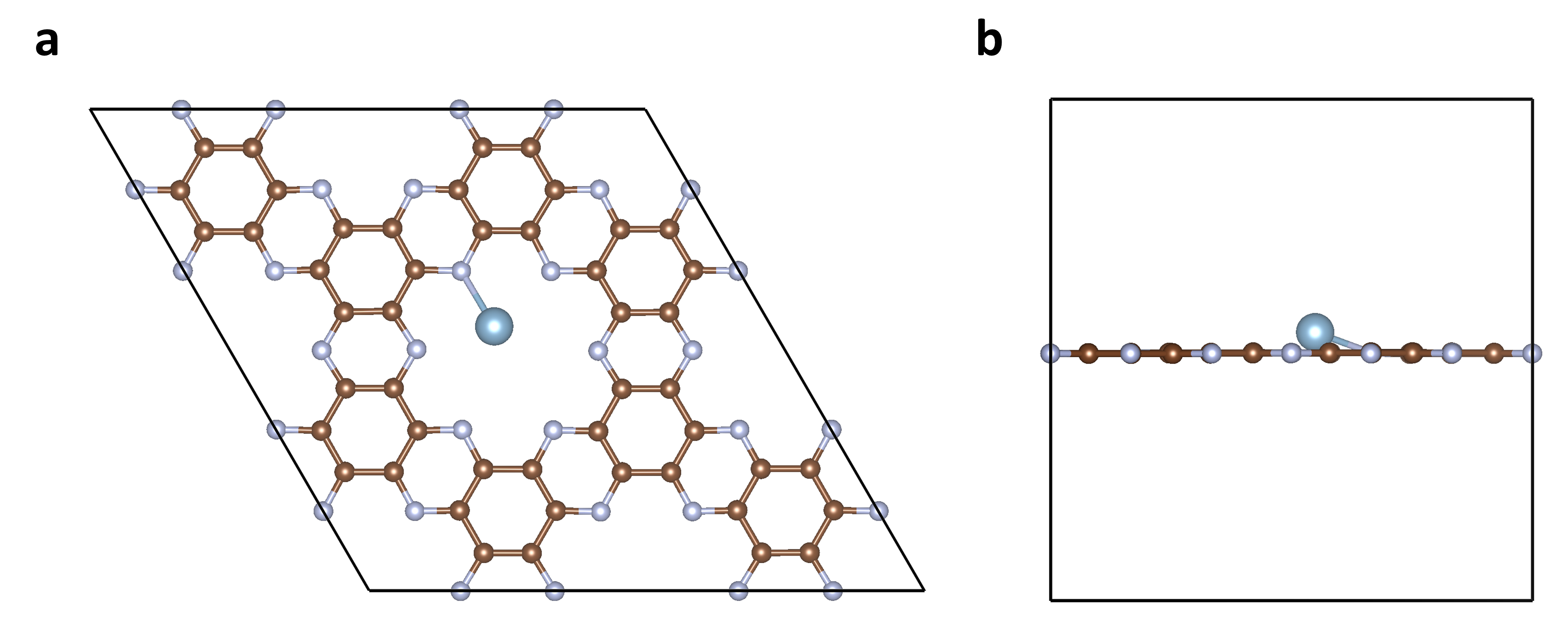}
  \caption{\textbf{Structure of single aluminum atom supported on C\textsubscript{2}N.}
  (\textbf{a}) View along the z-axis and (\textbf{b}) view along the x-axis
  of the Al@C\textsubscript{2}N structure.
  In the illustrations, silver spheres represent N atoms, brown spheres
  signify C atoms, and blue sphere indicates Al atom.}
  \label{supp_fig7:Al-C2N}
\end{figure}

Representative structures for each substrate are depicted in \cref{supp_fig2:Ge_g-C3N4} through \cref{supp_fig7:Al-C2N}. For the adsorption process, the initial state is defined by positioning the optimized adsorbates at a distance of 6.5 \text{\AA} from the single-metal atoms supported on the substrate, oriented along the z-axis.

% SI Figure 1: Periodic table of investigated metals
\begin{figure}[htbp]
  \centering
  \includegraphics[width=\textwidth]{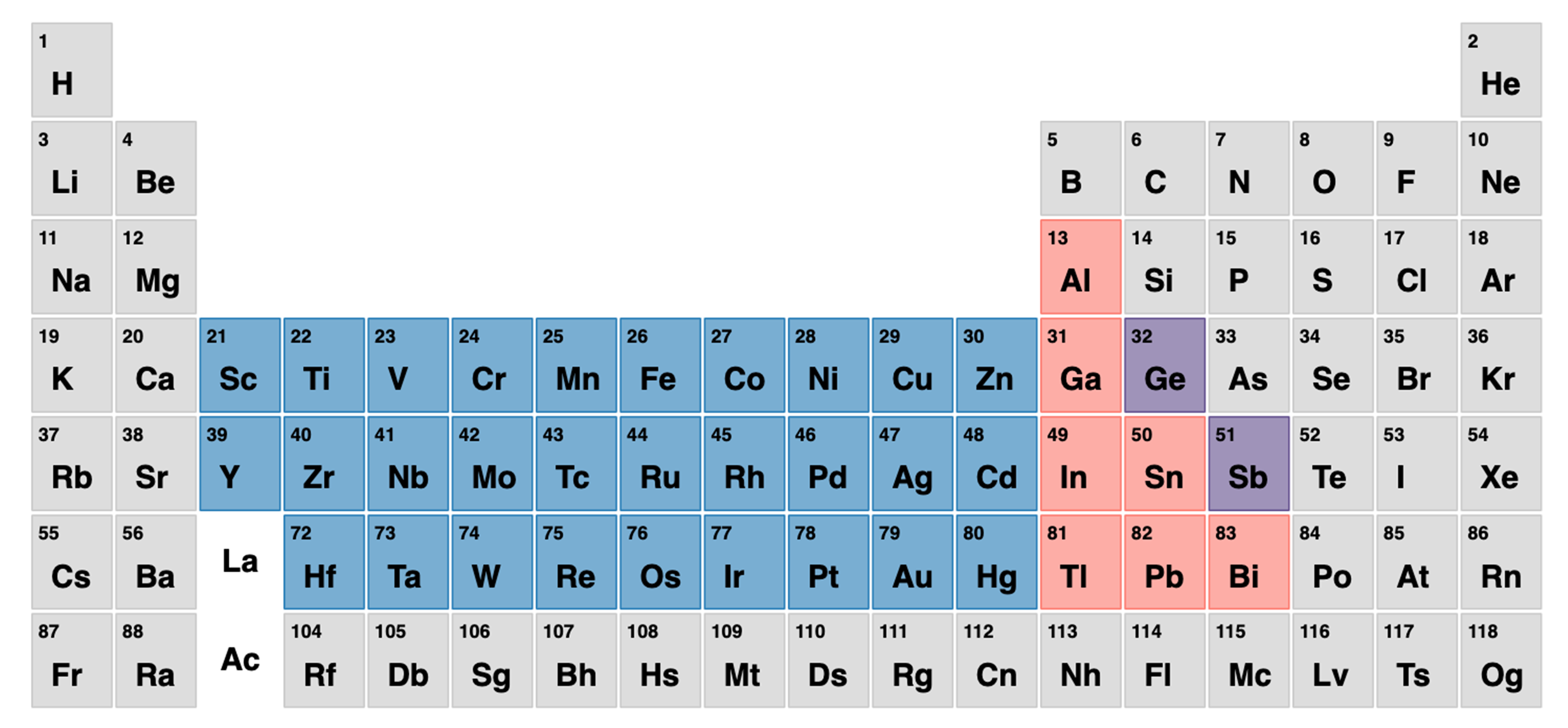}
  \caption{\textbf{Periodic table of investigated metals.}
  Periodic table highlighting investigated elements: transition metals in blue, metalloids in purple, and other metals in orange.}
  \label{supp_fig1:ptable}
\end{figure}

% SI Figure 2: Structure of Ge atom supported on g-C3N4

% SI Figure 3: Structure of Cr atom supported on nitrogen-doped graphene

% SI Figure 4: Structure of Os atom supported on graphene with dual-vacancy

% SI Figure 5: Structure of Cr atom supported on black phosphorous

% SI Figure 6: Structure of Y atom supported on boron nitride

% SI Figure 7: Structure of Al atom supported on C2N

\subsection[CO2RR pathways]{CO\textsubscript{2}RR pathways}
\label{supp_sec2.2_co2rr_paths}

\begin{figure}[htbp]
  \centering
  \includegraphics[width=0.8\textwidth]{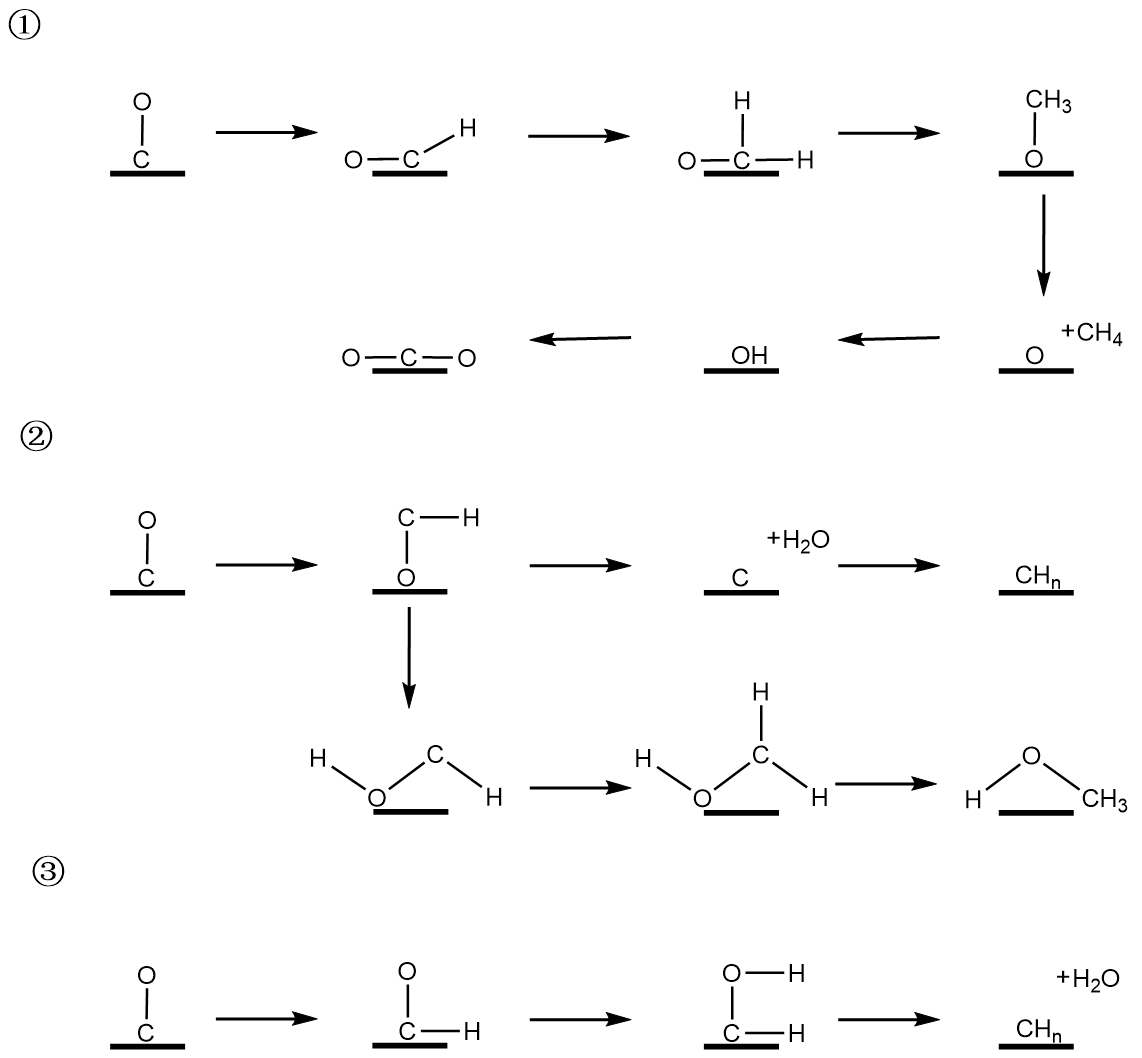}
  \caption{\textbf{Three investigated CO\textsubscript{2}RR pathways \cite{durand2011structure, nie2014reaction, peterson2010copper} investigated in this work.}}
  \label{supp_fig8:co2rr_paths}
\end{figure}

In this study, we explored three reaction pathways that have been previously documented in the literature \cite{durand2011structure, nie2014reaction, peterson2010copper}, as depicted in \cref{supp_fig8:co2rr_paths}.
The adsorption configuration of the *CHO intermediate is pivotal in each of these pathways for determining the reaction mechanism in the CO\textsubscript{2}RR process.
We calculated the energies associated with the *CHO intermediate for catalysts
supported on g-C\textsubscript{3}N\textsubscript{4}, nitrogen-doped graphene, and dual-vacancy graphene-under each mechanism.
These calculations are summarized in \cref{supp_table1:e_3_co2rr_paths}.
Our results indicate that Pathway 1 is the most energetically favorable for most catalysts we investigated.
Therefore, to streamline our analysis, we focused exclusively on Mechanism 1 across all catalysts examined.

% SI Figure 8: Three investigated CO2RR pathways

% Supplementary Table 1: DFT-calculated energies for protonated *CO intermediate for 3 pathways
\begin{table}[htbp]
\label{supp_table1:e_3_co2rr_paths}
  \caption{DFT-calculated energies in eV for protonated *CO intermediate
    for three investigated CO\textsubscript{2}RR pathways.}
  \resizebox{\textwidth}{!}{%
    \begin{tabular}{l *{9}{l}}
      \toprule
      \multirow{2}{*}{Metal}
      & \multicolumn{3}{c}{g-C\textsubscript{3}N\textsubscript{4}}
      & \multicolumn{3}{c}{nitrogen-doped graphene}
      & \multicolumn{3}{c}{graphene with dual-vacancy}  \\ \cline{2-10}
      & Pathway-1 & Pathway-2 & Pathway-3
      & Pathway-1 & Pathway-2 & Pathway-3
      & Pathway-1 & Pathway-2 & Pathway-3  \\
      \midrule
      Al & \textbf{-499.3494} & -497.8343 & -499.3005 & \textbf{-670.3489} & -668.5142 & -670.3093 & -667.2538 & -666.6725 & \textbf{-667.2665} \\
      Co & -502.5946 & -501.1257 & \textbf{-502.6139} & -672.8847 & -670.8013 & \textbf{-672.9020} & \textbf{-671.6284} & -670.4388 & -671.3827 \\
      Cr & -505.6012 & -504.2022 & \textbf{-505.6029} & -675.3126 & -674.1640 & \textbf{-675.3480} & \textbf{-673.4283} & -672.1919 & -673.3818 \\
      Cu & \textbf{-495.8913} & -493.6853 & -495.8820 & \textbf{-666.9894} & -664.9810 & -666.9705 & -666.9876 & -664.7684 & \textbf{-667.1395} \\
      Fe & \textbf{-501.3325} & -499.3969 & -501.3221 & -673.8242 & -672.0887 & \textbf{-673.8478} & \textbf{-672.7810} & -671.5694 & -672.4960 \\
      Ga & -498.2365 & -496.3967 & \textbf{-498.2912} & \textbf{-667.3963} & -665.4677 & -667.3482 & -665.7780 & -664.8573 & \textbf{-665.8030} \\
      Ge & -500.2052 & -497.9780 & \textbf{-500.2156} & \textbf{-668.1785} & -666.3484 & -668.1463 & \textbf{-667.7387} & -665.7394 & -667.7324 \\
      Mn & \textbf{-502.6355} & -500.8006 & -502.6294 & -674.6848 & -672.9574 & \textbf{-674.7030} & -673.2243 & -672.6495 & \textbf{-673.2361} \\
      Ni & \textbf{-500.4085} & -498.5165 & -500.3985 & \textbf{-670.0979} & -667.9544 & -670.0832 & -670.1619 & -668.3684 & \textbf{-670.1756} \\
      Sc & -503.4428 & -501.6129 & \textbf{-503.4437} & \textbf{-673.9590} & -671.7624 & -673.3818 & \textbf{-670.0841} & -668.1044 & -669.6050 \\
      Ti & -503.7848 & -502.6148 & \textbf{-503.8033} & \textbf{-675.1006} & -673.2417 & -674.6257 & \textbf{-672.7127} & -670.6787 & -672.2491 \\
      V  & \textbf{-503.6115} & -502.6188 & -503.5956 & \textbf{-675.0533} & -674.0394 & -675.0136 & \textbf{-673.1855} & -671.5382 & -673.1583 \\
      Zn & \textbf{-493.9854} & -491.5010 & -493.9743 & \textbf{-664.9935} & -663.0932 & -664.9758 & -663.7735 & -661.8097 & \textbf{-663.7959} \\
      Ag & \textbf{-494.4055} & -492.4511 & -494.3905 & -664.6322 & -662.9060 & \textbf{-664.6353} & -664.3972 & -662.3588 & \textbf{-665.0570} \\
      Cd & \textbf{-493.3084} & -491.1781 & -493.2951 & \textbf{-663.8832} & -661.6272 & -663.8653 & -661.8518 & -659.9601 & \textbf{-661.8631} \\
      In & -498.5272 & -496.6593 & \textbf{-498.5908} & \textbf{-666.3968} & -663.7707 & -666.3845 & -663.6532 & -661.5581 & \textbf{-663.6913} \\
      Mo & \textbf{-504.5248} & -503.3763 & -504.5148 & -675.3472 & -674.7118 & \textbf{-675.4389} & -674.7186 & -672.9335 & \textbf{-674.7397} \\
      Nb & -505.3967 & -504.7113 & \textbf{-505.3982} & \textbf{-676.0492} & -674.1997 & -675.4107 & \textbf{-674.6606} & -672.8268 & -674.5488 \\
      Pd & -500.0370 & -498.3310 & \textbf{-500.0455} & \textbf{-669.0243} & -667.0759 & -669.0205 & \textbf{-669.1169} & -667.3223 & -669.1119 \\
      Rh & \textbf{-499.2225} & -497.3162 & -499.2170 & -672.7920 & -670.7853 & \textbf{-672.8099} & \textbf{-671.4286} & -670.3036 & -671.4181 \\
      Ru & \textbf{-503.9986} & -502.1908 & -503.9640 & -674.0273 & -672.4060 & \textbf{-674.0339} & \textbf{-672.9956} & -672.1332 & -672.9909 \\
      Sb & \textbf{-497.3836} & -495.1348 & -497.3442 & \textbf{-666.7198} & -664.9611 & -666.7190 & -666.8632 & -664.8131 & \textbf{-666.8670} \\
      Sn & \textbf{-499.4952} & -497.2007 & -499.1116 & \textbf{-667.1459} & -665.3835 & -667.0999 & \textbf{-665.9821} & -663.8277 & -665.9755 \\
      Tc & \textbf{-503.0602} & -501.1754 & -503.0532 & -674.4342 & -673.8164 & \textbf{-674.4559} & -674.0615 & -672.9747 & \textbf{-674.0738} \\
      Y  & \textbf{-503.9232} & -502.0670 & -503.8988 & \textbf{-673.7620} & -671.6132 & -673.2241 & -669.4633 & -668.0403 & \textbf{-669.5030} \\
      Zr & \textbf{-505.2907} & -503.9148 & -505.1817 & \textbf{-675.6874} & -673.5456 & -675.0519 & \textbf{-673.1694} & -671.1993 & -672.8072 \\
      Au & \textbf{-497.0784} & -495.1720 & -497.0175 & \textbf{-665.1479} & -663.3239 & -665.1370 & -666.4185 & -664.2892 & \textbf{-666.5182} \\
      Bi & \textbf{-496.9946} & -494.7679 & -496.9934 & -665.9563 & -664.2409 & \textbf{-666.0512} & \textbf{-665.0896} & -663.2974 & -665.0887 \\
      Hf & \textbf{-506.4837} & -504.9734 & -506.4777 & \textbf{-677.3077} & -675.1797 & -676.6502 & \textbf{-674.7209} & -672.6650 & -674.2149 \\
      Hg & \textbf{-491.0035} & -488.7951 & -490.9728 & \textbf{-661.2458} & -659.2406 & -661.2398 & -661.7192 & -659.8398 & \textbf{-661.7300} \\
      Ir & \textbf{-503.0227} & -501.1368 & -503.0217 & -673.9926 & -672.0068 & \textbf{-674.0087} & \textbf{-673.5129} & -672.6725 & -673.4975 \\
      Os & -504.7648 & -502.9878 & \textbf{-504.8105} & -675.2098 & -673.6653 & \textbf{-675.2235} & \textbf{-674.9483} & -674.1970 & -674.9223 \\
      Pb & \textbf{-499.1183} & -497.1173 & -499.0544 & -666.4446 & -664.6897 & \textbf{-666.4625} & \textbf{-664.0200} & -662.2222 & -664.0123 \\
      Pt & \textbf{-499.3356} & -497.5110 & -499.3215 & -670.3996 & -668.2412 & \textbf{-670.4147} & \textbf{-671.1408} & -669.2725 & -671.1370 \\
      Re & \textbf{-504.1830} & -502.4735 & -504.1746 & \textbf{-675.7670} & -675.0715 & -675.7486 & \textbf{-675.8701} & -674.9245 & -675.8233 \\
      Ta & -506.6180 & -505.7793 & \textbf{-506.6241} & \textbf{-677.7222} & -675.9907 & -677.0911 & \textbf{-676.4371} & -674.5434 & -676.4262 \\
      Tl & \textbf{-497.9559} & -496.1578 & -497.9382 & -664.5047 & -662.7051 & \textbf{-664.5051} & -662.6279 & -660.6086 & \textbf{-662.6317}\\
      W  & \textbf{-505.8134} & -504.9085 & -505.8030 & -676.5304 & -676.1384 & \textbf{-676.9965} & \textbf{-676.7669} & -675.4895 & -676.7356 \\
      \bottomrule
    \end{tabular}
  }

  \smallskip

  \footnotesize\textit{Note:} Energies of the most stable configurations are highlighted in bold.
\end{table}

\subsection{DFT energies and energy corrections}
\label{supp_sec2.3_energies}

% Supplementary Table 2: Free energies for isolated species.
\begin{table}[htbp]
\label{supp_table2:species_free_energies}
  \caption{Free energies at 298.15 K for isolated species.}
  \center
  \small
  \begin{tabularx}{0.5\textwidth}{@{}lXr@{}}
    \toprule
    Species               & Free Energy (eV)  \\
    \midrule
    CO\textsubscript{2}   & -23.3140          \\
    CO                    & -15.3336          \\
    CH\textsubscript{4}   & -23.4639          \\
    H\textsubscript{2}    & -6.9315           \\
    H\textsubscript{2}O   & -14.3239          \\
    COOH                  & -24.3963          \\
    CHO                   & -17.0983          \\
    CH\textsubscript{2}O  & -22.1607          \\
    OCH\textsubscript{3}  & -24.5075          \\
    O                     & -1.9007           \\
    OH                    & -7.7303           \\
    \bottomrule
  \end{tabularx}

  \smallskip

  \begin{flushright}
  \begin{minipage}{\textwidth}
    \footnotesize\textit{Note:} VASPKIT \cite{wang2021vaspkit} was employed to calculate free energies.
  \end{minipage}
  \end{flushright}
\end{table}

% Supplementary Table 3: Free energy corrections for adsorbed intermediates
\begin{table}[htbp]
\label{supp_table3:ads_e_correction}
  \caption{Free energy corrections at 298.15 K for adsorbed intermediates.}
  \centering
  \small
  \begin{tabularx}{0.5\textwidth}{@{}lXr@{}}
    \toprule
    Species                & Free energy correction (eV)  \\
    \midrule
    *CO\textsubscript{2}   &  0.0864                      \\
    *COOH                  &  0.3693                      \\
    *CO                    &  0.0111                      \\
    *CHO                   &  0.2716                      \\
    *CH\textsubscript{2}O  &  0.5595                      \\
    *OCH\textsubscript{3}  &  0.8738                      \\
    *O                     & -0.0193                      \\
    *OH                    &  0.1993                      \\
    *H                     &  0.1533                      \\
    \bottomrule
  \end{tabularx}

  \smallskip

  \begin{flushright}  % justify footnote
  \begin{minipage}{\textwidth}
    \footnotesize\textit{Note:} {Free energy corrections were obtained by averaging the
      DFT-computed corrections, which include zero-point energy and entropy, across three substrates:
      g-C\textsubscript{3}N\textsubscript{4}, nitrogen-doped graphene, and dual-vacancy graphene.}
  \end{minipage}
  \end{flushright}
\end{table}

% Supplementary Table 4: DFT calculated final state energies for adsorbates supported on g-c3n4
\begin{table}[htbp]
\label{supp_table4:e_g-c3n4}
  \caption{DFT calculated final state energies in eV for adsorbates
    supported on g-C\textsubscript{3}N\textsubscript{4}.}
  \resizebox{\textwidth}{!}{%
    \begin{tabular}{l *{9}{l}}
      \toprule
      Metal & *CO\textsubscript{2} & *COOH    & *CO       & *CHO
      & *CH\textsubscript{2}O  & *OCH\textsubscript{3}  & *O        & *OH       & *H        \\
      \midrule
      Al & -481.2292 & -504.7972 & -508.6897 & -496.4490 & -500.7640 & -503.7935 & -509.9998 & -488.6623 & -485.4855 \\
      Co & -483.2039 & -507.0765 & -510.0453 & -500.1329 & -503.9316 & -507.0552 & -510.9052 & -489.9487 & -486.6194 \\
      Cr & -486.7813 & -509.9599 & -513.3962 & -502.1493 & -505.7125 & -509.8851 & -514.1702 & -493.7457 & -489.9202 \\
      Cu & -477.5955 & -502.5914 & -505.9613 & -495.6511 & -498.0502 & -502.4590 & -506.2829 & -484.3166 & -481.9683 \\
      Fe & -482.0576 & -507.1508 & -510.2918 & -500.1101 & -502.1626 & -506.0115 & -511.2543 & -490.0116 & -486.4067 \\
      Ga & -480.8854 & -503.9317 & -506.4938 & -495.7379 & -498.7813 & -503.1345 & -507.5029 & -486.1702 & -483.4575 \\
      Ge & -481.5807 & -504.6467 & -508.4204 & -496.4421 & -500.5712 & -504.0218 & -509.4116 & -488.4168 & -485.2680 \\
      Mn & -483.7690 & -508.5410 & -511.6514 & -500.6279 & -502.6739 & -508.6433 & -512.8569 & -491.2239 & -488.1082 \\
      Ni & -482.0061 & -505.1988 & -508.4173 & -498.5054 & -501.0040 & -504.8945 & -508.6266 & -487.3684 & -484.8272 \\
      Sc & -483.9963 & -509.3147 & -512.9882 & -501.4505 & -505.2024 & -509.4054 & -514.5969 & -493.4105 & -489.2632 \\
      Ti & -484.2282 & -509.6510 & -512.8528 & -502.0410 & -505.3761 & -509.9729 & -514.0700 & -493.6478 & -489.2234 \\
      V  & -484.1990 & -508.6926 & -512.3009 & -501.6777 & -504.4297 & -509.3618 & -513.4561 & -493.8569 & -488.9474 \\
      Zn & -475.2375 & -500.3556 & -504.4103 & -492.4144 & -496.5426 & -499.4232 & -504.6297 & -482.6782 & -481.0438 \\
      Ag & -476.7049 & -501.6430 & -504.5861 & -493.9138 & -496.7220 & -500.9477 & -504.5408 & -482.6604 & -481.2659 \\
      Cd & -475.1288 & -500.1854 & -503.4108 & -491.8954 & -495.6486 & -499.4959 & -503.6209 & -483.3427 & -479.8278 \\
      In & -481.1330 & -504.1653 & -506.4422 & -496.0016 & -498.6643 & -503.3286 & -507.3780 & -485.6281 & -483.1445 \\
      Mo & -484.2892 & -509.4263 & -512.9103 & -503.0025 & -505.5534 & -509.6730 & -513.0136 & -493.5700 & -489.4994 \\
      Nb & -485.4055 & -510.7952 & -514.0380 & -503.5663 & -506.3671 & -511.1572 & -514.7845 & -495.5118 & -490.2111 \\
      Pd & -481.5399 & -504.7593 & -508.5173 & -498.0817 & -501.1620 & -504.0892 & -508.1976 & -486.1551 & -483.9553 \\
      Rh & -479.8884 & -505.2422 & -508.6754 & -498.7129 & -501.4594 & -504.5254 & -508.9536 & -487.8771 & -485.0738 \\
      Ru & -483.8526 & -507.5120 & -511.5881 & -500.6857 & -504.1497 & -506.8934 & -510.9524 & -490.6690 & -488.0169 \\
      Sb & -478.5169 & -502.9708 & -506.5211 & -494.8740 & -498.9453 & -502.5347 & -507.2111 & -486.4448 & -483.4418 \\
      Sn & -481.4238 & -504.4677 & -507.8470 & -496.2883 & -499.9258 & -503.6685 & -508.8152 & -487.6963 & -484.6351 \\
      Tc & -482.7091 & -507.9652 & -512.1064 & -501.9896 & -504.6034 & -508.3486 & -511.4777 & -492.0381 & -488.0897 \\
      Y  & -484.6612 & -509.8564 & -513.6851 & -501.9522 & -505.8238 & -509.8344 & -515.2400 & -493.4038 & -489.6506 \\
      Zr & -485.4694 & -511.1852 & -514.3901 & -503.1958 & -506.9346 & -511.5864 & -515.6720 & -496.3037 & -490.4785 \\
      Au & -479.1941 & -502.3245 & -506.9013 & -495.6782 & -499.0711 & -501.6877 & -506.5462 & -484.6198 & -483.6909 \\
      Bi & -478.2695 & -502.8490 & -506.2055 & -494.6965 & -498.5557 & -502.3398 & -506.7999 & -485.7536 & -483.1348 \\
      Hf & -486.5155 & -512.4535 & -515.5965 & -504.3393 & -508.1599 & -512.8332 & -517.2187 & -496.4195 & -491.7486 \\
      Hg & -472.8730 & -500.0491 & -502.1889 & -491.7544 & -494.1224 & -499.2181 & -501.8365 & -480.3761 & -479.2046 \\
      Ir & -483.7878 & -507.2024 & -510.6763 & -500.2051 & -503.2948 & -506.5209 & -509.8509 & -490.9790 & -487.3503 \\
      Os & -484.9279 & -508.6427 & -512.5954 & -501.8541 & -505.2080 & -507.8334 & -511.8734 & -491.0155 & -489.0898 \\
      Pb & -481.3587 & -504.4450 & -507.5634 & -496.2379 & -499.6114 & -503.5822 & -508.1698 & -486.8476 & -484.2646 \\
      Pt & -481.1676 & -504.3670 & -508.2170 & -499.0073 & -500.9446 & -504.2695 & -508.1465 & -487.3508 & -484.6644 \\
      Re & -483.6421 & -508.9474 & -512.8388 & -502.8952 & -505.3651 & -509.1168 & -512.6374 & -493.0621 & -489.2811 \\
      Ta & -486.3299 & -512.1371 & -515.1895 & -504.6585 & -507.5888 & -512.5295 & -516.0165 & -497.0134 & -490.5695 \\
      Tl & -480.7357 & -503.7632 & -505.9937 & -495.5894 & -497.7724 & -503.0564 & -506.7919 & -484.0732 & -482.3846 \\
      W  & -485.2596 & -510.8145 & -514.28   & -504.2714 & -506.7988 & -511.0643 & -514.2358 & -494.4330 & -490.6506 \\
      \bottomrule
    \end{tabular}
  }
\end{table}

% Supplementary Table 5: DFT calculated final state energies for adsorbates supported on n-doped graphene
\begin{table}[htbp]
\label{supp_table5:e_n_gra}
    \caption{DFT calculated final state energies in eV for adsorbates supported on nitrogen-doped graphene.}
    \resizebox{\textwidth}{!}{%
      \begin{tabular}{l *{9}{l}}
        \toprule
        Metal & *CO\textsubscript{2} & *COOH    & *CO       & *CHO
        & *CH\textsubscript{2}O  & *OCH\textsubscript{3}  & *O        & *OH       & *H        \\
        \midrule
        Al	& -652.7589	&-675.7997	&-679.7595	&-668.0296	&-671.7596	&-676.2161	&-681.1311	&-658.7849	&-656.4702 \\
        Co	& -654.3280 &-677.5394	&-681.2518	&-670.0860  &-673.9208	&-676.7761	&-680.7219	&-659.1714	&-657.8777 \\
        Cr	& -657.0479	&-680.2074	&-683.7982	&-673.0760  &-676.1859	&-679.4701	&-684.5295	&-664.7306	&-660.4141 \\
        Cu	& -650.2181	&-673.3928	&-675.5873	&-665.2174	&-668.0662	&-672.5797	&-675.7047	&-653.5770  &-652.1010 \\
        Fe	& -655.4034	&-678.5897	&-682.1186	&-671.9101	&-674.7438	&-677.8515	&-682.1216	&-661.5422	&-658.7383 \\
        Ga	& -649.9563	&-672.9807	&-676.9763	&-664.9727	&-669.0739	&-672.8536	&-677.7169	&-655.5345	&-653.9243 \\
        Ge	& -651.0391	&-674.1655	&-677.2582	&-665.9255	&-669.5282	&-673.3438	&-678.1585	&-657.4964	&-654.3444 \\
        Mn	& -656.5635	&-679.7128	&-683.1092	&-672.6753	&-675.6877	&-679.0710  &-683.3898	&-663.1584	&-659.7760 \\
        Ni	& -653.1461	&-676.3270  &-678.5166	&-668.1448	&-671.1502	&-675.5454	&-678.5411	&-656.3929	&-655.1238 \\
        Sc	& -655.2274	&-678.9985	&-682.8810  &-670.9489	&-674.9850  &-679.1986	&-684.3192	&-662.2442	&-659.0243 \\
        Ti	& -655.5937	&-680.2412	&-683.5514	&-671.9689	&-675.9724	&-680.4589	&-685.1366	&-665.2871	&-659.7132 \\
        V	  & -656.0751	&-680.2584	&-683.5739	&-672.6281	&-676.0229	&-680.5926	&-684.9400	&-665.6038	&-659.9017 \\
        Zn	& -648.2316	&-671.3824	&-674.1973	&-663.2623	&-666.3990 	&-670.5613	&-674.5436	&-652.2900  &-650.9227 \\
        Ag	& -646.9759	&-670.2666	&-673.2766	&-663.1058	&-665.7601	&-669.7801	&-673.0113	&-651.0760  &-649.9287 \\
        Cd	& -646.1687	&-669.2620  &-672.8043	&-661.4439	&-664.9473	&-668.8180  &-672.9588	&-650.7684	&-649.7670 \\
        In	& -648.6726	&-671.7218	&-675.3788	&-663.4885	&-667.5183	&-670.7536	&-675.8630  &-653.8535	&-652.4757 \\
        Mo	& -655.7944	&-680.2966	&-683.6262	&-672.7667	&-676.3955	&-680.6291	&-684.8234	&-665.8138	&-659.8169 \\
        Nb	& -656.0277	&-680.8814	&-684.2662	&-672.5576	&-676.9139	&-681.4203	&-685.7720  &-666.2363	&-660.3734 \\
        Pd	& -652.2091	&-675.3967	&-677.2760  &-667.1855	&-669.8609	&-674.6033	&-677.3498	&-655.1080  &-654.0027 \\
        Rh	& -653.6928	&-676.9529	&-680.9755	&-669.5137	&-673.6394	&-676.2879	&-680.2104	&-658.6953	&-657.6908 \\
        Ru	& -654.5425	&-678.2124	&-681.9561	&-672.0589	&-674.6177	&-678.0945	&-681.9008	&-661.3838	&-658.6572 \\
        Sb	& -649.9262	&-672.9182	&-675.6457	&-664.7128	&-667.9159	&-672.0849	&-675.8420 	&-655.1699	&-652.4240 \\
        Sn	& -650.5072	&-673.6118	&-675.7767	&-665.3842	&-668.0544	&-672.7597	&-676.5046	&-655.8947	&-652.9261 \\
        Tc	& -655.4025	&-679.3044	&-682.7454	&-672.5785	&-676.3294	&-679.4840  &-683.3209	&-664.2323	&-659.2703 \\
        Y	  & -655.2660 &-679.0430	&-682.7571	&-670.8208	&-674.7780  &-679.0775	&-684.1210  &-661.8034	&-658.8885 \\
        Zr	& -655.9118	&-680.6859	&-684.1920  &-672.1497	&-676.4014	&-680.7930  &-685.7940  &-665.4878	&-660.4163 \\
        Au	& -648.3344	&-671.3969	&-673.1469	&-663.1955	&-665.5964	&-670.5800	&-673.6092	&-651.2436	&-649.7275 \\
        Bi	& -649.1484	&-672.1225	&-675.1630  &-663.9594	&-667.2981	&-671.4617	&-675.6926	&-654.7684	&-651.8907 \\
        Hf	& -657.3364	&-682.3019	&-685.8119	&-673.6603	&-678.0090  &-682.4001	&-687.4607	&-666.9550  &-662.0651 \\
        Hg	& -644.7368	&-667.9224	&-671.2200	&-659.6906	&-663.5208	&-667.0466	&-670.9016	&-648.8944	&-648.3755 \\
        Ir	& -654.8564	&-678.1079	&-682.2865	&-670.9826	&-674.9484	&-677.4097	&-681.3088	&-660.1543	&-658.9895 \\
        Os	& -655.6722	&-679.5364	&-683.3240  &-673.4289	&-675.9254	&-679.4423	&-683.1300	&-663.0791	&-659.9901 \\
        Pb	& -649.8843	&-672.9588	&-674.5856	&-664.7553	&-667.1174	&-672.0886	&-675.8823	&-653.6827	&-651.5940 \\
        Pt	& -653.3281	&-676.5221	&-678.6094	&-668.2951	&-671.2463	&-675.7317	&-678.4334	&-656.4419	&-655.4733 \\
        Re	& -656.3492	&-680.7664	&-683.9596	&-673.9732	&-676.5164	&-681.0627	&-684.8096	&-665.9526	&-660.5620 \\
        Ta	& -657.1988	&-682.6048	&-685.8231	&-674.1666	&-678.5029	&-682.9709	&-687.2294	&-667.7321	&-661.8794 \\
        Tl	& -648.1346	&-671.1831	&-673.2953	&-662.9610  &-665.6240  &-670.3512	&-673.4567	&-651.7909	&-650.4750 \\
        W	  & -657.0204	&-682.0186	&-685.3735	&-674.3843	&-678.2483	&-682.4377	&-686.3613	&-667.4547	&-661.5686 \\
        \bottomrule
      \end{tabular}
    }
\end{table}

% Supplementary Table 6: DFT calculated final state energies for adsorbates supported on graphene with dual-vac
\begin{table}[htbp]
\label{supp_table6:e_vac_gra}
  \caption{DFT calculated final state energies in eV for adsorbates supported on graphene with dual-vacancy.}
  \resizebox{\textwidth}{!}{%
    \begin{tabular}{l *{9}{c}}
      \toprule
      Metal & *CO\textsubscript{2} & *COOH    & *CO       & *CHO
      & *CH\textsubscript{2}O  & *OCH\textsubscript{3}  & *O        & *OH       & *H        \\
      \midrule
      Al & -650.3496 & -673.3468 & -676.2273 & -665.6303 & -668.4112 & -672.5763 & -677.4453 & -655.0864 & -653.0166 \\
      Co & -652.5157 & -675.5864 & -679.3861 & -669.0614 & -672.8955 & -675.8508 & -679.5268 & -658.7929 & -655.9377 \\
      Cr & -655.0421 & -678.1256 & -681.8470 & -671.1282 & -674.2868 & -679.1554 & -683.4111 & -663.3554 & -658.4913 \\
      Cu & -649.8187 & -672.9161 & -674.9327 & -664.8190 & -669.0819 & -672.1515 & -674.7438 & -652.8789 & -653.6953 \\
      Fe & -653.7964 & -677.0127 & -680.3412 & -670.0219 & -673.9525 & -676.7631 & -681.1346 & -660.9239 & -656.9759 \\
      Ga & -649.0340 & -672.0433 & -674.9630 & -663.8505 & -667.2101 & -671.2541 & -675.5229 & -653.2086 & -651.8693 \\
      Ge & -651.0133 & -674.2532 & -677.2272 & -666.0554 & -669.5660 & -673.4485 & -677.8152 & -655.7843 & -654.2731 \\
      Mn & -654.8183 & -678.0584 & -681.5834 & -670.9593 & -674.0994 & -677.9260 & -682.6048 & -662.5702 & -658.1064 \\
      Ni & -651.5295 & -674.6692 & -677.8941 & -667.2937 & -670.7325 & -673.9392 & -677.5204 & -656.1026 & -655.6559 \\
      Sc & -652.1630 & -675.5777 & -678.6101 & -667.5024 & -670.7255 & -675.4361 & -680.0287 & -658.0834 & -654.6913 \\
      Ti & -654.2548 & -677.9326 & -681.1516 & -670.0780 & -673.2830 & -677.9113 & -682.6933 & -661.1957 & -657.4040 \\
      V  & -654.6939 & -678.0200 & -681.8606 & -670.8517 & -674.2298 & -679.2097 & -683.4928 & -663.1639 & -658.4619 \\
      Zn & -646.9805 & -670.1676 & -672.2508 & -662.0849 & -664.7199 & -669.3791 & -672.6028 & -650.3134 & -648.8961 \\
      Ag & -647.4840 & -670.5954 & -672.4765 & -662.3757 & -665.1014 & -669.8199 & -672.2893 & -650.2207 & -651.5223 \\
      Cd & -645.2038 & -668.3897 & -670.2685 & -660.2502 & -662.7479 & -667.5994 & -670.6023 & -648.2958 & -646.9658 \\
      In & -646.8189 & -669.9379 & -673.1820 & -661.7351 & -665.3833 & -669.1524 & -673.5557 & -651.3730 & -650.2004 \\
      Mo & -656.2058 & -679.7202 & -683.2424 & -672.3832 & -675.6459 & -679.9903 & -684.8002 & -664.6331 & -659.9673 \\
      Nb & -656.1010 & -679.4842 & -683.3623 & -672.0587 & -675.6683 & -679.5996 & -684.8935 & -664.2634 & -659.9617 \\
      Pd & -650.5405 & -673.6524 & -677.0119 & -666.0816 & -669.7529 & -672.9394 & -676.5341 & -655.1586 & -654.8764 \\
      Rh & -652.7998 & -675.8992 & -679.6974 & -669.1468 & -672.2062 & -676.2142 & -679.9592 & -659.4214 & -656.7388 \\
      Ru & -654.8166 & -678.0140 & -681.6090 & -671.1415 & -674.0502 & -677.9828 & -682.4415 & -662.4653 & -658.2092 \\
      Sb & -649.9882 & -673.0639 & -675.9906 & -664.8414 & -668.3621 & -672.2685 & -676.6274 & -655.1055 & -653.1345 \\
      Sn & -648.5997 & -671.7246 & -675.2765 & -663.7120 & -667.5399 & -670.8850 & -675.8484 & -653.8138 & -652.4041 \\
      Tc & -655.8173 & -679.2476 & -682.6345 & -672.1709 & -675.0076 & -679.2729 & -683.7569 & -664.1434 & -659.2908 \\
      Y  & -652.1517 & -675.4891 & -678.6682 & -667.4644 & -670.6934 & -675.1478 & -680.0013 & -658.3045 & -654.7163 \\
      Zr & -654.8760 & -678.4572 & -681.6839 & -670.6090 & -673.8291 & -678.4537 & -683.1390 & -661.4741 & -657.9892 \\
      Au & -649.3963 & -672.5116 & -674.3660 & -664.2836 & -667.3357 & -671.7273 & -674.1359 & -652.0358 & -653.1138 \\
      Bi & -648.8651 & -672.0004 & -673.7920 & -663.7934 & -666.2170 & -671.1801 & -674.0672 & -652.8628 & -650.9149 \\
      Hf & -656.1155 & -679.7805 & -683.2130 & -671.9925 & -675.3123 & -679.9256 & -684.7502 & -662.8630 & -659.4860 \\
      Hg & -645.0948 & -668.2801 & -669.9749 & -660.1062 & -662.5696 & -667.4822 & -670.2672 & -647.8582 & -646.7652 \\
      Ir & -654.4999 & -677.8310 & -681.7312 & -671.1936 & -674.7821 & -678.1749 & -681.9532 & -661.9662 & -658.5262 \\
      Os & -656.3635 & -679.8433 & -683.4557 & -673.1036 & -675.9479 & -679.6710 & -684.4704 & -664.7281 & -660.0873 \\
      Pb & -647.8007 & -670.9491 & -673.2996 & -662.7392 & -665.6502 & -670.1373 & -673.5969 & -651.6692 & -650.4432 \\
      Pt & -652.4961 & -675.6092 & -678.9183 & -668.2053 & -671.6867 & -674.8876 & -678.4355 & -657.4936 & -656.2759 \\
      Re & -657.2440 & -681.0777 & -684.5183 & -673.9603 & -676.8950 & -681.2383 & -685.8328 & -666.1615 & -661.2885 \\
      Ta & -657.4068 & -681.0711 & -685.1062 & -673.5947 & -677.3738 & -681.6985 & -686.7561 & -666.0327 & -661.7361 \\
      Tl & -645.7286 & -668.8734 & -671.7205 & -660.7259 & -664.0001 & -668.1110 & -672.0305 & -649.7198 & -648.7563 \\
      W  & -657.5901 & -681.6722 & -685.1010 & -674.1028 & -677.7828 & -682.4493 & -686.6385 & -666.9364 & -661.8641 \\
      \bottomrule
    \end{tabular}
  }
\end{table}

% Supplementary Table 7: ZPE for adsorbates supported on g-c3n4
\begin{table}[htbp]
\label{supp_table7:zpe_g-c3n4}
  \caption{Zero-point energies in eV for relaxed adsorbates supported
    on g-C\textsubscript{3}N\textsubscript{4} at 298.15 K.}
  \resizebox{\textwidth}{!}{%
    \begin{tabular}{l *{9}{l}}
      \toprule
      Metal & *CO\textsubscript{2} & *COOH & *CO & *CHO
      & *CH\textsubscript{2}O & *OCH\textsubscript{3} & *O & *OH    & *H     \\
      \midrule
      Al & 0.2905 & 0.6130 & 0.1743 & 0.4606 & 0.7456 & 1.0974 & 0.0629 & 0.3521 & 0.1925 \\
      Co & 0.3038 & 0.5895 & 0.2051 & 0.4186 & 0.6269 & 1.0663 & 0.0571 & 0.3248 & 0.1507 \\
      Cr & 0.3140 & 0.5964 & 0.1640 & 0.4374 & 0.7725 & 1.0779 & 0.0670 & 0.3286 & 0.1626 \\
      Cu & 0.3157 & 0.6016 & 0.2045 & 0.4498 & 0.7797 & 1.0751 & 0.0494 & 0.3332 & 0.1577 \\
      Fe & 0.3088 & 0.5859 & 0.1931 & 0.4437 & 0.7720 & 1.0726 & 0.0599 & 0.3326 & 0.1550 \\
      Ga & 0.3102 & 0.6131 & 0.1448 & 0.4665 & 0.7186 & 1.0680 & 0.0608 & 0.3202 & 0.1813 \\
      Ge & 0.3091 & 0.6069 & 0.1396 & 0.4540 & 0.7269 & 1.0881 & 0.0581 & 0.3506 & 0.1904 \\
      Mn & 0.3022 & 0.5876 & 0.1733 & 0.4333 & 0.7637 & 1.0622 & 0.0687 & 0.3219 & 0.1432 \\
      Ni & 0.3023 & 0.5934 & 0.2074 & 0.4546 & 0.7662 & 1.0623 & 0.0542 & 0.3313 & 0.1611 \\
      Sc & 0.2880 & 0.6039 & 0.1771 & 0.4544 & 0.7473 & 1.0806 & 0.0657 & 0.3156 & 0.1480 \\
      Ti & 0.3044 & 0.6088 & 0.1889 & 0.4664 & 0.7863 & 1.0862 & 0.0796 & 0.3417 & 0.1590 \\
      V  & 0.3081 & 0.6005 & 0.2031 & 0.4408 & 0.7868 & 1.0882 & 0.0792 & 0.3388 & 0.1660 \\
      Zn & 0.3243 & 0.6080 & 0.1729 & 0.4601 & 0.7277 & 1.0832 & 0.0514 & 0.3468 & 0.1726 \\
      Ag & 0.3167 & 0.5993 & 0.1815 & 0.4457 & 0.7264 & 1.0590 & 0.0390 & 0.3304 & 0.1576 \\
      Cd & 0.3242 & 0.6007 & 0.1533 & 0.4502 & 0.7292 & 1.0604 & 0.0983 & 0.3322 & 0.1439 \\
      In & 0.3116 & 0.5687 & 0.1399 & 0.4025 & 0.7158 & 1.0551 & 0.0523 & 0.3315 & 0.1141 \\
      Mo & 0.3109 & 0.6071 & 0.2163 & 0.4441 & 0.7970 & 1.0825 & 0.0765 & 0.3312 & 0.1851 \\
      Nb & 0.3101 & 0.5971 & 0.2013 & 0.4340 & 0.7960 & 1.0908 & 0.0798 & 0.3329 & 0.1598 \\
      Pd & 0.3148 & 0.6091 & 0.1996 & 0.4778 & 0.7402 & 1.0777 & 0.0525 & 0.3425 & 0.1508 \\
      Rh & 0.3064 & 0.6180 & 0.2215 & 0.4654 & 0.7649 & 1.0844 & 0.0615 & 0.3643 & 0.1946 \\
      Ru & 0.3060 & 0.6201 & 0.2104 & 0.4668 & 0.7761 & 1.0800 & 0.0663 & 0.3421 & 0.2026 \\
      Sb & 0.3050 & 0.6244 & 0.1461 & 0.4696 & 0.7473 & 1.1052 & 0.0689 & 0.3676 & 0.2043 \\
      Sn & 0.3115 & 0.5941 & 0.1441 & 0.4373 & 0.7202 & 1.0761 & 0.0591 & 0.3397 & 0.1601 \\
      Tc & 0.3170 & 0.6102 & 0.2178 & 0.4543 & 0.7890 & 1.0807 & 0.0698 & 0.3442 & 0.2008 \\
      Y  & 0.2768 & 0.5949 & 0.1712 & 0.4467 & 0.7278 & 1.0780 & 0.0573 & 0.3206 & 0.1349 \\
      Zr & 0.2939 & 0.6068 & 0.1888 & 0.4631 & 0.7737 & 1.0875 & 0.0744 & 0.3155 & 0.1592 \\
      Au & 0.3123 & 0.6245 & 0.2147 & 0.4761 & 0.7312 & 1.0845 & 0.0527 & 0.3432 & 0.2066 \\
      Bi & 0.3076 & 0.6148 & 0.1485 & 0.4612 & 0.7392 & 1.0935 & 0.0611 & 0.3554 & 0.1893 \\
      Hf & 0.2928 & 0.6104 & 0.1880 & 0.4658 & 0.7769 & 1.0954 & 0.0765 & 0.3218 & 0.1728 \\
      Hg & 0.3220 & 0.6221 & 0.1486 & 0.4707 & 0.7338 & 1.0809 & 0.0541 & 0.3469 & 0.1924 \\
      Ir & 0.3127 & 0.6311 & 0.2271 & 0.4818 & 0.7530 & 1.0740 & 0.1084 & 0.3314 & 0.2142 \\
      Os & 0.3122 & 0.6260 & 0.2258 & 0.4796 & 0.7531 & 1.0911 & 0.0670 & 0.3501 & 0.2154 \\
      Pb & 0.3152 & 0.5834 & 0.1401 & 0.4239 & 0.7113 & 1.0486 & 0.0513 & 0.3043 & 0.1352 \\
      Pt & 0.3144 & 0.6361 & 0.2225 & 0.4908 & 0.7569 & 1.0921 & 0.0603 & 0.3672 & 0.2058 \\
      Re & 0.3171 & 0.6175 & 0.2213 & 0.4665 & 0.8021 & 1.0985 & 0.0780 & 0.3597 & 0.2032 \\
      Ta & 0.3099 & 0.5977 & 0.2000 & 0.4414 & 0.7993 & 1.0880 & 0.0847 & 0.3410 & 0.1780 \\
      Tl & 0.3114 & 0.5614 & 0.1389 & 0.3494 & 0.7105 & 1.0410 & 0.0246 & 0.3210 & 0.0951 \\
      W  & 0.3146 & 0.6059 & 0.2160 & 0.4456 & 0.8033 & 1.0886 & 0.0725 & 0.3387 & 0.1834 \\
      \bottomrule
    \end{tabular}
  }
\end{table}

% Supplementary Table 8: Entropy corrections for adsorbate@g-C3N4
\begin{table}[htbp]
\label{supp_table8:s_g-c3n4}
  \caption{Entropy corrections ($-T \cdot S$) in eV for relaxed
    adsorbates supported on g-C\textsubscript{3}N\textsubscript{4} at 298.15 K.}
  \resizebox{\textwidth}{!}{%
    \begin{tabular}{l *{9}{l}}
      \toprule
      Metal & *CO\textsubscript{2} & *COOH& *CO     & *CHO
      & *CH\textsubscript{2}O& *OCH\textsubscript{3}& *O      & *OH     & *H      \\
      \midrule
      Al & -0.2195 & -0.2476 & -0.1874 & -0.1939 & -0.2359 & -0.1759 & -0.0894 & -0.1010 & -0.0124 \\
      Co & -0.2359 & -0.2665 & -0.1544 & -0.1572 & -0.1755 & -0.2095 & -0.1078 & -0.1510 & -0.0341 \\
      Cr & -0.1862 & -0.2605 & -0.2075 & -0.2069 & -0.1478 & -0.1733 & -0.0790 & -0.1437 & -0.0221 \\
      Cu & -0.3098 & -0.2403 & -0.1655 & -0.1803 & -0.1647 & -0.2378 & -0.0977 & -0.1375 & -0.0297 \\
      Fe & -0.2184 & -0.2028 & -0.1676 & -0.1797 & -0.1742 & -0.2640 & -0.0815 & -0.1285 & -0.0215 \\
      Ga & -0.2602 & -0.2558 & -0.2291 & -0.1881 & -0.1252 & -0.2576 & -0.0897 & -0.1441 & -0.0182 \\
      Ge & -0.1978 & -0.2423 & -0.1635 & -0.1802 & -0.2104 & -0.2364 & -0.1178 & -0.1052 & -0.0087 \\
      Mn & -0.2142 & -0.2678 & -0.1863 & -0.2059 & -0.1788 & -0.2752 & -0.0808 & -0.1425 & -0.0261 \\
      Ni & -0.2460 & -0.1948 & -0.1592 & -0.1224 & -0.1275 & -0.2506 & -0.0997 & -0.1384 & -0.0224 \\
      Sc & -0.2227 & -0.2369 & -0.1843 & -0.1667 & -0.1817 & -0.2035 & -0.0728 & -0.1416 & -0.0226 \\
      Ti & -0.1978 & -0.2263 & -0.1685 & -0.1389 & -0.1414 & -0.1795 & -0.0578 & -0.1013 & -0.0195 \\
      V  & -0.2603 & -0.2511 & -0.1452 & -0.2028 & -0.1360 & -0.2374 & -0.0612 & -0.1082 & -0.0186 \\
      Zn & -0.2928 & -0.2543 & -0.1909 & -0.1959 & -0.2202 & -0.2331 & -0.0910 & -0.1185 & -0.0228 \\
      Ag & -0.2458 & -0.2428 & -0.1910 & -0.1890 & -0.3097 & -0.2421 & -0.1060 & -0.1392 & -0.0276 \\
      Cd & -0.2961 & -0.2561 & -0.2368 & -0.2023 & -0.2076 & -0.2630 & -0.0270 & -0.1408 & -0.0337 \\
      In & -0.1978 & -0.2895 & -0.1071 & -0.1840 & -0.1338 & -0.2826 & -0.1011 & -0.1386 & -0.0405 \\
      Mo & -0.1912 & -0.2310 & -0.1310 & -0.1680 & -0.1269 & -0.2228 & -0.0641 & -0.1496 & -0.0155 \\
      Nb & -0.1896 & -0.2540 & -0.1460 & -0.1924 & -0.1290 & -0.2147 & -0.0575 & -0.1177 & -0.0261 \\
      Pd & -0.2405 & -0.2385 & -0.1649 & -0.1740 & -0.2538 & -0.2201 & -0.0900 & -0.1143 & -0.0518 \\
      Rh & -0.2408 & -0.2385 & -0.1355 & -0.1734 & -0.1853 & -0.2247 & -0.0863 & -0.0866 & -0.0222 \\
      Ru & -0.2197 & -0.2304 & -0.1450 & -0.1769 & -0.1655 & -0.1574 & -0.0832 & -0.1294 & -0.0138 \\
      Sb & -0.2354 & -0.2277 & -0.1320 & -0.1793 & -0.2211 & -0.1961 & -0.0616 & -0.0808 & -0.0090 \\
      Sn & -0.2606 & -0.2544 & -0.2227 & -0.2021 & -0.3045 & -0.2568 & -0.0787 & -0.1171 & -0.0160 \\
      Tc & -0.1875 & -0.2373 & -0.1347 & -0.1733 & -0.1485 & -0.2301 & -0.0697 & -0.1007 & -0.0121 \\
      Y  & -0.2557 & -0.2528 & -0.1929 & -0.1886 & -0.2130 & -0.2081 & -0.0824 & -0.1393 & -0.0285 \\
      Zr & -0.2083 & -0.2325 & -0.1624 & -0.1459 & -0.1574 & -0.2014 & -0.0640 & -0.1556 & -0.0208 \\
      Au & -0.1901 & -0.2267 & -0.1484 & -0.1846 & -0.2939 & -0.2307 & -0.0857 & -0.1271 & -0.0138 \\
      Bi & -0.2366 & -0.2392 & -0.1915 & -0.1899 & -0.2361 & -0.1986 & -0.0683 & -0.0934 & -0.0100 \\
      Hf & -0.2043 & -0.2219 & -0.1618 & -0.1418 & -0.1489 & -0.1981 & -0.0610 & -0.1414 & -0.0163 \\
      Hg & -0.2980 & -0.2371 & -0.2812 & -0.1995 & -0.3039 & -0.2465 & -0.0878 & -0.1245 & -0.0175 \\
      Ir & -0.2097 & -0.2290 & -0.1261 & -0.1803 & -0.1913 & -0.2221 & -0.0377 & -0.1379 & -0.0159 \\
      Os & -0.1795 & -0.2235 & -0.1294 & -0.1519 & -0.1609 & -0.2027 & -0.0680 & -0.1093 & -0.0124 \\
      Pb & -0.2546 & -0.2691 & -0.1666 & -0.2165 & -0.2641 & -0.2520 & -0.1007 & -0.1097 & -0.0257 \\
      Pt & -0.2411 & -0.2029 & -0.1496 & -0.1593 & -0.2009 & -0.2246 & -0.0950 & -0.0937 & -0.0137 \\
      Re & -0.1869 & -0.2262 & -0.1293 & -0.1697 & -0.1266 & -0.2165 & -0.0644 & -0.0953 & -0.0146 \\
      Ta & -0.1872 & -0.2532 & -0.1523 & -0.1847 & -0.1265 & -0.1798 & -0.0535 & -0.1040 & -0.0177 \\
      Tl & -0.2599 & -0.2833 & -0.1807 & -0.1538 & -0.2422 & -0.1501 & -0.0126 & -0.1338 & -0.0533 \\
      W  & -0.1791 & -0.2368 & -0.1335 & -0.1723 & -0.1225 & -0.2253 & -0.0673 & -0.1180 & -0.0201 \\
      \bottomrule
    \end{tabular}
  }
\end{table}

% Supplementary Table 9: ZPE for adsorbates supported on N-doped graphene
\begin{table}[htbp]
\label{supp_table9:zpe_n_gra}
  \caption{Zero-point energies in eV for relaxed adsorbates
    supported on nitrogen-doped graphene at 298.15 K.}
  \resizebox{\textwidth}{!}{%
    \begin{tabular}{l *{9}{c}}
      \toprule
      Metal & *CO\textsubscript{2} & *COOH & *CO & *CHO
      & *CH\textsubscript{2}O & *OCH\textsubscript{3} & *O & *OH    & *H     \\
      \midrule
      Al & 0.3093 & 0.6024 & 0.1683 & 0.4395 & 0.7509 & 1.0855 & 0.0552 & 0.3401 & 0.1864 \\
      Co & 0.3140 & 0.6247 & 0.2044 & 0.4787 & 0.7345 & 1.0615 & 0.0606 & 0.3315 & 0.2006 \\
      Cr & 0.3108 & 0.5983 & 0.2007 & 0.4410 & 0.7211 & 1.0824 & 0.0828 & 0.3253 & 0.1625 \\
      Cu & 0.3145 & 0.5943 & 0.1460 & 0.4377 & 0.7196 & 1.0160 & 0.0407 & 0.3104 & 0.1523 \\
      Fe & 0.3117 & 0.6148 & 0.2195 & 0.4634 & 0.7877 & 1.0678 & 0.0703 & 0.3163 & 0.1858 \\
      Ga & 0.3129 & 0.6080 & 0.1561 & 0.4537 & 0.7495 & 1.0826 & 0.0573 & 0.3420 & 0.1897 \\
      Ge & 0.3131 & 0.6206 & 0.1402 & 0.4629 & 0.7281 & 1.1014 & 0.0686 & 0.3538 & 0.2110 \\
      Mn & 0.3120 & 0.6111 & 0.2128 & 0.4601 & 0.8078 & 1.0640 & 0.0793 & 0.3148 & 0.2040 \\
      Ni & 0.3110 & 0.6059 & 0.1485 & 0.4602 & 0.7283 & 1.0467 & 0.0402 & 0.3168 & 0.1620 \\
      Sc & 0.2741 & 0.5910 & 0.1620 & 0.4456 & 0.7019 & 1.0758 & 0.0575 & 0.3166 & 0.1338 \\
      Ti & 0.2920 & 0.6020 & 0.1781 & 0.4525 & 0.7593 & 1.0831 & 0.0731 & 0.2981 & 0.1481 \\
      V  & 0.2984 & 0.6055 & 0.1865 & 0.4609 & 0.7756 & 1.0843 & 0.0786 & 0.3171 & 0.1525 \\
      Zn & 0.3143 & 0.5928 & 0.1554 & 0.4357 & 0.7116 & 1.0521 & 0.0394 & 0.3248 & 0.1640 \\
      Ag & 0.3127 & 0.5862 & 0.1862 & 0.4307 & 0.7518 & 1.0419 & 0.0356 & 0.3231 & 0.1308 \\
      Cd & 0.3055 & 0.5910 & 0.1518 & 0.4402 & 0.7312 & 1.0524 & 0.0366 & 0.3189 & 0.1483 \\
      In & 0.3096 & 0.5986 & 0.1414 & 0.4477 & 0.7152 & 1.0724 & 0.0437 & 0.3300 & 0.1609 \\
      Mo & 0.2949 & 0.6111 & 0.1914 & 0.4728 & 0.7806 & 1.0887 & 0.0795 & 0.3253 & 0.1500 \\
      Nb & 0.2946 & 0.5996 & 0.1767 & 0.4601 & 0.7874 & 1.0896 & 0.0742 & 0.3048 & 0.1515 \\
      Pd & 0.3122 & 0.5993 & 0.1444 & 0.4490 & 0.7303 & 1.0342 & 0.0300 & 0.3047 & 0.1558 \\
      Rh & 0.3115 & 0.6288 & 0.1969 & 0.4826 & 0.7388 & 1.0867 & 0.0602 & 0.3453 & 0.2025 \\
      Ru & 0.3152 & 0.6122 & 0.2180 & 0.4569 & 0.7918 & 1.0947 & 0.0698 & 0.3483 & 0.1873 \\
      Sb & 0.3115 & 0.5874 & 0.1397 & 0.4349 & 0.7184 & 1.0467 & 0.0646 & 0.3351 & 0.1752 \\
      Sn & 0.3112 & 0.6002 & 0.1400 & 0.4367 & 0.7239 & 1.0836 & 0.0568 & 0.3359 & 0.1710 \\
      Tc & 0.3053 & 0.6110 & 0.2105 & 0.4251 & 0.7934 & 1.0870 & 0.0807 & 0.3496 & 0.1741 \\
      Y  & 0.2709 & 0.5835 & 0.1580 & 0.4307 & 0.7059 & 1.0703 & 0.0500 & 0.3200 & 0.1226 \\
      Zr & 0.2719 & 0.5963 & 0.1723 & 0.4430 & 0.7327 & 1.0846 & 0.0669 & 0.3176 & 0.1472 \\
      Au & 0.3137 & 0.5480 & 0.1442 & 0.3577 & 0.7209 & 0.9864 & 0.0174 & 0.2881 & 0.1027 \\
      Bi & 0.3111 & 0.5778 & 0.1378 & 0.4190 & 0.7133 & 1.0475 & 0.0550 & 0.3241 & 0.1533 \\
      Hf & 0.2699 & 0.6002 & 0.1714 & 0.4452 & 0.7372 & 1.0904 & 0.0687 & 0.3242 & 0.1569 \\
      Hg & 0.3113 & 0.6082 & 0.1397 & 0.4569 & 0.7258 & 1.0720 & 0.0412 & 0.3304 & 0.1625 \\
      Ir & 0.3132 & 0.6309 & 0.2106 & 0.4856 & 0.7390 & 1.0888 & 0.0626 & 0.3456 & 0.2022 \\
      Os & 0.3156 & 0.6148 & 0.2179 & 0.4592 & 0.7971 & 1.0984 & 0.0768 & 0.3463 & 0.1812 \\
      Pb & 0.3116 & 0.5352 & 0.1385 & 0.3677 & 0.7147 & 1.0532 & 0.0474 & 0.3172 & 0.1302 \\
      Pt & 0.3139 & 0.6134 & 0.1439 & 0.4671 & 0.7359 & 1.0406 & 0.0482 & 0.3114 & 0.1736 \\
      Re & 0.3035 & 0.6045 & 0.2095 & 0.4566 & 0.7995 & 1.0957 & 0.0846 & 0.3493 & 0.1818 \\
      Ta & 0.2751 & 0.6019 & 0.1869 & 0.4630 & 0.7724 & 1.0937 & 0.0751 & 0.3060 & 0.1669 \\
      Tl & 0.3131 & 0.5988 & 0.1402 & 0.4439 & 0.7193 & 1.0169 & 0.0573 & 0.3244 & 0.1506 \\
      W  & 0.2889 & 0.6117 & 0.1927 & 0.4777 & 0.7852 & 1.0941 & 0.0810 & 0.3273 & 0.1832 \\
      \bottomrule
    \end{tabular}
  }
\end{table}

% Supplementary Table 10: Entropy corrections for adsorbate@N-doped graphene
\begin{table}[htbp]
\label{supp_table10:s_n_gra}
  \caption{Entropy corrections ($-T \cdot S$) in eV for relaxed
    adsorbates supported on nitrogen-doped graphene at 298.15 K.}
  \resizebox{\textwidth}{!}{%
    \begin{tabular}{l *{9}{l}}
      \toprule
      Metal & *CO\textsubscript{2} & *COOH& *CO     & *CHO
      & *CH\textsubscript{2}O& *OCH\textsubscript{3}& *O      & *OH     & *H      \\
      \midrule
      Al & -0.1981 & -0.2564 & -0.1979 & -0.1424 & -0.2051 & -0.1839 & -0.0794 & -0.1211 & -0.0113 \\
      Co & -0.2358 & -0.2213 & -0.1589 & -0.1584 & -0.2025 & -0.2223 & -0.0701 & -0.0817 & -0.0123 \\
      Cr & -0.2593 & -0.1973 & -0.1568 & -0.2037 & -0.2291 & -0.1679 & -0.0569 & -0.0875 & -0.0228 \\
      Cu & -0.2457 & -0.2636 & -0.2132 & -0.2040 & -0.1267 & -0.2137 & -0.1188 & -0.1711 & -0.0226 \\
      Fe & -0.1936 & -0.2385 & -0.1356 & -0.1758 & -0.1985 & -0.1910 & -0.0647 & -0.0892 & -0.0172 \\
      Ga & -0.2565 & -0.1937 & -0.1600 & -0.2025 & -0.2139 & -0.1766 & -0.0727 & -0.1269 & -0.0126 \\
      Ge & -0.2469 & -0.2413 & -0.1027 & -0.1937 & -0.1687 & -0.1551 & -0.0690 & -0.1091 & -0.0080 \\
      Mn & -0.1929 & -0.2402 & -0.1409 & -0.1736 & -0.1949 & -0.2011 & -0.0563 & -0.1631 & -0.0160 \\
      Ni & -0.1239 & -0.2440 & -0.2027 & -0.1785 & -0.2318 & -0.1792 & -0.0929 & -0.1520 & -0.0170 \\
      Sc & -0.1938 & -0.2647 & -0.1418 & -0.2024 & -0.2291 & -0.2139 & -0.0790 & -0.1516 & -0.0293 \\
      Ti & -0.2093 & -0.2423 & -0.1798 & -0.1640 & -0.1765 & -0.1433 & -0.0692 & -0.1462 & -0.0262 \\
      V  & -0.2112 & -0.2373 & -0.1734 & -0.1578 & -0.1642 & -0.0767 & -0.0621 & -0.1522 & -0.0273 \\
      Zn & -0.2567 & -0.2001 & -0.2491 & -0.2135 & -0.1581 & -0.1396 & -0.1091 & -0.1668 & -0.0211 \\
      Ag & -0.2600 & -0.2035 & -0.1802 & -0.2231 & -0.1723 & -0.1531 & -0.1293 & -0.1487 & -0.0596 \\
      Cd & -0.2615 & -0.2042 & -0.1827 & -0.2185 & -0.2727 & -0.2094 & -0.1192 & -0.1850 & -0.0320 \\
      In & -0.1886 & -0.2021 & -0.1593 & -0.2221 & -0.1949 & -0.2061 & -0.1158 & -0.1603 & -0.0238 \\
      Mo & -0.2049 & -0.2075 & -0.1659 & -0.1327 & -0.1650 & -0.2041 & -0.0605 & -0.1258 & -0.0437 \\
      Nb & -0.2072 & -0.2300 & -0.1145 & -0.1439 & -0.1403 & -0.2086 & -0.0684 & -0.1863 & -0.0251 \\
      Pd & -0.1843 & -0.2620 & -0.1476 & -0.2000 & -0.1645 & -0.2057 & -0.1131 & -0.1688 & -0.0223 \\
      Rh & -0.2986 & -0.2220 & -0.1669 & -0.1667 & -0.1953 & -0.1423 & -0.0693 & -0.1229 & -0.0149 \\
      Ru & -0.1998 & -0.1779 & -0.1375 & -0.1827 & -0.1455 & -0.2017 & -0.0653 & -0.1103 & -0.0202 \\
      Sb & -0.1938 & -0.1984 & -0.1084 & -0.2024 & -0.1970 & -0.0996 & -0.0780 & -0.1183 & -0.0109 \\
      Sn & -0.1931 & -0.2040 & -0.1708 & -0.2214 & -0.1801 & -0.2006 & -0.0911 & -0.1474 & -0.0185 \\
      Tc & -0.2038 & -0.2277 & -0.1483 & -0.1505 & -0.1515 & -0.1810 & -0.0569 & -0.1008 & -0.0234 \\
      Y  & -0.1994 & -0.2704 & -0.1507 & -0.1605 & -0.2919 & -0.2162 & -0.0981 & -0.1548 & -0.0344 \\
      Zr & -0.2233 & -0.2608 & -0.1845 & -0.1849 & -0.2016 & -0.2144 & -0.0776 & -0.1500 & -0.0242 \\
      Au & -0.1872 & -0.1923 & -0.2237 & -0.1738 & -0.1907 & -0.2639 & -0.1497 & -0.1276 & -0.0301 \\
      Bi & -0.1978 & -0.2736 & -0.1731 & -0.1581 & -0.2369 & -0.0995 & -0.0937 & -0.1386 & -0.0171 \\
      Hf & -0.2235 & -0.2562 & -0.1836 & -0.1839 & -0.1987 & -0.2082 & -0.0759 & -0.1417 & -0.0211 \\
      Hg & -0.1907 & -0.1929 & -0.1053 & -0.0955 & -0.1765 & -0.2558 & -0.1157 & -0.1694 & -0.0357 \\
      Ir & -0.2967 & -0.2218 & -0.1486 & -0.1659 & -0.2038 & -0.2134 & -0.0723 & -0.1326 & -0.0200 \\
      Os & -0.1948 & -0.2462 & -0.1422 & -0.1945 & -0.1448 & -0.2046 & -0.0598 & -0.1205 & -0.0322 \\
      Pb & -0.1979 & -0.2320 & -0.1108 & -0.2414 & -0.1437 & -0.2639 & -0.1126 & -0.1465 & -0.0238 \\
      Pt & -0.1770 & -0.2479 & -0.1481 & -0.1915 & -0.2820 & -0.2030 & -0.0791 & -0.1666 & -0.0228 \\
      Re & -0.1958 & -0.2535 & -0.1508 & -0.1867 & -0.1462 & -0.2483 & -0.0561 & -0.1018 & -0.0249 \\
      Ta & -0.2078 & -0.2197 & -0.1675 & -0.1373 & -0.1542 & -0.2084 & -0.0698 & -0.1979 & -0.0195 \\
      Tl & -0.2573 & -0.2043 & -0.1719 & -0.2270 & -0.1288 & -0.2326 & -0.0688 & -0.1876 & -0.0332 \\
      W  & -0.2007 & -0.2007 & -0.1642 & -0.1194 & -0.1565 & -0.2032 & -0.0607 & -0.1275 & -0.0202 \\
      \bottomrule
    \end{tabular}
  }
\end{table}

% Supplementary Table 11: ZPE for adsorbates supported on graphene with dual-vac
\begin{table}[htbp]
\label{supp_table11:zpe_vac_gra}
  \caption{Zero-point energies in eV for relaxed adsorbates
    supported on graphene with dual-vacancy at 298.15 K.}
  \resizebox{\textwidth}{!}{%
    \begin{tabular}{l *{9}{l}}
      \toprule
      Metal & *CO\textsubscript{2} & *COOH & *CO & *CHO
      & *CH\textsubscript{2}O & *OCH\textsubscript{3}          & *O     & *OH    & *H     \\
      \midrule
      Al & 0.3140 & 0.5988 & 0.1888 & 0.4386 & 0.7218 & 1.0769 & 0.0472 & 0.3366 & 0.1729 \\
      Co & 0.3080 & 0.6179 & 0.2173 & 0.4899 & 0.7928 & 1.0616 & 0.0573 & 0.3458 & 0.1726 \\
      Cr & 0.3318 & 0.5862 & 0.1957 & 0.4277 & 0.8826 & 1.0733 & 0.0792 & 0.3294 & 0.1560 \\
      Cu & 0.3140 & 0.6023 & 0.1769 & 0.5229 & 0.7172 & 1.0131 & 0.0323 & 0.3298 & 0.2857 \\
      Fe & 0.3177 & 0.6110 & 0.2083 & 0.5038 & 0.7828 & 1.0802 & 0.0776 & 0.3515 & 0.1987 \\
      Ga & 0.3126 & 0.6056 & 0.1492 & 0.4494 & 0.7196 & 1.0788 & 0.0446 & 0.3370 & 0.1810 \\
      Ge & 0.3138 & 0.6171 & 0.1471 & 0.4657 & 0.7196 & 1.0935 & 0.0540 & 0.3514 & 0.2028 \\
      Mn & 0.3167 & 0.6141 & 0.2050 & 0.4803 & 0.8009 & 1.0809 & 0.0756 & 0.3411 & 0.1594 \\
      Ni & 0.3134 & 0.6251 & 0.1942 & 0.4571 & 0.7230 & 1.0738 & 0.0496 & 0.3521 & 0.2877 \\
      Sc & 0.3229 & 0.5874 & 0.1668 & 0.4367 & 0.7358 & 1.0673 & 0.0533 & 0.3129 & 0.1288 \\
      Ti & 0.3240 & 0.6049 & 0.1749 & 0.4562 & 0.7202 & 1.0694 & 0.0670 & 0.3143 & 0.1559 \\
      V  & 0.2937 & 0.6042 & 0.1933 & 0.4593 & 0.8437 & 1.0745 & 0.0753 & 0.3319 & 0.1687 \\
      Zn & 0.3123 & 0.5888 & 0.1564 & 0.4317 & 0.7111 & 1.0363 & 0.0324 & 0.3157 & 0.1363 \\
      Ag & 0.3172 & 0.5984 & 0.1399 & 0.4312 & 0.7199 & 1.0034 & 0.0258 & 0.3141 & 0.2932 \\
      Cd & 0.3120 & 0.5796 & 0.1487 & 0.4176 & 0.7144 & 1.0198 & 0.0278 & 0.3084 & 0.1280 \\
      In & 0.3134 & 0.5973 & 0.1458 & 0.4443 & 0.7170 & 1.0666 & 0.0418 & 0.3236 & 0.1610 \\
      Mo & 0.3052 & 0.5923 & 0.1942 & 0.4668 & 0.7718 & 1.0844 & 0.0745 & 0.3350 & 0.1755 \\
      Nb & 0.2952 & 0.6071 & 0.1862 & 0.4605 & 0.7505 & 1.0749 & 0.0721 & 0.3274 & 0.1717 \\
      Pd & 0.3151 & 0.6191 & 0.1872 & 0.4609 & 0.7187 & 1.0525 & 0.0397 & 0.3371 & 0.2963 \\
      Rh & 0.2870 & 0.6096 & 0.2076 & 0.4297 & 0.7850 & 1.0628 & 0.0604 & 0.3590 & 0.2678 \\
      Ru & 0.3000 & 0.6040 & 0.2097 & 0.4645 & 0.7812 & 1.0710 & 0.0790 & 0.3319 & 0.1758 \\
      Sb & 0.3122 & 0.6143 & 0.1436 & 0.4604 & 0.7258 & 1.0938 & 0.0608 & 0.3481 & 0.1912 \\
      Sn & 0.3132 & 0.6077 & 0.1493 & 0.4550 & 0.7182 & 1.0807 & 0.0517 & 0.3338 & 0.1770 \\
      Tc & 0.3051 & 0.6025 & 0.2086 & 0.4585 & 0.7777 & 1.0690 & 0.0799 & 0.3285 & 0.2898 \\
      Y  & 0.3219 & 0.5847 & 0.1632 & 0.4299 & 0.7245 & 1.0646 & 0.0557 & 0.3182 & 0.1164 \\
      Zr & 0.3213 & 0.5966 & 0.1669 & 0.4527 & 0.7200 & 1.0620 & 0.0568 & 0.3151 & 0.1510 \\
      Au & 0.3172 & 0.5927 & 0.1448 & 0.4677 & 0.7261 & 0.9827 & 0.0291 & 0.2904 & 0.2898 \\
      Bi & 0.3108 & 0.6062 & 0.1406 & 0.4521 & 0.7185 & 1.0309 & 0.0528 & 0.3289 & 0.1696 \\
      Hf & 0.3241 & 0.5981 & 0.1713 & 0.4543 & 0.7236 & 1.0751 & 0.0586 & 0.3142 & 0.1550 \\
      Hg & 0.3130 & 0.5763 & 0.1421 & 0.3959 & 0.7161 & 0.9951 & 0.0177 & 0.2917 & 0.1258 \\
      Ir & 0.3054 & 0.6129 & 0.2147 & 0.4199 & 0.7939 & 1.0680 & 0.0712 & 0.3603 & 0.1863 \\
      Os & 0.3033 & 0.6068 & 0.2155 & 0.4695 & 0.7674 & 1.0779 & 0.0823 & 0.3330 & 0.1401 \\
      Pb & 0.3118 & 0.6044 & 0.1424 & 0.4518 & 0.7199 & 1.0740 & 0.0502 & 0.3250 & 0.1623 \\
      Pt & 0.3124 & 0.6242 & 0.1985 & 0.4680 & 0.7263 & 1.0514 & 0.0477 & 0.3406 & 0.2725 \\
      Re & 0.3033 & 0.6034 & 0.2108 & 0.4601 & 0.7864 & 1.0807 & 0.0780 & 0.3317 & 0.2191 \\
      Ta & 0.2938 & 0.6078 & 0.1872 & 0.4601 & 0.7707 & 1.0754 & 0.0732 & 0.3243 & 0.1783 \\
      Tl & 0.3091 & 0.5948 & 0.1424 & 0.4406 & 0.7123 & 1.0568 & 0.0342 & 0.3173 & 0.1569 \\
      W  & 0.3066 & 0.6062 & 0.1912 & 0.3984 & 0.8010 & 1.0777 & 0.0787 & 0.3333 & 0.1815 \\
      \bottomrule
    \end{tabular}
  }
\end{table}

% Supplementary Table 12: Entropy corrections for adsorbate@graphene with dual-vac
\begin{table}[htbp]
\label{supp_table12:s_vac_gra}
  \caption{Entropy corrections ($-T \cdot S$) in eV for relaxed
    adsorbates supported on graphene with dual-vacancy at 298.15 K.}
  \resizebox{\textwidth}{!}{%
    \begin{tabular}{l *{9}{l}}
      \toprule
        Metal & *CO\textsubscript{2} & *COOH& *CO     & *CHO
        & *CH\textsubscript{2}O& *OCH\textsubscript{3}& *O      & *OH     & *H      \\
        \midrule
        Al & -0.2501 & -0.2622 & -0.1799 & -0.2106 & -0.3101 & -0.2050 & -0.0980 & -0.1303 & -0.0165 \\
        Co & -0.1962 & -0.2352 & -0.1529 & -0.1520 & -0.1515 & -0.0821 & -0.0427 & -0.1165 & -0.0376 \\
        Cr & -0.2178 & -0.2281 & -0.1603 & -0.2297 & -0.0849 & -0.0961 & -0.0623 & -0.1200 & -0.0250 \\
        Cu & -0.1907 & -0.2528 & -0.1276 & -0.1239 & -0.1974 & -0.2279 & -0.0711 & -0.1357 & -0.0044 \\
        Fe & -0.2328 & -0.2473 & -0.0848 & -0.1300 & -0.1611 & -0.1248 & -0.0629 & -0.0996 & -0.0292 \\
        Ga & -0.1842 & -0.2577 & -0.2003 & -0.2061 & -0.1912 & -0.2332 & -0.0956 & -0.1395 & -0.0145 \\
        Ge & -0.2491 & -0.1850 & -0.1442 & -0.1942 & -0.1974 & -0.1522 & -0.0886 & -0.1119 & -0.0101 \\
        Mn & -0.2384 & -0.2493 & -0.1559 & -0.1654 & -0.1993 & -0.1304 & -0.0599 & -0.1090 & -0.0275 \\
        Ni & -0.2497 & -0.2279 & -0.1683 & -0.1836 & -0.2402 & -0.2254 & -0.1159 & -0.1027 & -0.0021 \\
        Sc & -0.2252 & -0.1996 & -0.1449 & -0.2023 & -0.2586 & -0.2259 & -0.1128 & -0.1670 & -0.0322 \\
        Ti & -0.2137 & -0.2297 & -0.1918 & -0.1703 & -0.2184 & -0.2455 & -0.0745 & -0.1598 & -0.0186 \\
        V  & -0.2319 & -0.2441 & -0.1693 & -0.1713 & -0.0933 & -0.0952 & -0.0673 & -0.1171 & -0.0164 \\
        Zn & -0.1971 & -0.2071 & -0.1798 & -0.2147 & -0.1577 & -0.2074 & -0.1203 & -0.1637 & -0.0293 \\
        Ag & -0.2465 & -0.2579 & -0.1696 & -0.1739 & -0.1930 & -0.2345 & -0.0117 & -0.1000 & -0.0038 \\
        Cd & -0.1963 & -0.2145 & -0.2124 & -0.2350 & -0.2120 & -0.2341 & -0.1338 & -0.1754 & -0.0354 \\
        In & -0.2567 & -0.2013 & -0.1464 & -0.1558 & -0.2003 & -0.1995 & -0.1043 & -0.1065 & -0.0228 \\
        Mo & -0.2106 & -0.2086 & -0.1656 & -0.1746 & -0.1628 & -0.2038 & -0.0668 & -0.1129 & -0.0144 \\
        Nb & -0.2307 & -0.2394 & -0.1752 & -0.1660 & -0.1679 & -0.1668 & -0.0688 & -0.1288 & -0.0138 \\
        Pd & -0.3134 & -0.2362 & -0.1751 & -0.1707 & -0.1779 & -0.0819 & -0.0677 & -0.1222 & -0.0039 \\
        Rh & -0.2128 & -0.2468 & -0.1542 & -0.1986 & -0.1595 & -0.0720 & -0.0949 & -0.0933 & -0.0021 \\
        Ru & -0.2275 & -0.2613 & -0.1426 & -0.1760 & -0.1693 & -0.0786 & -0.0577 & -0.1332 & -0.0175 \\
        Sb & -0.1891 & -0.1940 & -0.2266 & -0.1526 & -0.2398 & -0.2284 & -0.0788 & -0.1157 & -0.0132 \\
        Sn & -0.2529 & -0.1968 & -0.2028 & -0.2165 & -0.1898 & -0.1845 & -0.0886 & -0.1568 & -0.0170 \\
        Tc & -0.2158 & -0.2603 & -0.1459 & -0.1731 & -0.1578 & -0.0973 & -0.0574 & -0.1222 & -0.0647 \\
        Y  & -0.2962 & -0.2657 & -0.1496 & -0.2075 & -0.2598 & -0.2231 & -0.1096 & -0.1662 & -0.0384 \\
        Zr & -0.1548 & -0.2417 & -0.1360 & -0.1770 & -0.2745 & -0.2015 & -0.0919 & -0.1551 & -0.0185 \\
        Au & -0.2413 & -0.2076 & -0.2877 & -0.1138 & -0.2448 & -0.2634 & -0.0095 & -0.1447 & -0.0041 \\
        Bi & -0.1970 & -0.2006 & -0.2337 & -0.1648 & -0.1315 & -0.2394 & -0.0927 & -0.1399 & -0.0183 \\
        Hf & -0.2728 & -0.2398 & -0.1920 & -0.1784 & -0.2711 & -0.3070 & -0.0834 & -0.1620 & -0.0186 \\
        Hg & -0.1953 & -0.2848 & -0.1710 & -0.1987 & -0.2065 & -0.1285 & -0.1491 & -0.1469 & -0.0370 \\
        Ir & -0.2144 & -0.1796 & -0.1476 & -0.1609 & -0.1474 & -0.0751 & -0.0699 & -0.0953 & -0.0258 \\
        Os & -0.2163 & -0.2409 & -0.1420 & -0.1599 & -0.1776 & -0.0816 & -0.0565 & -0.1218 & -0.0724 \\
        Pb & -0.1964 & -0.2001 & -0.2339 & -0.1644 & -0.1894 & -0.1890 & -0.0888 & -0.1086 & -0.0227 \\
        Pt & -0.1825 & -0.2340 & -0.1657 & -0.1759 & -0.2307 & -0.1921 & -0.0612 & -0.1216 & -0.0022 \\
        Re & -0.2117 & -0.2566 & -0.1472 & -0.1661 & -0.1530 & -0.1479 & -0.0619 & -0.1223 & -0.0261 \\
        Ta & -0.2269 & -0.2462 & -0.1744 & -0.1715 & -0.1696 & -0.1787 & -0.0694 & -0.1428 & -0.0135 \\
        Tl & -0.1981 & -0.2044 & -0.2304 & -0.2198 & -0.2048 & -0.2595 & -0.1146 & -0.1764 & -0.0256 \\
        W  & -0.2006 & -0.2433 & -0.1670 & -0.1453 & -0.1295 & -0.1025 & -0.0613 & -0.1185 & -0.0157 \\
        \bottomrule
    \end{tabular}
  }
\end{table}

\subsection{Correlation analysis of adsorption energies}
\label{supp_sec2.4_corr_analysis}

\begin{figure}[htbp]
  \centering
  \includegraphics[width=0.7\textwidth]{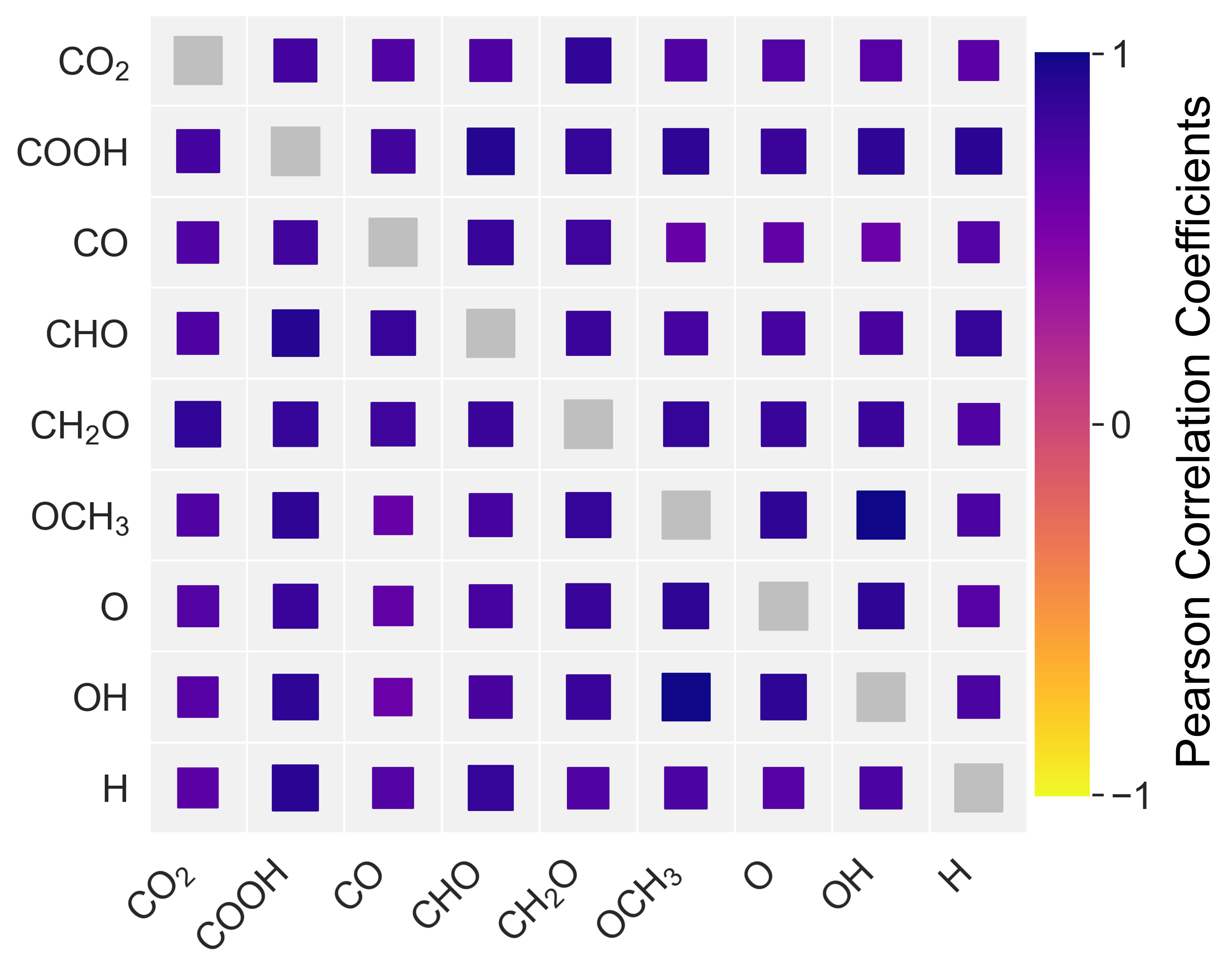}
  \caption{\textbf{Pearson correlation map for adsorption energies of various adsorbates for all six substrates.}}
  \label{supp_fig9:corr_eads}
\end{figure}

\cref{supp_fig9:corr_eads} presents a Pearson correlation map that illustrates the relationships between the adsorption energies of various intermediates. This analysis encompasses all six substrates under consideration. The mean Pearson correlation coefficient is 0.6023 with a variance of 0.0109, signifying a robust correlation between the adsorption energies of different intermediates, accompanied by low variability.

% SI Figure 9: Pearson correlation map for Eads of adsorbates for all substrates

\subsection{The scaling relation scheme and the hybrid scaling relation}
\label{supp_sec2.5_scaling}

In line with the scaling relation framework proposed by Abild-Pedersen et al. \cite{abild2007scaling},
and later applied to the CO\textsubscript{2} reduction to methane process by Peterson \& Nørskov \cite{peterson2012activity},
adsorbates in the CO\textsubscript{2}RR process can be categorized into C-centered or O-centered species.
In each group, a representative species is designated as the ``descriptor'', allowing approximating the adsorption energies of other species within the same group using scaling relations.
The following equations capture this relationship:

\begin{align}
  G_{\text{ads}}X &= a_X \cdot G_{\text{ads}}{\text{CO}} + c_X
  && \text{(where  $X$ = COOH, CHO, CH\textsubscript{2}O)}  \label{supp_eq1:scaling_relation_x_co}  \\
  G_{\text{ads}}Y &= a_Y \cdot G_{\text{ads}}{\text{OH}} + c_Y
  && \text{(where  $Y$ = O, OCH\textsubscript{3})}          \label{supp_eq2:scaling_relation_y_oh}
\end{align}

Here, CO and OH serve as descriptors for C-centered ($X$) and O-centered ($Y$) groups, respectively, and $a_{\mathrm{X}}, c_{\mathrm{X}}, a_{\mathrm{Y}}, c_{\mathrm{Y}}$ are species-specific scaling parameters.
The scaling relations simplify the adsorption energy landscape to a two-dimensional space, thereby enabling visualization.

However, it's crucial to note that in the CO\textsubscript{2}RR process, most species are not solely C- or O-centered.
Thus, it becomes advantageous to incorporate both types of species in the scaling relation.
Building upon the hybrid scaling relation introduced in the main text, the free energy change ($\Delta\mathit{G}$) for any given reaction step can be readily calculated using \cref{main_eq14:general_scaling_relation} from the main text.
Consider the generic reaction step:
\begin{align}
  \ce{\text{*}A + H+ + e- -> \text{*}B}  \label{supp_eq3:generic_reaction}
\end{align}

The free energy change $\Delta\mathit{G}$ is defined as:
\begin{align}
  \Delta G = G(\text{*}\mathrm{B}) - G(\text{*}\mathrm{A}) - G(\mathrm{H}^+) - G(e^-)  \label{supp_eq4:free_energy_change}
\end{align}

Focusing on any reaction step delineated by \cref{main_eq1:co2rr1} to \cref{main_eq8:co2rr8} in the main text,
\cref{main_eq15:limiting_potential} shows that limiting potential $U_{\mathrm{L}}$ at a specific external potential $U$ is fully determined by the adsorption energies of two descriptors.
For instance, for the reaction described by \cref{main_eq2:co2rr2}:
\begin{align}
  \Delta G &= G(\text{*COOH}) - G(\text{*}) - G(\text{CO}_2) - G(\text{H}^+) - G(\text{e}^-)  \\
  &= G(\text{COOH}) + G(\text{*}) + G_{\text{ads}}\text{COOH} -
  G(\text{*}) - G(\text{CO}_2) -G(\text{H}^+) - G(\text{e}^-) \\
  &= G_{\text{ads}}\text{COOH} + G(\text{COOH}) - G(\text{CO}_2) -G(\text{H}^+) - G(\text{e}^-)  \label{supp_eq5:exp_reaction}
\end{align}

From adsorption energy linear relations, we have:
\begin{align}
  G_{ads}{\mathrm{COOH}} = a_{\text{COOH}} \cdot G_{ads}{\mathrm{CO}} +
  b_{\text{COOH}} \cdot G_{ads}{\mathrm{OH}} + c_{\text{COOH}}  \label{supp_eq6:eads}
\end{align}

which then reduces Supplementary \cref{supp_eq5:exp_reaction} to:
\begin{align}
  \begin{split}
    \Delta G = &(a_{\text{COOH}} \cdot G_{\text{ads}}{\mathrm{CO}} +
    b_{\text{COOH}} \cdot G_{\text{ads}}{\mathrm{OH}}) + \\
    &[c_{\text{COOH}} + G(\mathrm{COOH}) -
    G(\mathrm{CO}_2) - G(\mathrm{H}^+) - G(\mathrm{e}^-)]  \label{supp_eq7:reduced_eads}
  \end{split}
\end{align}
In this case, the first term depends solely on the descriptors, while the second term remains constant at a given potential $U$.

% Supplementary Table 13: Eads scaling relation params from hybrid descriptor method
\begin{table}[htbp]
\label{supp_table13:scaling_params}
  \caption{Adsorption free energy scaling relation parameters
    determined by the hybrid descriptor method.}
  \center
  \small
  \begin{tabularx}{0.75\textwidth}{@{}l *{3}{X} @{}}
    \toprule
    Adsorbate             & $a_Z$   & $b_Z$   & $c_Z$    \\
    \midrule
    COOH                  & 0.4032  & 0.5132  &  0.0899  \\
    CHO                   & 0.5977  & 0.3664  & -2.6763  \\
    CH\textsubscript{2}O  & 0.5011  & 0.5011  & -0.9806  \\
    OCH\textsubscript{3}  & 0.0532  & 1.0114  &  0.8489  \\
    O                     & 0.3758  & 1.2580  &  0.0910  \\
    H                     & 0.3164  & 0.3568  & -0.6764  \\
    \bottomrule
  \end{tabularx}

  \smallskip

  \begin{flushright}
    \begin{minipage}{\textwidth}
      \footnotesize\textit{Note:} $a_{Z}$, $b_{Z}$ and $c_{Z}$ correspond
      to coefficients for $G_{\text{ads}}\text{CO}$, $G_{\text{ads}}\text{OH}$
      and constant terms, where $Z$ stands for the adsorbed species of interest.
  \end{minipage}
  \end{flushright}
\end{table}

% SI Figure 10: Improvement in R2 from hybrid-descriptor method
\begin{figure}[htbp]
  \centering
  \includegraphics[width=\textwidth]{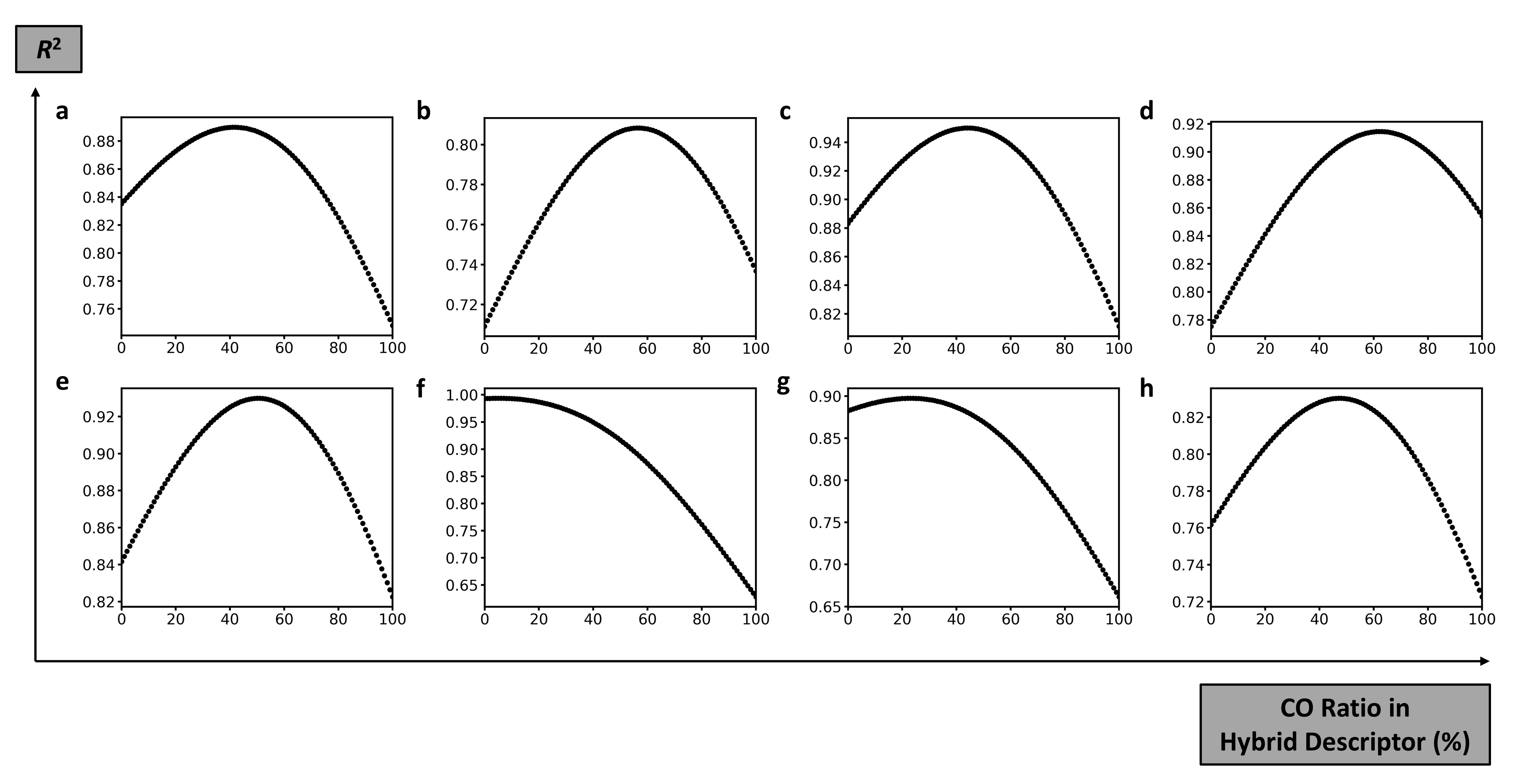}
  \caption{\textbf{$R^2$ improvement from hybrid-descriptor method.}
  Improvement in the coefficient of determination ($R^2$) with x-axis
  as CO ratio in CO/OH hybrid descriptors.
  (\textbf{a}) averaged $R^2$ and those of (\textbf{b}) CO\textsubscript{2},
  (\textbf{c}) COOH,
  (\textbf{d}) CHO, (\textbf{e}) CH\textsubscript{2}O, (\textbf{f})
  OCH\textsubscript{3}, (\textbf{g}) O and (\textbf{h}) H are shown.}
  \label{supp_fig10:r2_hyb_des}
\end{figure}

% Supplementary Table 14: Performance comparison of hybrid/single descriptors
\begin{table}[htbp]
\label{supp_table14:hybrid_des_perf}
  \caption{Performance comparison of hybrid descriptor and single
    descriptor for linear scaling relations.}
  \small
  \center
  \begin{tabularx}{0.75\textwidth}{@{}l *{3}{X} @{}}
    \toprule
    \multirow{2}{*}{Adsorbate} & Hybrid Descriptor $R^2$ & Single Descriptor $R^2$ & \multirow{2}{*}{$R^2$} \\
                               & (Optimal CO Ratio)      & Reference Species       &                        \\
    \midrule
    CO\textsubscript{2}   & 0.8083 (57\%)  & 0.7368 (CO)  & 0.0715  \\
    COOH                  & 0.9500 (44\%)  & 0.8113 (CO)  & 0.1387  \\
    CHO                   & 0.9146 (62\%)  & 0.8541 (CO)  & 0.0605  \\
    CH\textsubscript{2}O  & 0.9299 (50\%)  & 0.8226 (CO)  & 0.1073  \\
    OCH\textsubscript{3}  & 0.9936 (5\%)   & 0.9929 (OH)  & 0.0007  \\
    O                     & 0.8974 (23\%)  & 0.8825 (OH)  & 0.0149  \\
    H                     & 0.8303 (47\%)  & 0.7617 (OH)  & 0.0686  \\
    \bottomrule
  \end{tabularx}
\end{table}

\subsection{Volcano plots}
\label{supp_sec2.6_volcano}

\begin{figure}[htbp]
  \centering
  \includegraphics[width=\textwidth]{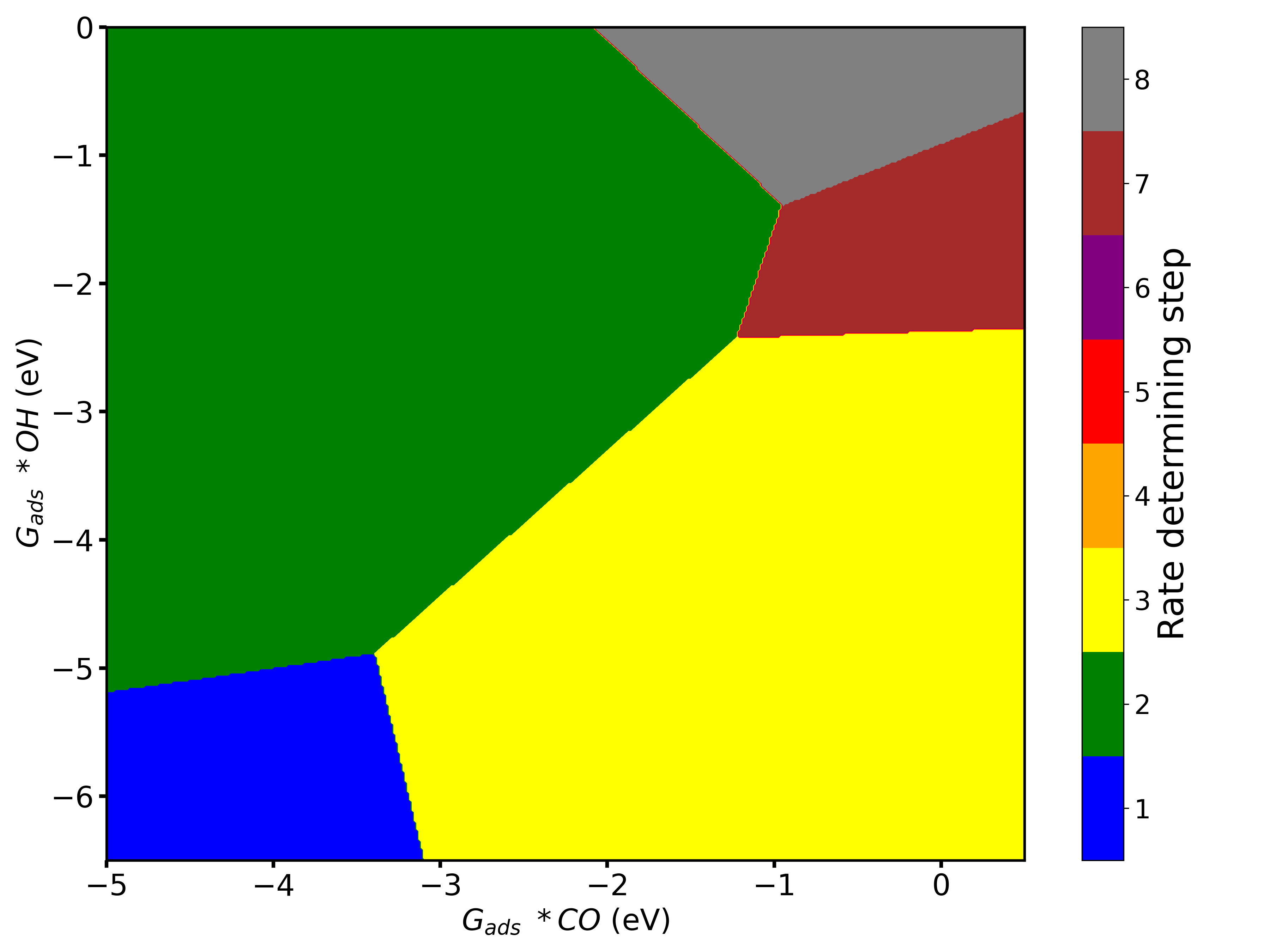}
  \caption{\textbf{CO\textsubscript{2}RR rate determining step volcano plot.}
  This graph illustrates the rate-determining step associated with each region within the CO\textsubscript{2}RR process.}
  \label{supp_fig11:rds_volcano}
\end{figure}

\begin{figure}[htbp]
  \centering
  \includegraphics[width=\textwidth]{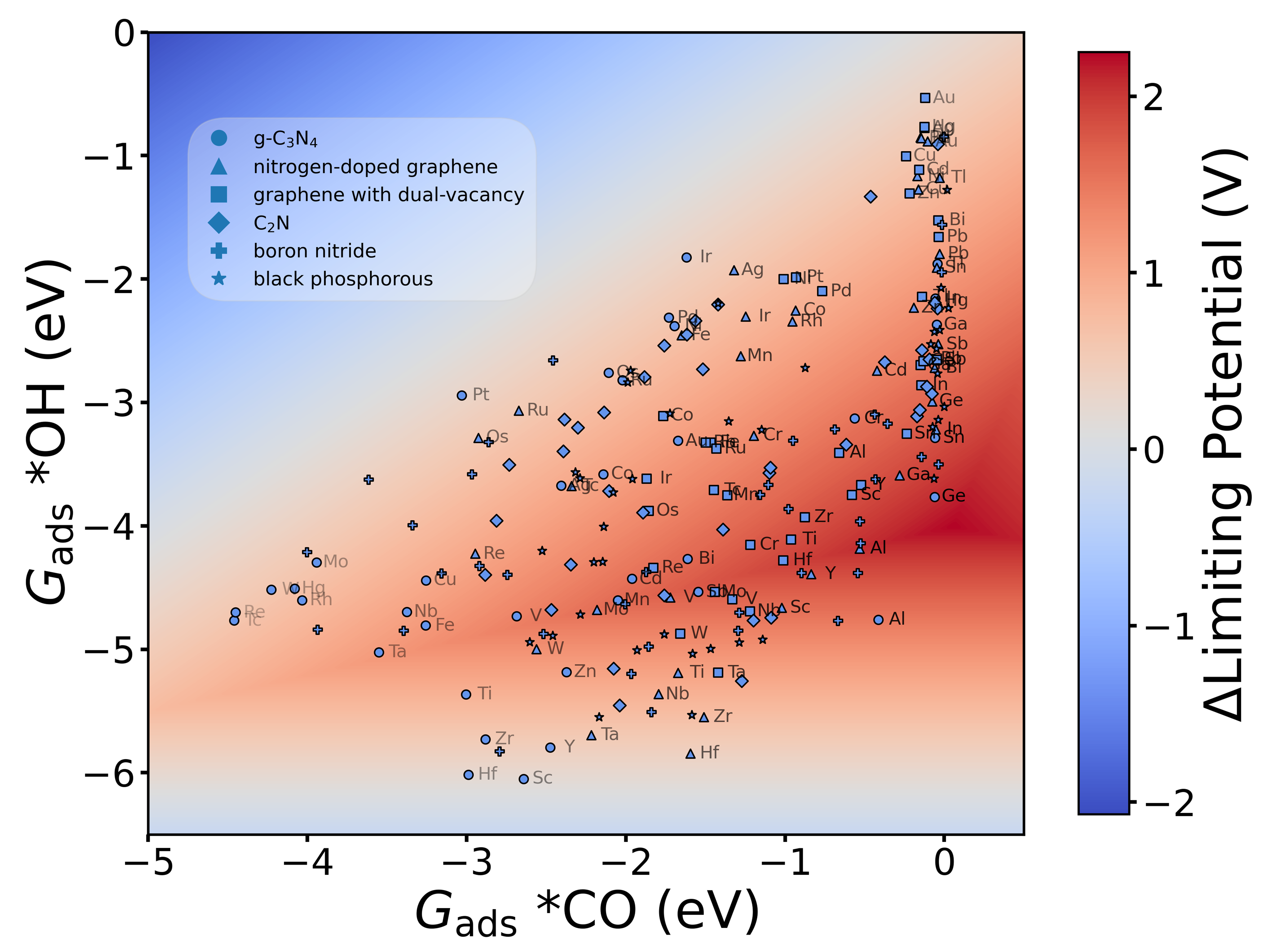}
  \caption{\textbf{CO\textsubscript{2}RR selectivity volcano plot.}
  Selectivity volcano plot computed from linear scaling relations.
  In the plot, the circle, triangle, square, diamond, plus, and star symbols correspond to
  metals supported on g-C\textsubscript{3}N\textsubscript{4}, nitrogen-doped graphene, graphene with dual-vacancy,
  black phosphorous, BN, and C\textsubscript{2}N, respectively.}
  \label{supp_fig12:sel_volc}
\end{figure}

In addition to the selectivity volcano plot featured in the main text, we provide a comprehensive plot capturing all rate-determining steps for CO\textsubscript{2}RR in \cref{supp_fig11:rds_volcano}. A separate plot focusing on selectivity over HER is also presented in \cref{supp_fig12:sel_volc}.

% SI Figure 11: CO2RR rate determining step volcano plot

% SI Figure 12: CO2RR selectivity volcano plot

\subsection{Activity periodic tables}
\label{supp_sec2.7_ptable}

% SI Figure 13: Limiting potential periodic table for nitrogen-doped graphene

% SI Figure 14: Limiting potential periodic table for graphene with dual-vacancy
\begin{figure}[htbp]
  \centering
  \includegraphics[width=\textwidth]{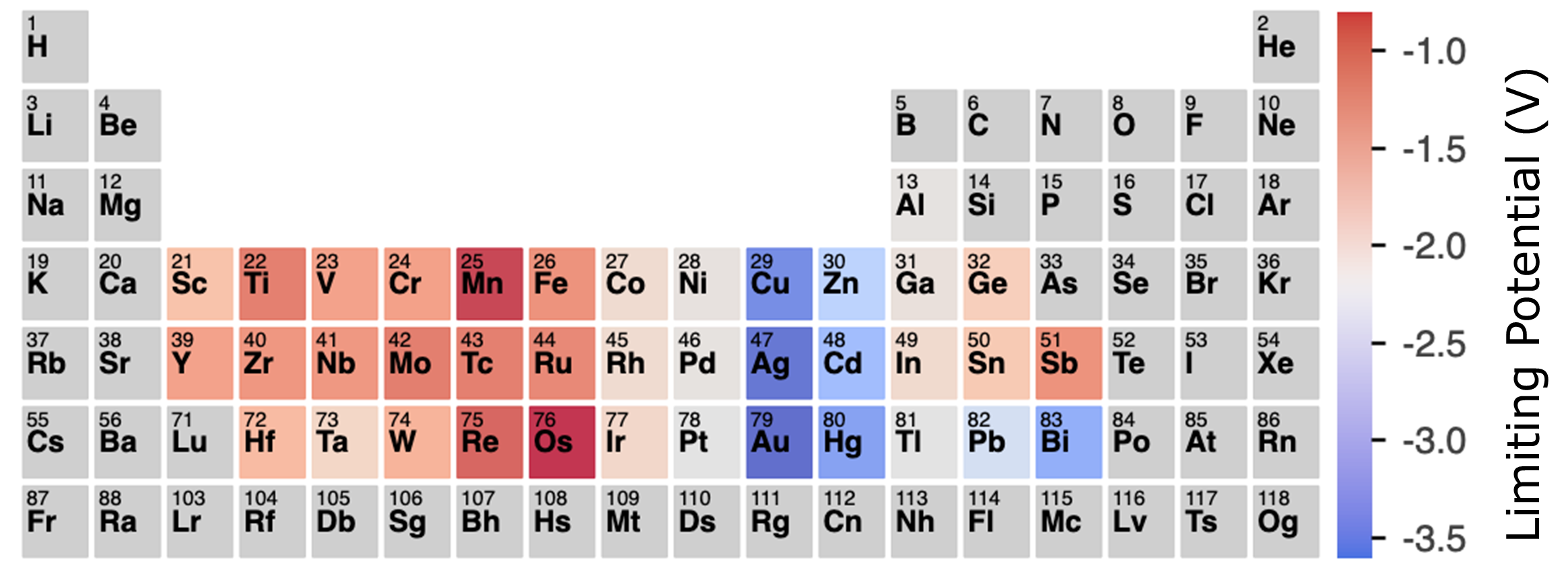}
  \caption{\textbf{Limiting potential periodic table for graphene with dual-vacancy.}}
  \label{supp_fig14:gra-vac_ptable}
\end{figure}

% SI Figure 15: Limiting potential periodic table for black phosphorus
\begin{figure}[htbp]
  \centering
  \includegraphics[width=\textwidth]{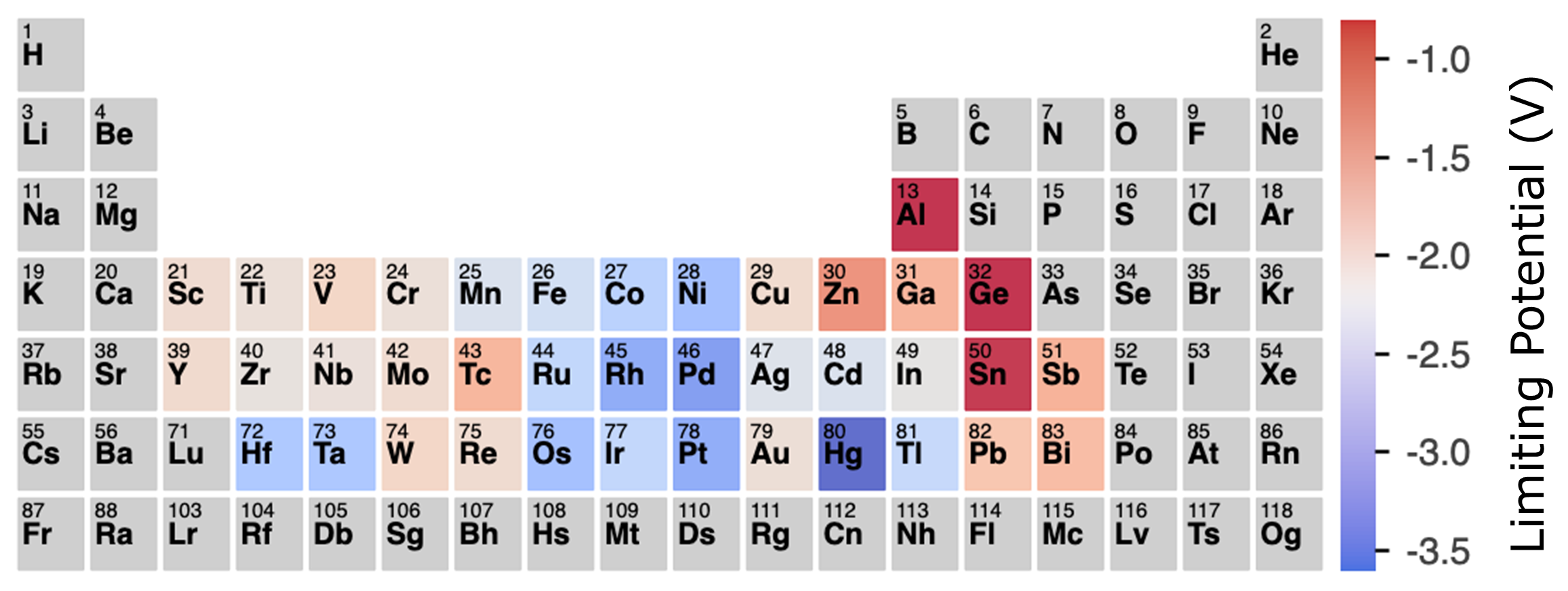}
  \caption{\textbf{Limiting potential periodic table for black phosphorus.}}
  \label{supp_fig15:BP_ptable}
\end{figure}

% SI Figure 16: Limiting potential periodic table for boron nitride
\begin{figure}[htbp]
  \centering
  \includegraphics[width=\textwidth]{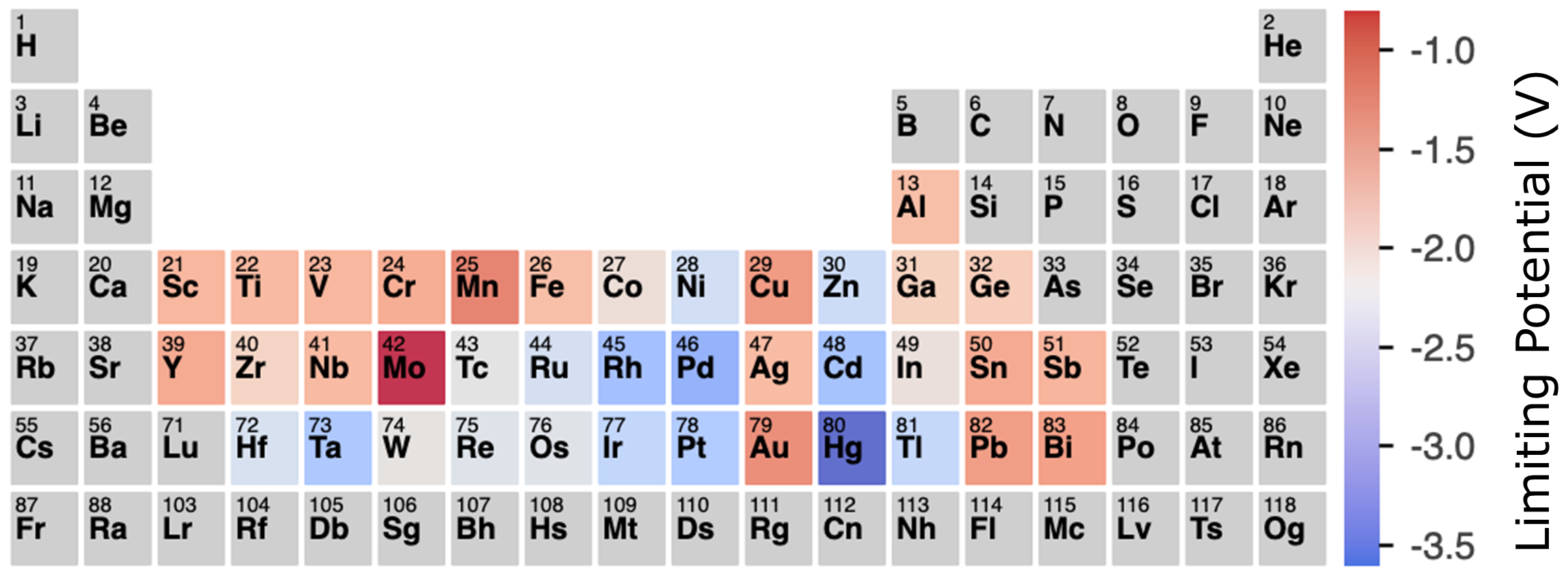}
  \caption{\textbf{Limiting potential periodic table for boron nitride.}}
  \label{supp_fig16:BN_ptable}
\end{figure}

% SI Figure 17: Limiting potential periodic table for monolayer C2N

\newpage
% SI Section Two: Additional details on Machine Learning

\section{Additional details on machine learning method}

\subsection{d-band center and adsorption energy relations}
\label{supp_sec3.1_dband_eads}

% SI Figure 18: d-band venter vs adsorption energy scatter plot

\subsection{Feature correlation analysis}
\label{supp_sec3.2_feature_corr}

\begin{figure}[htbp]
  \centering
  \includegraphics[width=\textwidth]{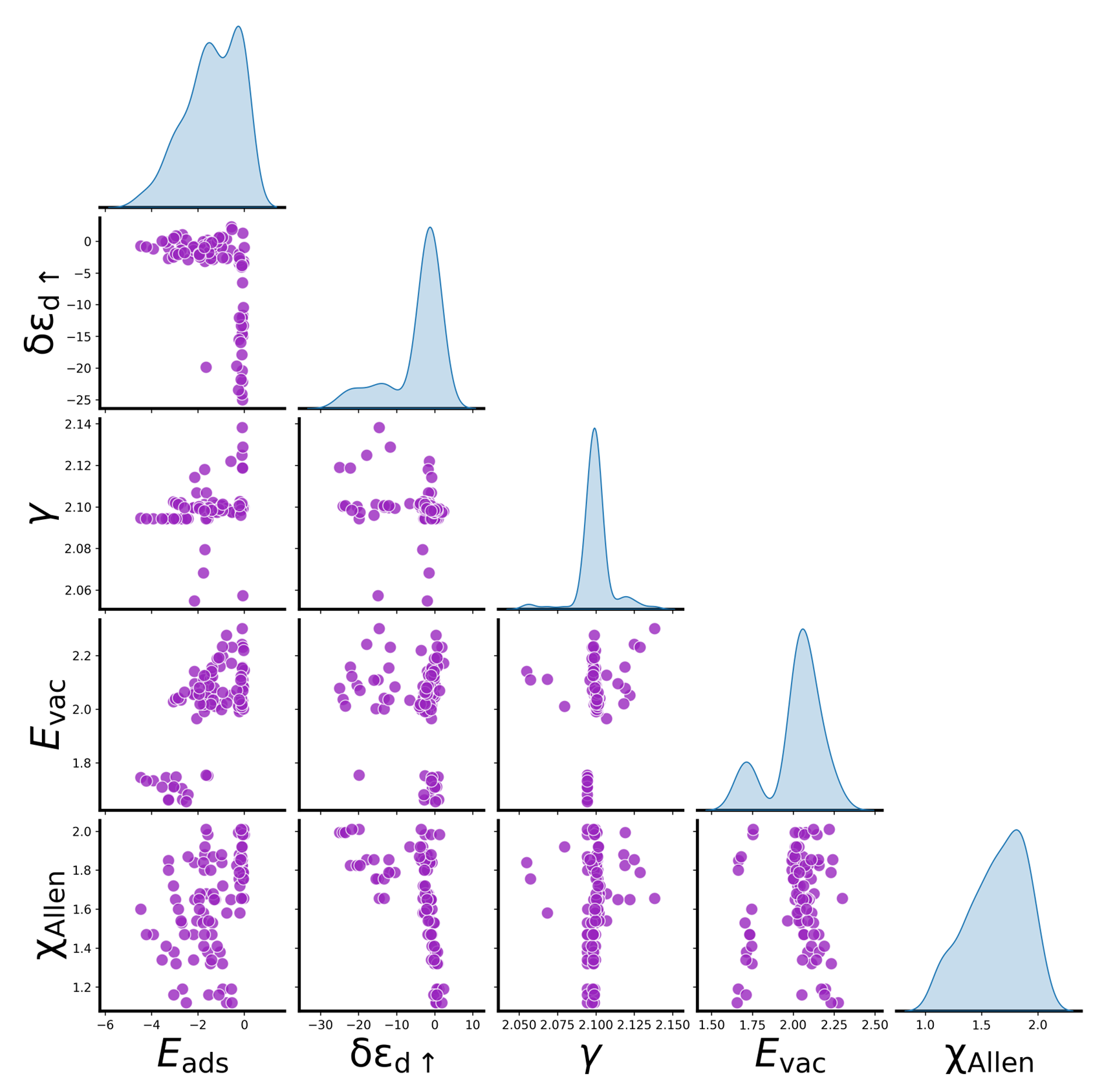}
  \caption{\textbf{Pairwise Plot of Adsorption Energy and Key Descriptors.}
  This chart displays the pairwise relationships between CO adsorption energy and
  its four most correlated descriptors: d-band center (spin-up) $\delta\epsilon_{\text{d}}\uparrow$,
  lattice parameter $\gamma$, vacuum level $E_\text{vac}$, and electronegativity
  on the Allen scale $\chi_\text{Allen}$, identified by Kendall rank correlation analysis.
  Data from all six substrates investigated in this study are included.}
  \label{supp_fig19:pairwise_eads_des}
\end{figure}

\cref{supp_fig19:pairwise_eads_des} displays the pairwise relationships between CO adsorption energy
and its four most correlated descriptors: d-band center (spin-up) $\delta\epsilon_{\text{d}}\uparrow$,
lattice parameter $\gamma$, vacuum level $E_\text{vac}$, and electronegativity
on the Allen scale $\chi_\text{Allen}$, identified by Kendall rank correlation analysis.
\cref{supp_table15:descriptor_notations} lists notations for elementary descriptors and electronic descriptors.

% SI Figure 19: Pairwise plot of adsorption energy vs key descriptors

% Supplementary Table 15: Notations for elementary descriptors and electronic descriptors
\begin{table}[htbp]
\label{supp_table15:descriptor_notations}
  \caption{Notations for elementary descriptors and electronic descriptors.}
  \resizebox{\textwidth}{!}{%
    \begin{tabular}{llll}
      \toprule
      \multicolumn{2}{c}{Elementary Descriptors}       & \multicolumn{2}{c}{Electronic Descriptors}            \\
      \midrule
      $R$            & atomic radius                   & $\alpha$, $\beta$, $\gamma$  & lattice parameters     \\
      $E_\text{ea}$  & electron affinity               & $\Phi_{\text{DFT}}$    & DFT calculated work function \\
      $\Phi$         & work function                   & $E_\text{vac}$         & vacuum level                 \\
      $\chi_\text{Allen}$, $\chi_\text{P}$, $\chi_\text{RevP}$
      & \parbox[t]{4cm}{electronegativity in Allen, Pauling and revised Pauling scales}
      & $E_\text{g}$, $E_\text{g}\uparrow$, $E_\text{g}\uparrow$
      & \parbox[t]{4cm}{average, spin-up and spin-down band gaps}                                              \\
      $A_\text{r}$   & relative atomic mass            & $e_\text{d}$           & number of d-electrons        \\
      $E_\text{i}$   & ionization energy & $\delta\epsilon_{\text{d}}\uparrow$  & d-band centre (spin-up)      \\
      $G$            & group number                    & $W_\text{d}$           & d-band width                 \\
      $P$            & period number                   & $e_{\text{Bader}}$     & Bader charge                 \\
      $V$            & number of valence electrons     &                        &                              \\
      \bottomrule
    \end{tabular}
  }
\end{table}

\subsection{Input pipeline construction}
\label{supp_sec3.3_input_pipeline}

\begin{figure}[htbp]
  \centering
  \includegraphics[width=0.3\textwidth]{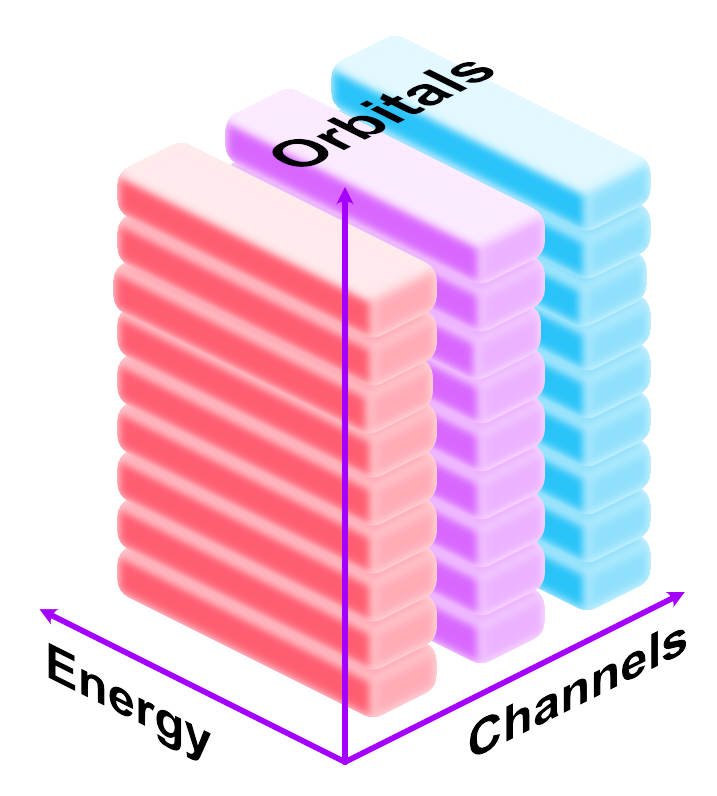}
  \caption{\textbf{Schematic representation of the input eDOS data structure.}}
  \label{supp_fig20:input_dos}
\end{figure}

The format of the 2D array for the eDOS is naturally well-suited for use as neural network inputs.
In this study, the array adopts a shape of [numSamplings, numOrbitals, numChannels], as illustrated in \cref{supp_fig20:input_dos}.
Here, ``numSamplings'' represents the number of data points sampled across the energy spectrum,
``numOrbitals'' denotes the number of DOS orbitals,
and ``numChannels'' indicates the number of atomic channels.

% SI Figure 20: Representation of the input eDOS data structure

For practical purposes, certain spatial hierarchies are expected to remain invariant,
such as the energy range over which the eDOS is sampled.
In this work, we maintain a consistent energy range of -14 eV to 6 eV,
sampled at a density of 50 \si{\text{points} \cdot \text{eV}^{-1}}.
In addition to maintaining a fixed energy range,
the following steps are taken to standardize the input eDOS data:
  1.	For atoms lacking d-electrons, the eDOS is zero-padded.
  2.	f-orbital eDOS data are excluded.
  3.	The eDOS of adsorbates is zero-padded along the Channels axis to align with the adsorbate that contains the maximum number of atoms, which, in this study, is the *OCH\textsubscript{3} species. This padded eDOS is then appended to the metal eDOS.

As a result of these preprocessing steps, the input eDOS arrays consistently have a shape of [4000, 9, 6].
This corresponds to 4000 sampling points across the energy range, 9 eDOS orbitals, and 6 atomic channels.

\subsection{Hyperparameter tuning for CNN}
\label{supp_sec3.4_hyperparam}

% Supplementary Table 16: The hyperparameter search space
\begin{table}[htbp]
\label{supp_table16:hyperparam_space}
  \caption{The hyperparameter search space.}
  \resizebox{\textwidth}{!}{%
    \begin{tabular}{lll}
      \toprule
      Hyperparameter& Search Space              & Description                                                       \\
      \midrule
      learningRate  & [10\textsuperscript{-4}, 10\textsuperscript{-2}], log scale & Learning rate of Adam \cite{kingma2017adam} optimizer \\
      dropoutRate   & [0, 1]                    & Dropout rate                                                      \\
      root\_fc0     & 128, 256, 512             & Output dimensionality of 1\textsuperscript{st} fully connected layer in root           \\
      root\_fc1     & 256, 512, 1024, 2048      & Output dimensionality of 2\textsuperscript{nd} fully connected layer in root           \\
      root\_act     & ``tanh'', ``relu'', ``sigmoid'' & Activation function of fully connected layers in root             \\
      kernelSize    & [2, 32]                   & Width of the convolution window                                   \\
      numFilters    & [8, 64]                   & Number of filters in the convolution                              \\
      numConvLayers & [1, 16]                   & Number of convolution layers                                      \\
      numConvBlocks & [1, 16]                   & Number of convolution blocks                                      \\
      br\_fc0       & 16, 32, 64, 128, 256, 512 & Output dimensionality of 1\textsuperscript{st} fully connected layer in branch      \\
      br\_fc1       & 8, 16, 32, 64, 128, 256   & Output dimensionality of 2\textsuperscript{nd} fully connected layer in branch      \\
      br\_act       & ``tanh'', ``relu'', ``sigmoid'' & Activation function of fully connected layers in branch           \\
      \bottomrule
    \end{tabular}
  }
\end{table}

% Supplementary Table 17: Optimal hyperparameters identified with Hyperband algorithm
\begin{table}[htbp]
\label{supp_table17:opt_hyperparam}
  \caption{Optimal hyperparameters identified with Hyperband algorithm.}
  \resizebox{\textwidth}{!}{%
    \begin{tabular}{lll}
      \toprule
      Hyperparameter& Optimal Value & Description                                              \\
      \midrule
      learningRate  & $10^{-3}$ & Learning rate of Adam \cite{kingma2017adam} optimizer                              \\
      dropoutRate   & 0.3       & Dropout rate                                                 \\
      root\_fc0     & 256       & Output dimensionality of 1\textsuperscript{st} fully connected layer in root   \\
      root\_fc1     & 512       & Output dimensionality of 2\textsuperscript{nd} fully connected layer in root   \\
      root\_act     & ``relu''    & Activation function of fully connected layers in root        \\
      kernelSize    & 10        & Width of convolution window                                  \\
      numFilters    & 17        & Number of filters in the convolution                         \\
      numConvLayers & 3         & Number of convolution layers                                 \\
      numConvBlocks & 1         & Number of convolution blocks                                 \\
      br\_fc0       & 256       & Output dimensionality of 1\textsuperscript{st} fully connected layer in branch \\
      br\_fc1       & 128       & Output dimensionality of 2\textsuperscript{nd} fully connected layer in branch \\
      br\_act       & ``tanh''    & Activation function of fully connected layers in branch      \\
      \bottomrule
    \end{tabular}
  }
\end{table}

% Supplementary Table 18: Prediction MAEs of the CNN model
\begin{table}[htbp]
\label{supp_table18:cnn_mae}
  \caption{Prediction mean absolute errors of the CNN model.}
  \small
  \center
  \begin{tabularx}{0.75\textwidth}{@{}lXX@{}}
    \toprule
    Adsorbate             & Original dataset (eV)  & Augmented dataset (eV)  \\
    \midrule
    CO\textsubscript{2}   & 0.0447                 & 0.0336                  \\
    COOH                  & 0.0436                 & 0.0581                  \\
    CO                    & 0.0402                 & 0.0402                  \\
    CHO                   & 0.0614                 & 0.0778                  \\
    CH\textsubscript{2}O  & 0.0553                 & 0.0632                  \\
    OCH\textsubscript{3}  & 0.0605                 & 0.0636                  \\
    O                     & 0.1135                 & 0.1081                  \\
    OH                    & 0.0547                 & 0.0602                  \\
    H                     & 0.0703                 & 0.0646                  \\
    \bottomrule
  \end{tabularx}

  \smallskip

  \begin{flushright}
  \begin{minipage}{\textwidth}
    \footnotesize\textit{Note:} The same CNN model, trained on the
      augmented dataset, was used for all predictions. The term ``original dataset''
      refers to evaluations conducted on the original dataset without any augmented data;
      it does not imply the use of a different CNN model trained solely on the original dataset.
      Similarly, ``augmented dataset'' refers to evaluations on the dataset with
      both original data and augment data, not a distinct model.
  \end{minipage}
  \end{flushright}
\end{table}

\subsection{Occlusion experiment}
\label{supp_sec3.5_occlusion}

% SI Figure 21: Illustration of the eDOS occlusion experiment
\begin{figure}[htbp]
  \centering
  \includegraphics[width=\textwidth]{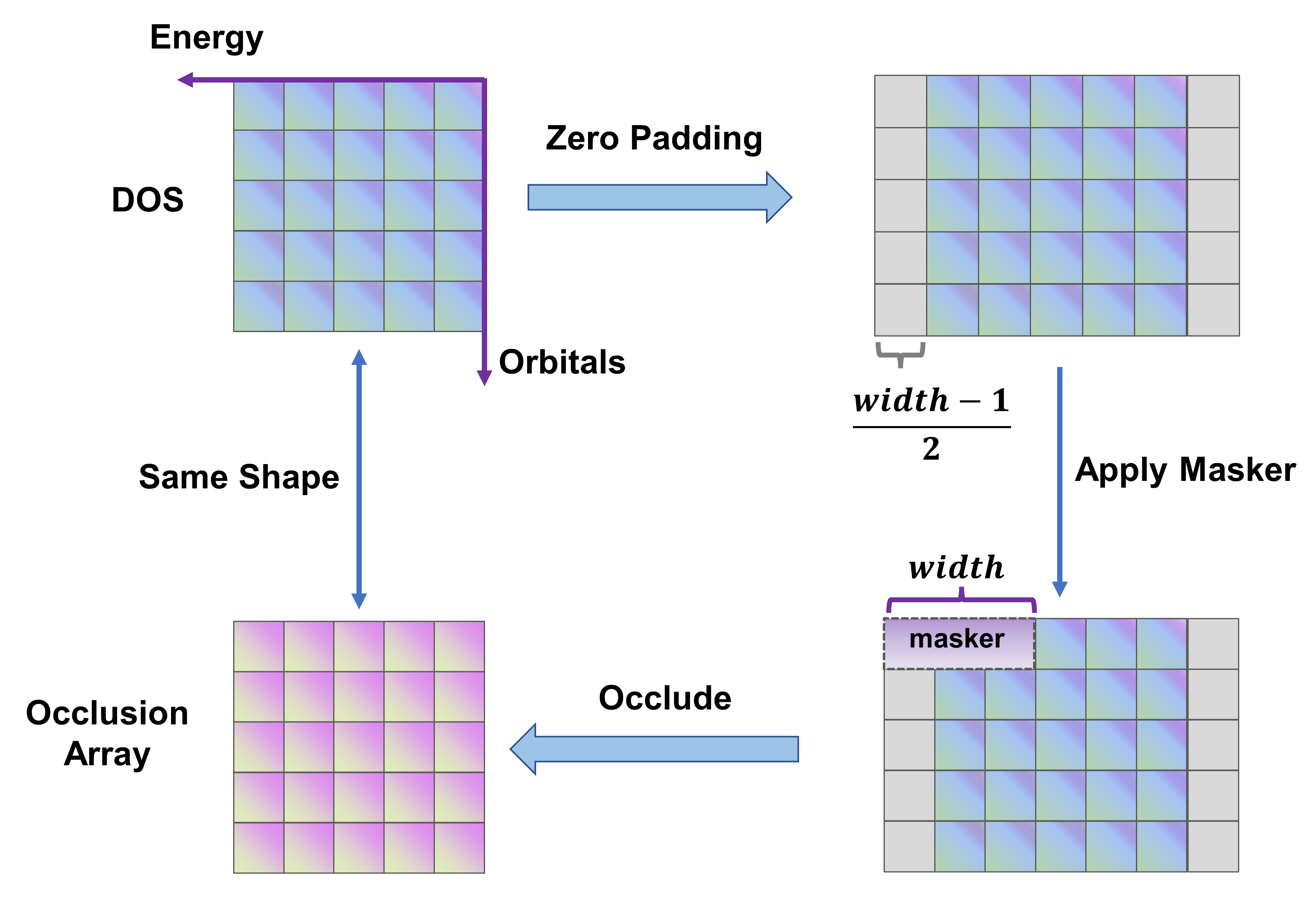}
  \caption{\textbf{Illustration of the eDOS occlusion experiment.}
  This figure demonstrates the step-by-step process of the occlusion experiment.
  The input eDOS array first undergoes zero-padded, and then a masker of width ``width'' is applied,
  resulting in an occlusion array that preserves the shape of the input array.}
  \label{supp_fig21:occlusion}
\end{figure}

% SI Figure 22: Occlusion experiment with a masker width of 1
\begin{figure}[htbp]
  \centering
  \includegraphics[width=0.6\textwidth]{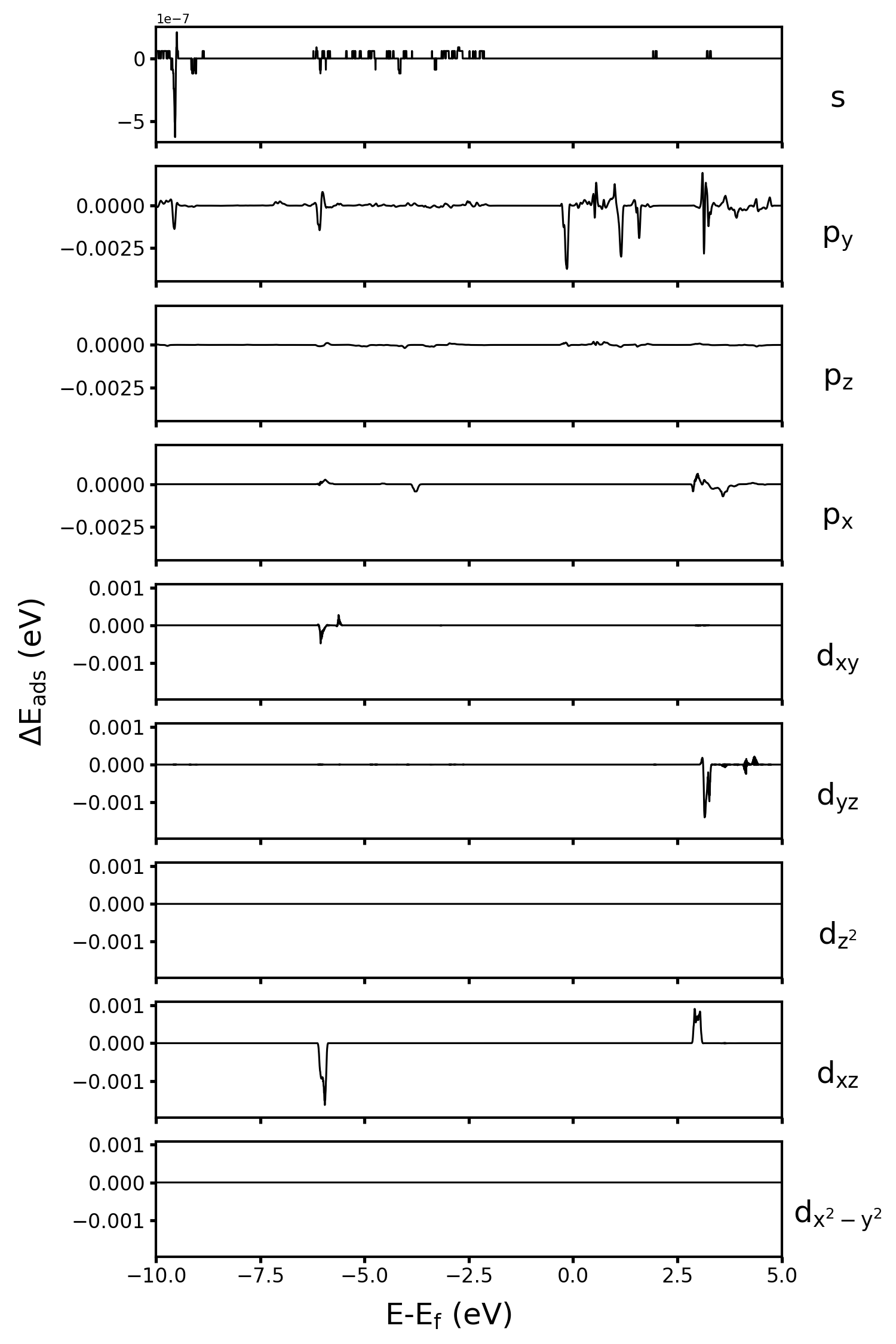}
  \caption{\textbf{Occlusion Experiment with a masker width of ``1''.}
  This figure shows an occlusion experiment on the Ge atom in
  the initial state of CO adsorption on Ge@g-C\textsubscript{3}N\textsubscript{4}.}
  \label{supp_fig22:occl_wid1}
\end{figure}

% SI Figure 23: Occlusion experiment with a masker width of 3
\begin{figure}[htbp]
  \centering
  \includegraphics[width=0.6\textwidth]{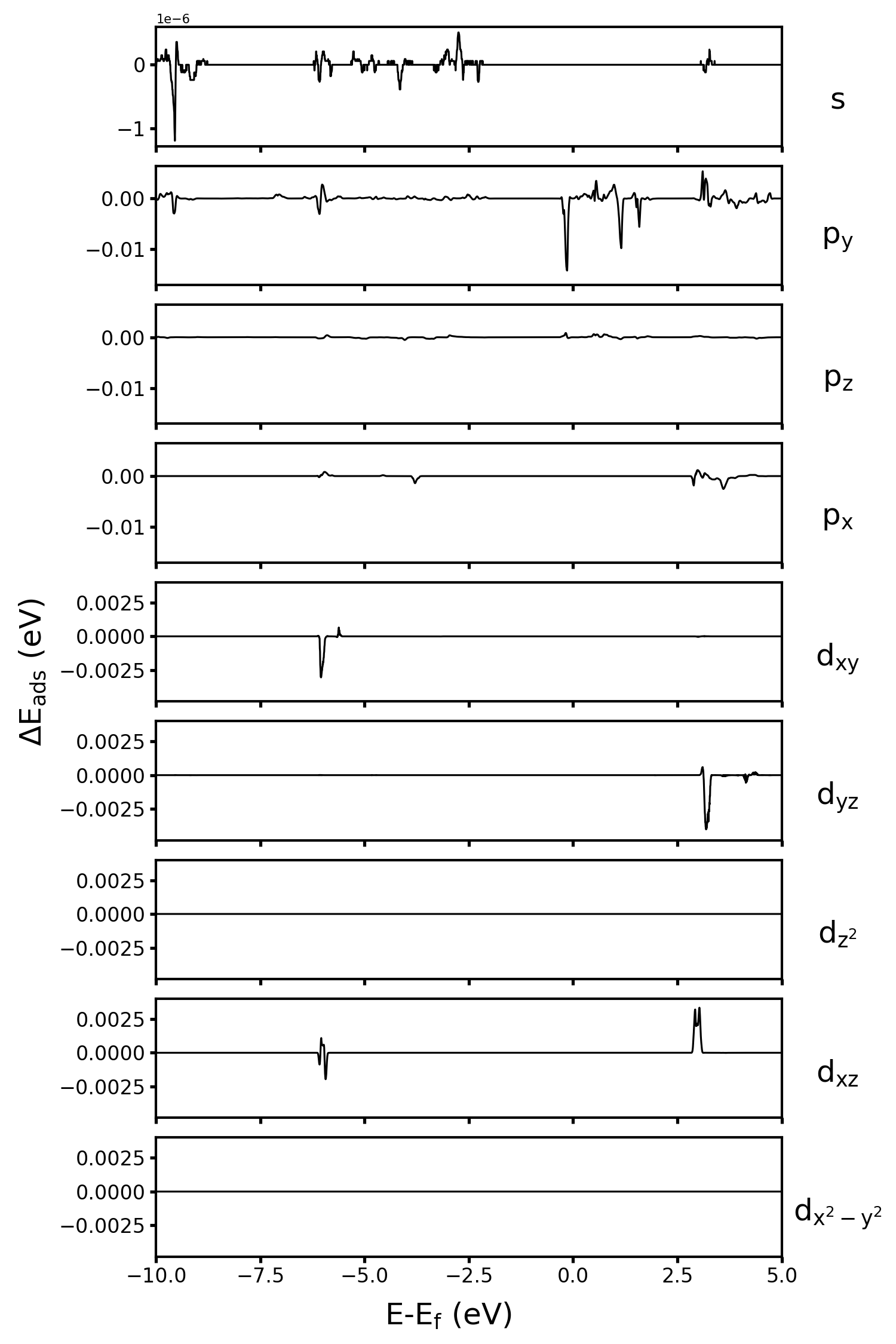}
  \caption{\textbf{Occlusion Experiment with a masker width of ``3''.}
  This figure shows an occlusion experiment on the Ge atom in
  the initial state of CO adsorption on Ge@g-C\textsubscript{3}N\textsubscript{4}.}
  \label{supp_fig23:occl_wid3}
\end{figure}

% SI Figure 24: Occlusion experiment with a masker width of 5
\begin{figure}[htbp]
  \centering
  \includegraphics[width=0.6\textwidth]{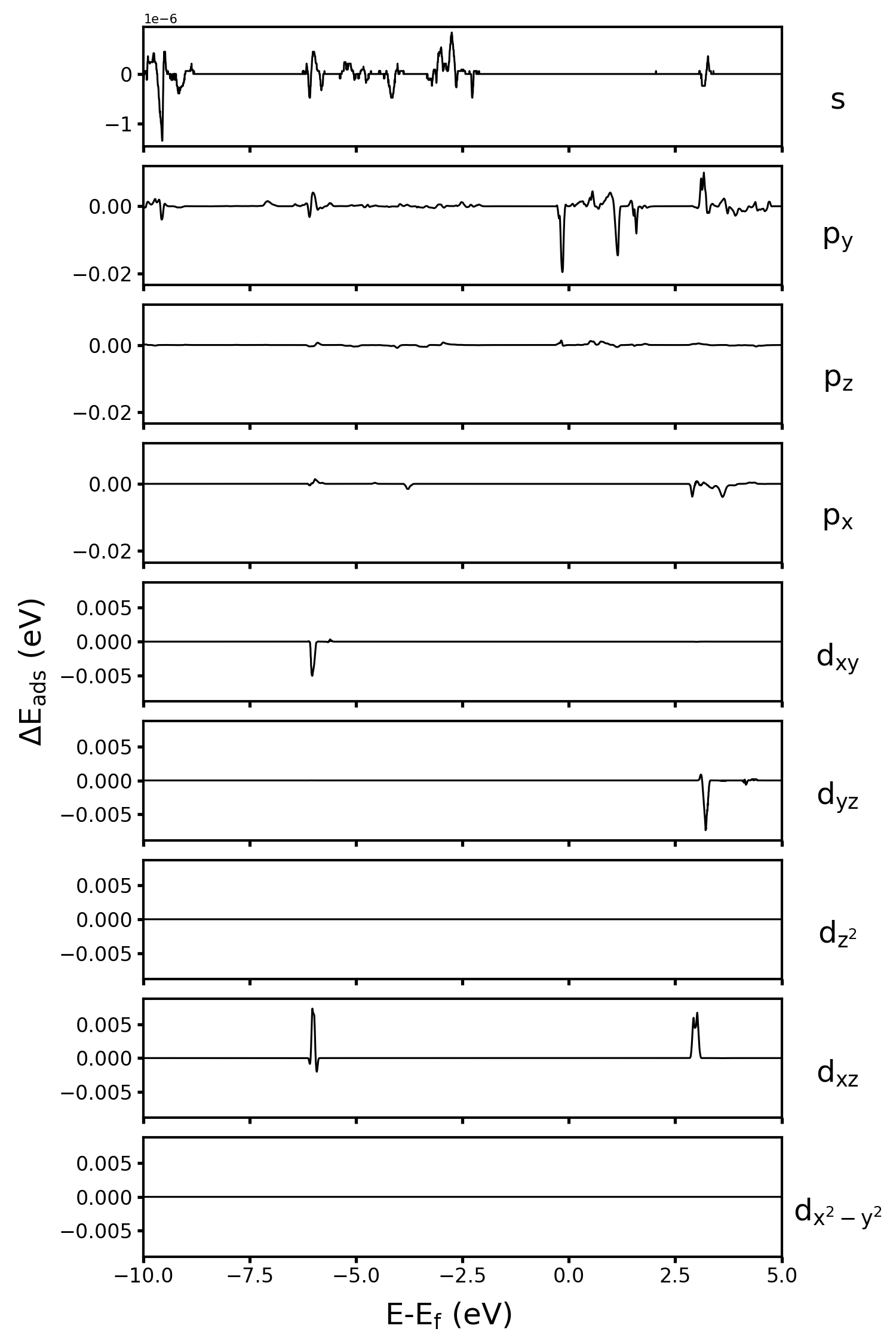}
  \caption{\textbf{Occlusion Experiment with a masker width of ``5''.}
  This figure shows an occlusion experiment on the Ge atom in
  the initial state of CO adsorption on Ge@g-C\textsubscript{3}N\textsubscript{4}.}
  \label{supp_fig24:occl_wid5}
\end{figure}

% SI Figure 25: Occlusion experiment with a masker width of 11
\begin{figure}[htbp]
  \centering
  \includegraphics[width=0.6\textwidth]{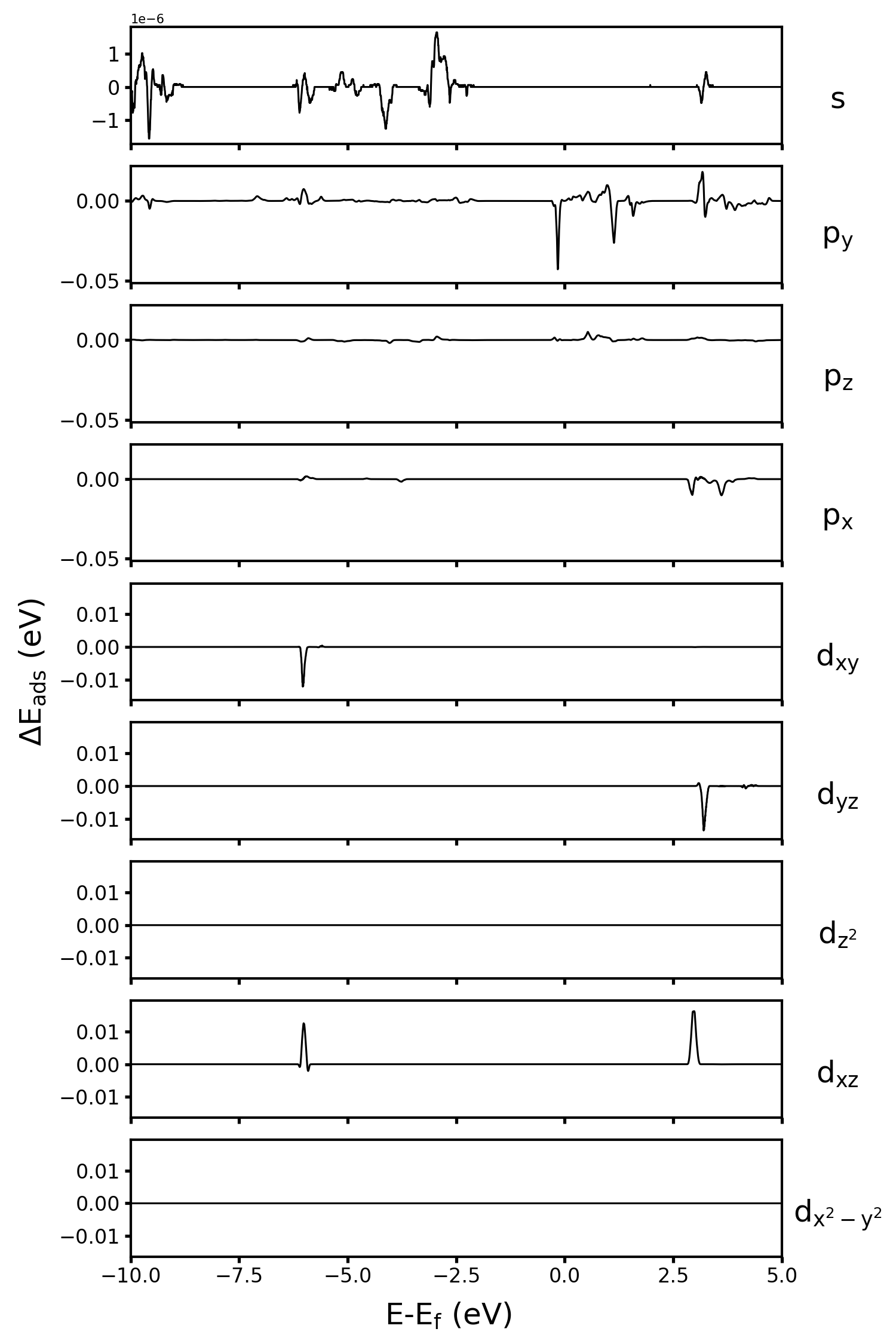}
  \caption{\textbf{Occlusion Experiment with a masker width of ``11''.}
  This figure shows an occlusion experiment on the Ge atom in
  the initial state of CO adsorption on Ge@g-C\textsubscript{3}N\textsubscript{4}.}
  \label{supp_fig25:occl_wid11}
\end{figure}

% SI Figure 26: Occlusion experiment with a masker width of 21
\begin{figure}[htbp]
  \centering
  \includegraphics[width=0.6\textwidth]{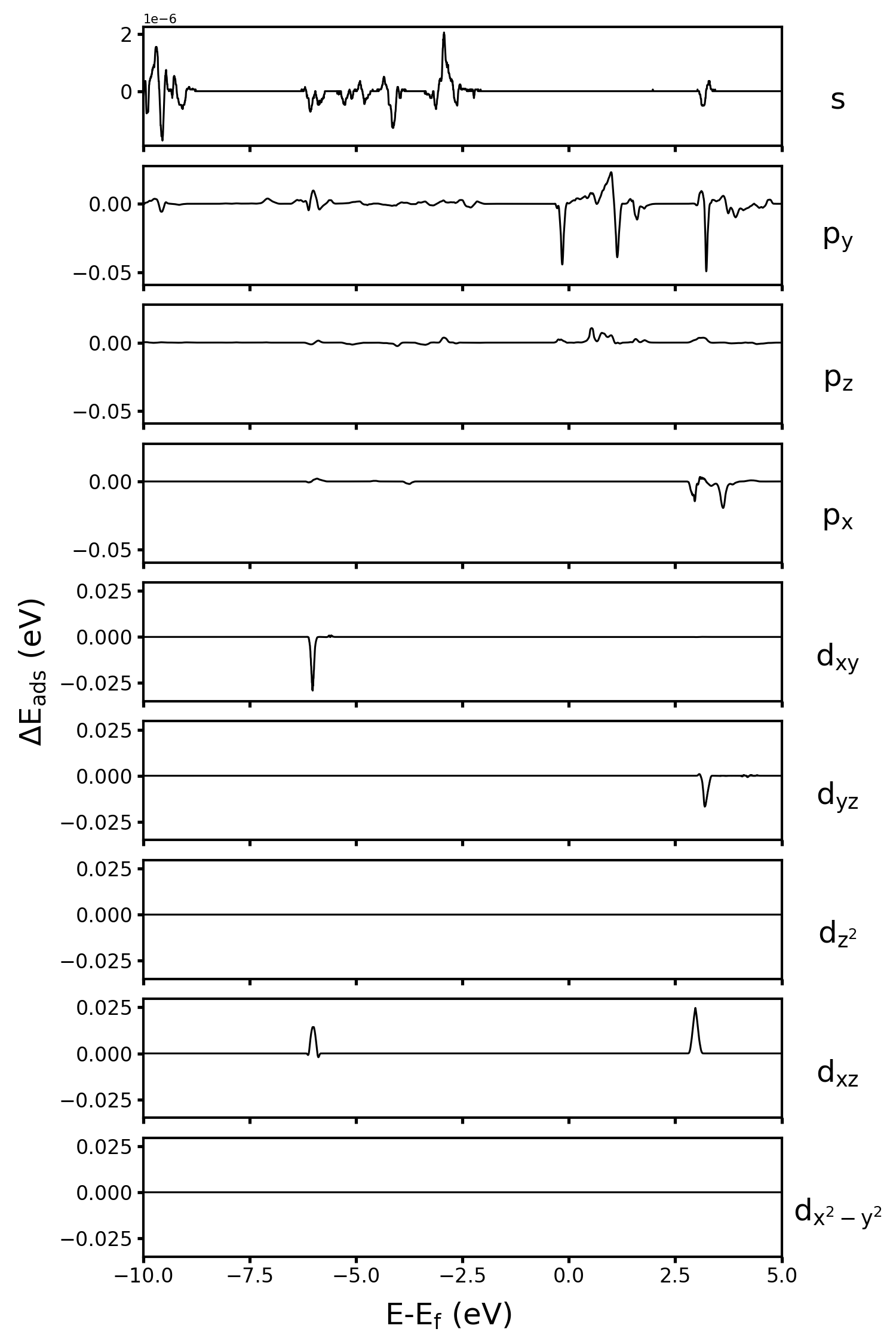}
  \caption{\textbf{Occlusion Experiment with a masker width of ``21''.}
  This figure shows an occlusion experiment on the Ge atom in
  the initial state of CO adsorption on Ge@g-C\textsubscript{3}N\textsubscript{4}.}
  \label{supp_fig26:occl_wid21}
\end{figure}

% SI Figure 27: Occlusion experiment with a masker width of 31
\begin{figure}[htbp]
  \centering
  \includegraphics[width=0.6\textwidth]{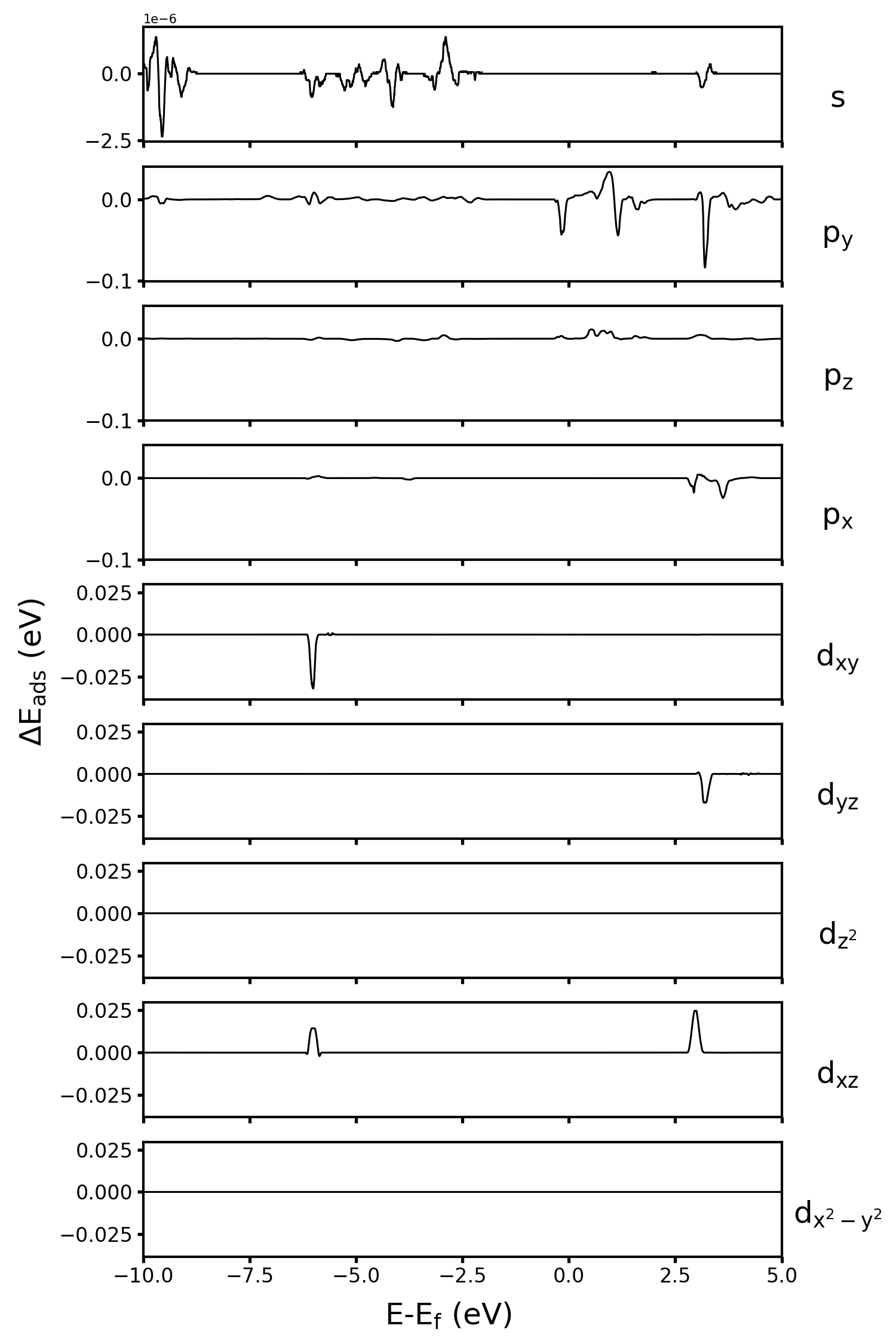}
  \caption{\textbf{Occlusion Experiment with a masker width of ``31''.}
  This figure shows an occlusion experiment on the Ge atom in
  the initial state of CO adsorption on Ge@g-C\textsubscript{3}N\textsubscript{4}.}
  \label{supp_fig27:occl_wid31}
\end{figure}

% SI Figure 28: Occlusion experiment with a masker width of 41
\begin{figure}[htbp]
  \centering
  \includegraphics[width=0.6\textwidth]{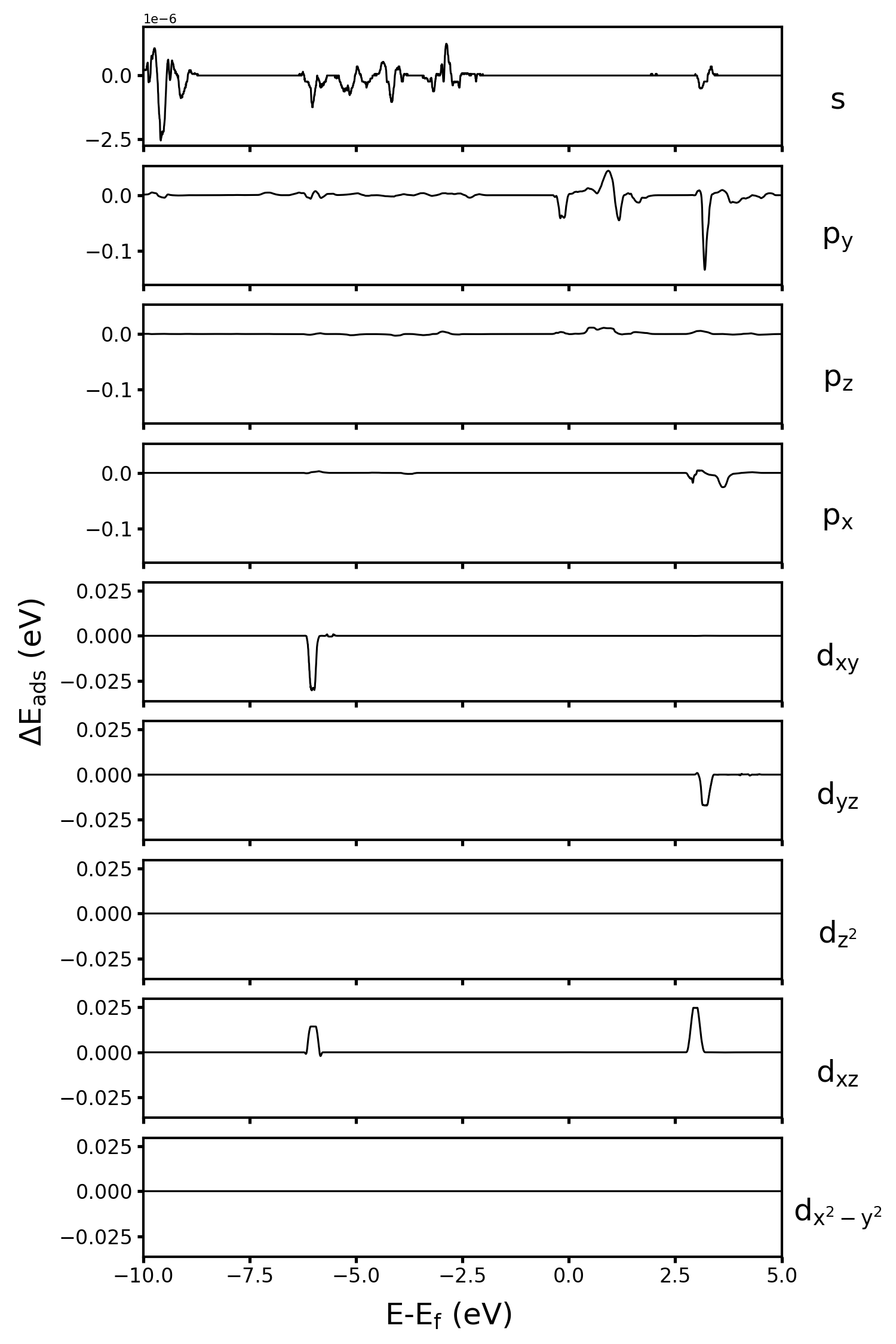}
  \caption{\textbf{Occlusion Experiment with a masker width of ``41''.}
  This figure shows an occlusion experiment on the Ge atom in
  the initial state of CO adsorption on Ge@g-C\textsubscript{3}N\textsubscript{4}.}
  \label{supp_fig28:occl_wid41}
\end{figure}

% SI Figure 29: Occlusion experiment with a masker width of 51
\begin{figure}[htbp]
  \centering
  \includegraphics[width=0.6\textwidth]{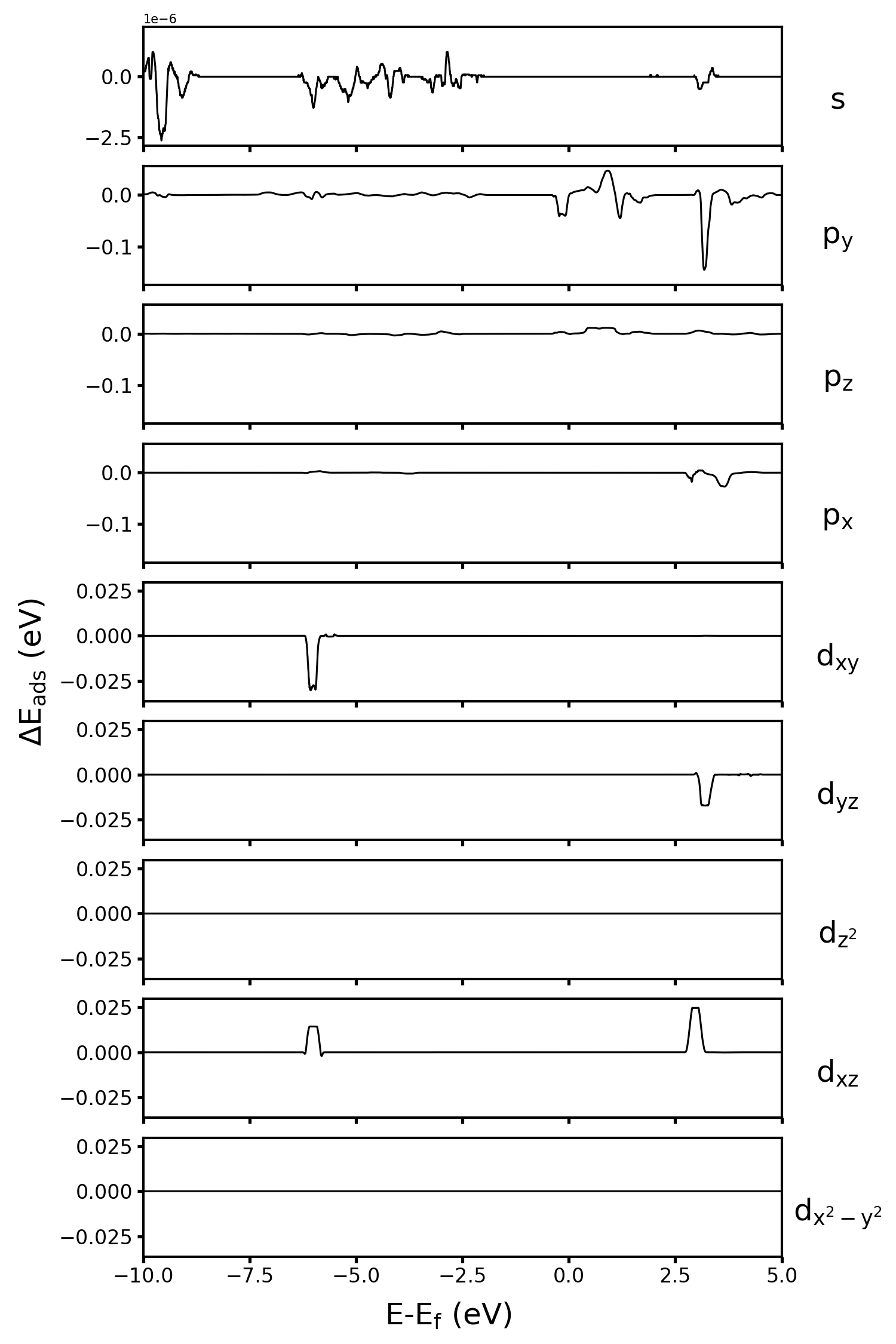}
  \caption{\textbf{Occlusion Experiment with a masker width of ``51''.}
  This figure shows an occlusion experiment on the Ge atom in
  the initial state of CO adsorption on Ge@g-C\textsubscript{3}N\textsubscript{4}.}
  \label{supp_fig29:occl_wid51}
\end{figure}

\subsection{Shifting experiment}
\label{supp_sec3.6_shifting}

% SI Figure 30: Illustration of the eDOS shifting experiment
\begin{figure}[htbp]
  \centering
  \includegraphics[width=\textwidth]{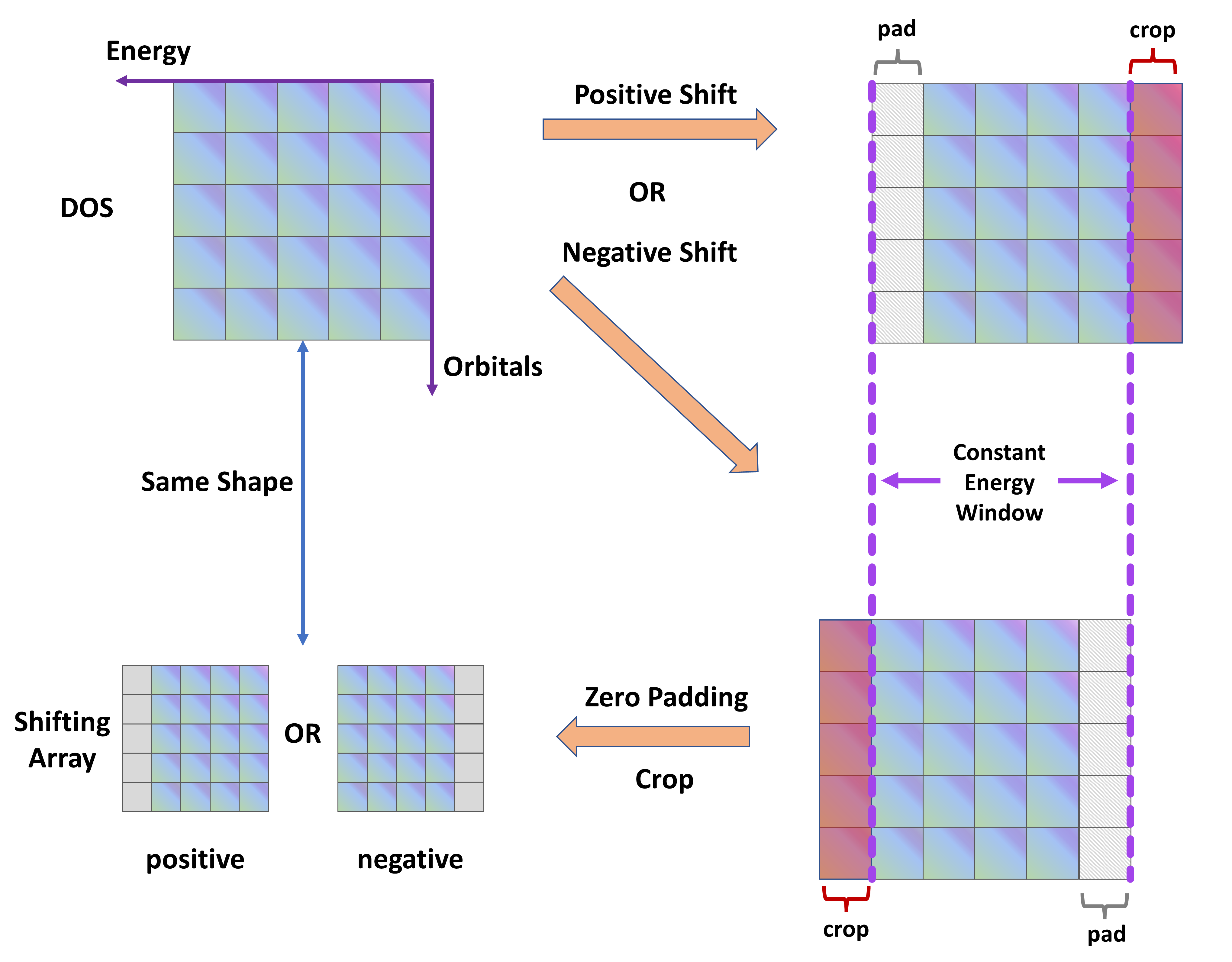}
  \caption{\textbf{Illustration of the eDOS shifting experiment.}
  This figure illustrates the sequential stages of the shifting experiment.
  Initially, the input eDOS array is shifted along the Energy axis.
  Subsequently, it undergoes zero-padding and cropping,
  ensuring a consistent shape based on the shifting direction, to match the input eDOS array.
  The resulting shifted array is then processed by the CNN model to
  assess the disturbance in adsorption energy caused by the shifting operation.}
  \label{supp_fig30:shifting}
\end{figure}

% SI Figure 31: Shifting experiment on p orbital
\begin{figure}[htbp]
  \centering
  \includegraphics[width=0.75\textwidth]{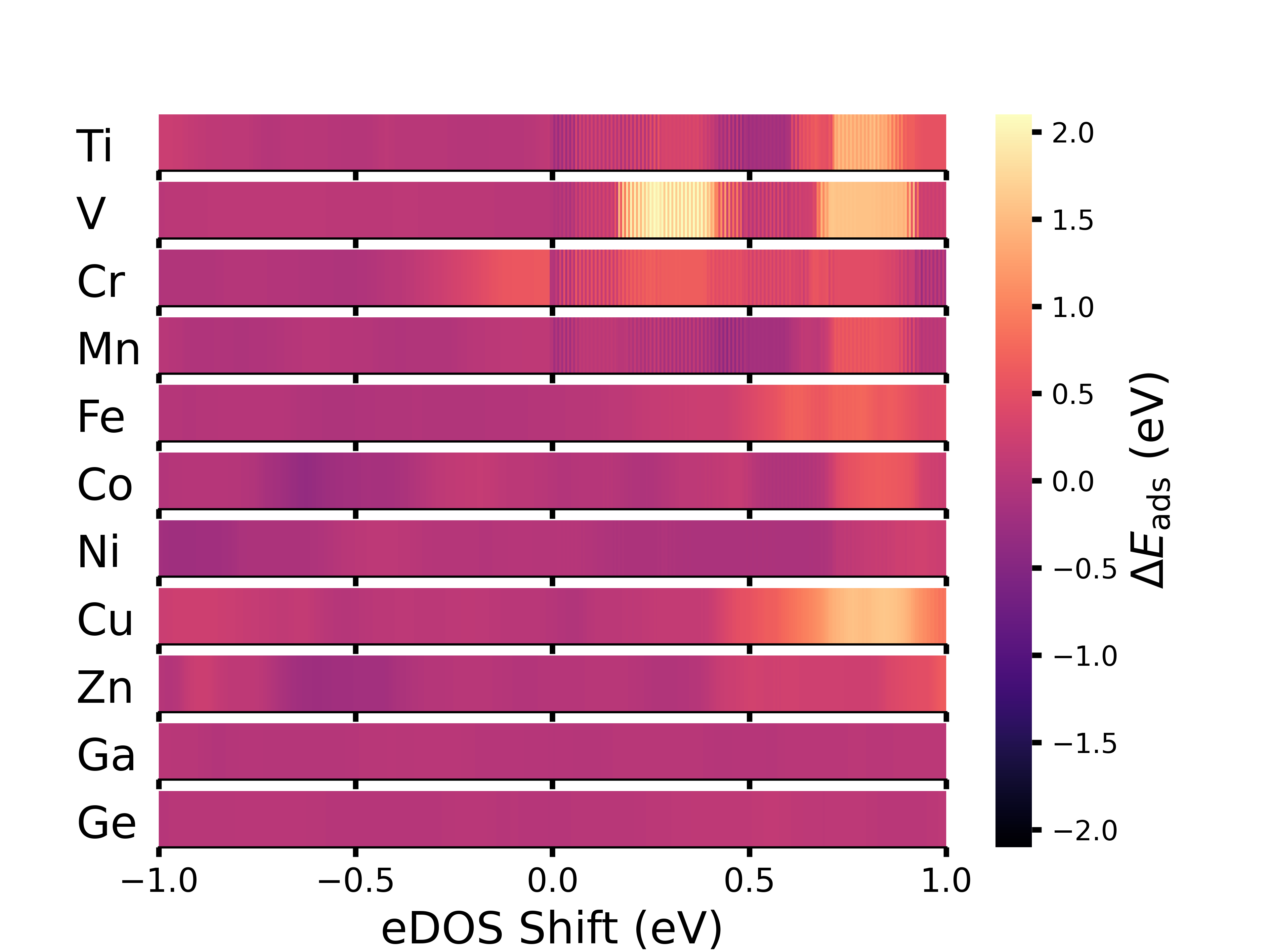}
  \caption{\textbf{Shifting experiment on p orbital.}
  Effect of orbital shifting on CO adsorption energy, predicted by the CNN model,
  for single metal atom catalysts supported on g-C\textsubscript{3}N\textsubscript{4}.
  The perturbations caused by shifting of entire p orbital are presented.
  The shifting step size corresponds to the energy resolution of the eDOS, set at 0.005 eV in this study.
  A positive eDOS shift indicates a shift towards higher energy levels, and vice versa.}
  \label{supp_fig31:p_shifting}
\end{figure}

\newpage
% SI Section Three: Other Additional Information

\section{Other additional information}
\label{supp_sec4_env}

\subsection{Additional information on machine learning environment}
\label{supp_sec4.1_ml_env}

The hyperparameter optimization for the proposed CNN model was performed using NVIDIA\textsuperscript{\textregistered} V100 Tensor Core GPUs,
courtesy of the Australian National Computational Infrastructure's (NCI) HPC Gadi system.
The HyperBand algorithm \cite{li2018hyperband}, as implemented in
KerasTuner \cite{omalley2019kerastuner}, facilitated the tuning process.

% Supplementary Table 19: Packages in the DL env
\begin{table}[htbp]
\label{supp_table19:pack_dl_env}
  \caption{Core packages in the deep learning environment.}
  \small
  \center
  \begin{tabularx}{0.9\textwidth}{@{}l *{3}{X} @{}}
    \toprule
    Name                 & Version  & Build                         & Channel      \\
    \midrule
    conda                & 4.13.0   & py39h06a4308\textunderscore0  & anaconda     \\
    cuda-nvcc            & 11.8.89  & 0                             & nvidia       \\
    cuda-toolkit         & 11.8.0   & 0                             & nvidia       \\
    cudatoolkit          & 11.6.0   & habf752d\textunderscore9      & nvidia       \\
    ipykernel            & 6.9.1    & py39h06a4308\textunderscore0  & anaconda     \\
    ipython              & 7.34.0   & pypi\textunderscore0          & pypi         \\
    jupyter\_client      & 7.2.2    & py39h06a4308\textunderscore0  & anaconda     \\
    jupyterlab           & 3.4.4    & py39h06a4308\textunderscore0  & anaconda     \\
    keras                & 2.9.0    & pypi\textunderscore0          & pypi         \\
    keras-preprocessing  & 1.1.2    & pypi\textunderscore0          & pypi         \\
    keras-tuner          & 1.3.5    &                               & pypi         \\
    matplotlib           & 3.5.1    & py39h06a4308\textunderscore1  & anaconda     \\
    numpy                & 1.22.4   & pypi\textunderscore0          & pypi         \\
    pandas               & 1.3.4    & py39h8c16a72\textunderscore0  & anaconda     \\
    python               & 3.9.12   & h12debd9\textunderscore1      & anaconda     \\
    scikit-learn         & 1.1.1    & py39h6a678d5\textunderscore0  & anaconda     \\
    scipy                & 1.8.1    & py39hddc5342\textunderscore3  & conda-forge  \\
    tensorboard          & 2.9.1    & pypi\textunderscore0          & pypi         \\
    tensorflow           & 2.9.3    & pypi\textunderscore0          & pypi         \\
    tensorflow-gpu       & 2.9.3    & pypi\textunderscore0          & pypi         \\
    xlrd                 & 2.0.1    & pyhd3eb1b0\textunderscore0    & anaconda     \\
    yaml                 & 0.2.5    & h7b6447c\textunderscore0      & anaconda     \\
    \bottomrule
  \end{tabularx}

  \smallskip

  \begin{flushright}
  \begin{minipage}{\textwidth}
    \footnotesize\textit{Note:} The complete list of packages is
      provided along with the source code.
  \end{minipage}
  \end{flushright}
\end{table}

\subsection{Additional information on data analysis and visualization environment}
\label{supp_sec4.2_vis_env}

Data analysis and visualization tasks were executed on an Apple MacBook Air featuring an octa-core M2 chip.
For accuracy, the TensorFlow-Metal plugin for Mac-based GPU acceleration was deliberately excluded,
as it could potentially yield erroneous CNN model predictions.

The illustrations of CNN architecture displayed in \cref{main_fig1:pipeline} of the main text,
as well as \cref{supp_fig21:occlusion} and \cref{supp_fig30:shifting},
were obtained from the GitHub repository dair-ai/ml-visuals \cite{Saravia_ML_Visuals_2021}.
Blender was used to create and render the visual representation of the
catalyst analysis pipeline shown in \cref{main_fig1:pipeline} of the main text.

% Supplementary Table 20: Packages in data analysis and vis env
\begin{table}[htbp]
\label{supp_table20:pack_vis_env}
  \caption{Core packages in data analysis and visualization environment.}
  \small
  \center
  \begin{tabularx}{0.9\textwidth}{@{}l *{3}{X} @{}}
    \toprule
    Name              & Version  & Build                     & Channel  \\
    \midrule
    bokeh             & 3.2.2    & pypi\textunderscore0      & pypi     \\
    colorcet          & 3.0.1    & pypi\textunderscore0      & pypi     \\
    heatmapz          & 0.0.4    & pypi\textunderscore0      & pypi     \\
    keras             & 2.13.1   & pypi\textunderscore0      & pypi     \\
    keras-tuner       & 1.3.5    & pypi\textunderscore0      & pypi     \\
    matplotlib        & 3.7.2    & pypi\textunderscore0      & pypi     \\
    numpy             & 1.24.3   & pypi\textunderscore0      & pypi     \\
    pandas            & 2.0.3    & pypi\textunderscore0      & pypi     \\
    python            & 3.11.4   & hb885b13\textunderscore0  &          \\
    pyyaml            & 6.0.1    & pypi\textunderscore0      & pypi     \\
    scikit-learn      & 1.3.0    & pypi\textunderscore0      & pypi     \\
    scipy             & 1.11.1   & pypi\textunderscore0      & pypi     \\
    seaborn           & 0.12.2   & pypi\textunderscore0      & pypi     \\
    tensorboard       & 2.13.0   & pypi\textunderscore0      & pypi     \\
    tensorflow        & 2.13.0   & pypi\textunderscore0      & pypi     \\
    tensorflow-macos  & 2.13.0   & pypi\textunderscore0      & pypi     \\
    \bottomrule
  \end{tabularx}

  \smallskip

  \begin{flushright}
  \begin{minipage}{\textwidth}
    \footnotesize\textit{Note:} The complete list of packages
      is provided along with the source code.
  \end{minipage}
  \end{flushright}
\end{table}

% End of Supplementary Information

% Bibliography
\newpage
\bibliographystyle{plain}
\bibliography{references}

\end{document}